\documentclass[preprint]{aastex}
\usepackage{emulateapj5}
\usepackage{apjfonts}
\usepackage{amssymb, amsmath}
\usepackage{natbib}
\usepackage{graphicx}
\usepackage{CJK}

\input psfig.tex

\newdimen\digitwidth      
\setbox1=\hbox{0}       
\digitwidth=\wd1        
\catcode`"=\active      

\def"{\kern\digitwidth}

\def\aa{{A\&A}}
\def\aas{{ A\&AS}}
\def\aj{{AJ}}

\def\amin{$^\prime$}
\def\annrev{{ARA\&A}}
\def\apj{{ApJ}}
\def\apjs{{ApJS}}
\def\asec{$^{\prime\prime}$}
\def\deg{$^{\circ}$}

\def\lax{{$\mathrel{\hbox{\rlap{\hbox{\lower4pt\hbox{$\sim$}}}\hbox{$<$}}}$}}
\def\gax{{$\mathrel{\hbox{\rlap{\hbox{\lower4pt\hbox{$\sim$}}}\hbox{$>$}}}$}}
\def\simlt{\lower.5ex\hbox{$\; \buildrel < \over \sim \;$}}
\def\simgt{\lower.5ex\hbox{$\; \buildrel > \over \sim \;$}}

\def\mnras{{MNRAS}}
\def\nat{{Nature}}

\def\sb{mag~arcsec$^{-2}$}

\def\solmass{$M_\odot$}

\def\galfit{{\tt GALFIT}}
\def\ser{{S\'{e}rsic}}

\slugcomment{Draft version 6}
\shorttitle{THE CARNEGIE-IRVINE GALAXY SURVEY. III.}
\shortauthors{HUANG ET AL.}

\begin{document}

\begin{CJK*}{UTF8}{gbsn}

\title{The Carnegie-Irvine Galaxy Survey. III. The Three-Component Structure 
of Nearby Elliptical Galaxies}

\author{Song Huang (黄崧)\altaffilmark{1,2,3} Luis C. Ho\altaffilmark{2}, 
Chien Y. Peng\altaffilmark{4}, Zhao-Yu Li (李兆聿)\altaffilmark{5}, and 
Aaron J. Barth\altaffilmark{6} }
\date{}                                          

\altaffiltext{1}{School of Astronomy and Space Science, Nanjing University,
Nanjing 210093, China}

\altaffiltext{2}{The Observatories of the Carnegie Institution for Science, 
813 Santa Barbara Street, Pasadena, CA 91101, USA}

\altaffiltext{3}{Key Laboratory of Modern Astronomy and Astrophysics, Nanjing
University, Nanjing 210093, China}

\altaffiltext{4}{Giant Magellan Telescope Organization, 251 South Lake Avenue,
Suite 300, Pasadena, CA 91101, USA}

\altaffiltext{5}{Key Laboratory for Research in Galaxies and Cosmology,
Shanghai Astronomical Observatory, Chinese Academy of Sciences,
80 Nandan Road, Shanghai 200030, China}

\altaffiltext{6}{Department of Physics and Astronomy, 4129 Frederick Reines 
Hall, University of California, Irvine, CA 92697-4575, USA}

\begin{abstract}

Motivated by recent developments in our understanding of the formation and 
evolution of massive galaxies, we explore the detailed photometric structure 
of a representative sample of 94 bright, nearby elliptical galaxies, using 
high-quality optical images from the Carnegie-Irvine Galaxy Survey.  The 
sample spans a range of environments and stellar masses, from $M_* = 
10^{10.2}$ to $10^{12.0} \, M_{\odot}$.  We exploit the unique capabilities 
of two-dimensional image decomposition to explore the possibility that local 
elliptical galaxies may contain photometrically distinct substructure that 
can shed light on their evolutionary history. Compared with the traditional 
one-dimensional approach, these two-dimensional models are capable of 
consistently recovering the surface brightness distribution and the systematic 
radial variation of geometric information at the same time. Contrary to 
conventional perception, we find that the global light distribution of the 
majority (\gax 75\%) of elliptical galaxies is not well described by a single 
\ser\ function.  Instead, we propose that local elliptical galaxies generically 
contain three subcomponents: a compact ($R_e$ \lax $1$ kpc) inner component with
luminosity fraction $f \approx 0.1-0.15$; an intermediate-scale 
($R_e \approx 2.5$ kpc) middle component with $f \approx 0.2-0.25$; and a 
dominant ($f = 0.6$), extended ($R_e \approx 10$ kpc) outer envelope.  All
subcomponents have average \ser\ indices $n \approx 1-2$, significantly lower 
than the values typically obtained from single-component fits.  The individual 
subcomponents follow well-defined photometric scaling relations and the stellar
mass-size relation.  We discuss the physical nature of the substructures and
their implications for the formation of massive elliptical galaxies.

\end{abstract}

\maketitle

\section{Introduction}

The formation and evolution of massive elliptical galaxies, as fundamentally 
important areas in astrophysics, are extensively studied using different 
observational constraints.  Among them, the photometric analysis of their
morphological structure remains one of the most efficient and straightforward 
methods.  Since different physical processes can result in different stellar 
configurations, the surface brightness distribution of a galaxy, as the 
first-order proxy of the underlying stellar mass distribution, provides an 
important archaeological record of its evolutionary history.

With their nearly featureless morphologies, elliptical galaxies traditionally 
have been viewed as structurally simple objects.  Historically, the surface 
brightness profile of luminous elliptical galaxies was described by an $R^{1/4}$
law (de~Vaucouleurs 1948, 1953). This simple profile reflects the high central 
concentration and large radial extension of the light distribution of massive 
elliptical galaxies. As an empirical surface brightness model, de~Vaucouleurs 
profile was embraced by different theories for elliptical galaxy formation. Both 
the dissipational monolithic collapse model (e.g., Larson 1974) and the 
initially dissipationless disk merger model (e.g., van Albada 1982; Carlberg 
1986) predicted that the density profile of the final product should be very
similar to the $R^{1/4}$ law.  With the advent of more accurate photometric 
data, however, the surface brightness profile of elliptical galaxies was found 
to deviate systematically from de~Vaucouleurs' profile (Kormendy 1977; King 
1978; Lauer 1985). These observations showed that elliptical galaxies have a 
variety of inner profiles that cannot be accommodated by de~Vaucouleurs' law. 

Beginning in the late 1980s, it was realized that the more flexible $R^{1/n}$ 
\ser\ (1968) profile does a much better job of describing the surface brightness 
profile of elliptical galaxies (Capaccioli 1989; Caon et al. 1993). Over a large 
radial range, nearby elliptical galaxies were found to be well fit by \ser\ 
profiles with $2.5$ \lax\ $n$ \lax\ 10 (e.g., Caon et al. 1994; Bertin et al. 
2002; Kormendy et al. 2009, hereinafter K09).  Compared to other types of 
galaxies, ellipticals have significantly larger \ser\ indices, reflecting their
high degree of central concentration.  As the detailed properties of the \ser\ 
function became better understood, both theoretically (Ciotti 1991; Hjorth \& 
Madsen 1995; Ciotti \& Bertin 1999; Lima Neto et al. 1999) and observationally 
(Graham \& Driver 2005), a variety of studies have explored empirical 
correlations between the \ser\ index and other key structural parameters (e.g., 
Caon et al. 1993; Trujillo et al. 2001; Graham et al. 2005).  Modern numerical
simulations of galaxy formation that invoke either dissipational monolithic 
collapse (Nipoti et al. 2006) or dissipational merging (Aceves et al. 2006; 
Naab \& Trujillo 2006) can recover the variety of \ser\ shapes observed in 
ellipticals.

Neither monolithic collapse nor dissipational mergers provides an adequate 
description of the full evolutionary process of massive elliptical galaxies. 
Recent observations show that massive, quiescent early-type galaxies have been 
in place since $z \approx 2.5$ (Cimatti et al. 2004, 2008; McCarthy et al.
2004). High-redshift massive galaxies have a great diversity in properties.  Of 
particular interest is the population of ``red nuggets,'' which appear to be 
nascent, evolutionary precursors of local giant ellipticals. On average, these
quiescent galaxies have smaller size (Daddi et al. 2005; Trujillo et al. 2006;
Toft et al. 2007; van~der~Wel et al. 2008; van~Dokkum et al. 2008, 2009; 
Damjanov et al. 2009; Cassata et al. 2010, 2011; Szomoru et al. 2012) and 
possibly higher central stellar velocity dispersion (Cappellari et al. 2009; 
Cenarro \& Trujillo 2009; van~Dokkum et al. 2009; Onodera et al. 2010; 
van~de~Sande et al. 2011) than local elliptical galaxies of the same stellar 
mass.  On average the massive early-type galaxy population appears to require a
doubling in stellar mass and tripling in size during the epoch between $z = 2.0$ 
and 1.0 (van~Dokkum et al. 2010; see also Cimatti et al. 2012).  Stacking 
analysis of large samples of early-type galaxies at different redshift bins 
(Bezanson et al. 2009; van~Dokkum et al. 2010) and deep near-infrared images of
individual objects (Szomoru et al. 2012) reveal that high-redshift red nuggets 
differ from local ellipticals in two ways: they lack an extended outer envelope 
while at the same time their inner stellar mass density is higher.  These 
observational constraints strongly challenge theoretical scenarios wherein 
either dissipational collapse or gas-rich major mergers dominate during the main 
phases of elliptical galaxy formation. Dissipational processes lead to greater 
central concentration, not envelope building as required by the observed 
dramatic size growth. Whatever the mechanism, there does not appear to be a high
enough rate of major mergers to do the job since $z\approx2.0$ (Williams et al.
2011; Nipoti et al. 2012)

A variety of different physical explanations have been proposed to explain 
the above recent observations, including dry (minor) mergers (Naab et al. 
2007, 2009), fading gas-rich mergers (Hopkins et al. 2009b), mergers plus 
selection effects (van~der~Wel et al. 2009), and feedback by active galactic 
nuclei (Fan et al. 2008).  Among these models, the two-phase formation 
scenario of Oser et al. (2010, 2012; see also Johansson et al. 2012) seems 
particularly promising in matching a number of observational constraints 
(Coccato et al. 2010; Greene et al. 2012; Papovich et al. 2012; Romanowsky 
et al. 2012), numerical simulations (Gabor \& Dav{\'e} 2012), and 
semi-analytic models (Khochfar et al. 2011). Under this picture, the initial
evolution phase at high redshifts ($z \le 3$) is dominated by in situ star 
formation induced by cold-phase accretion and gas-rich mergers, while 
later evolution is instead dominated by dissipationless processes such as 
dry (minor) mergers and hot-phase accretion.  It is the second phase that is 
responsible for the build-up of the low-density outer envelope of local 
massive ellipticals. 

Elliptical galaxies are not as simple as they seem.  Fine structures such as 
shells and tidal features have been identified in some ellipticals even 
before the era of CCD photometry (Malin 1979; Schweizer 1980, 1982; Malin \& 
Carter 1983). Deep, modern images of nearby elliptical galaxies show that the 
majority of them, in fact, contain extended, low-surface brightness features 
consistent with a tidal origin (Tal et al. 2009).  On smaller scales, dust 
absorption features, including regular dust lanes, are frequently visible 
(Lauer 1985; Ebneter et al. 1988), as are embedded nuclear stellar disks 
(Carter 1978; Bender 1988; Franx et al. 1989; Peletier et al. 1990; Ledo et al. 
2010).  These substructures are often attributed to merger or accretion 
events.  Even when the overall light distribution of ellipticals appears smooth,
it is not uncommon to see isophotal twists (Kormendy 1982; Fasano \& Bonoli 
1989), which often cannot be explained as an artifact of dust absorption.  Nieto 
et al. (1992) suggested that structural subcomponents may give rise to the 
isophotal twists observed in ellipticals.

The innermost regions of nearby elliptical galaxies contain photometricially
distinct components.  This was recognized even from ground-based images 
(Lauer 1985; Kormendy 1985) and has been fully elaborated with higher 
resolution data from \emph{HST} (e.g., Jaffe et al. 1994; Lauer et al. 1995, 
2007; C\^{o}t\'e et al. 2006; Ferrarese et al. 2006). Depending on the 
relative steepness of the inner luminosity profile compared to the global 
one, the central regions of nearby elliptical galaxies have been classified 
into ``core'' and ``power law'' (or ``extra-light'') types, whose incidence 
depends strongly on galaxy luminosity (Faber et al. 1997; Lauer et al. 2007; 
K09): steep power-law cusps are found mostly in lower-luminosity systems, 
whereas lower density cores reside exclusively in the most massive, luminous
ellipticals.  Various empirical functions have been used to describe these 
inner ``deviations'' of the surface brightness profile.  The most popular 
has been the ``Nuker'' law (Lauer et al. 1995), but other forms have been 
used (Ferrarese et al. 1994; Graham et al. 2003; Trujillo et al. 2004; 
C\^{o}t\'e et al. 2006; Hopkins et al. 2009a).  

At the another extreme end in scale, the most luminous members of the 
elliptical galaxy family are known to have extended stellar envelopes (Morgan 
\& Lesch 1965; Oemler 1974; Schombert 1986).  This phenomenon is especially 
common for the brightest cluster galaxies. 
It is also known that in certain cases these extended envelopes 
around the brightest cluster galaxies can have a multiple-component structure
(e.g., NGC 3311, the brightest cluster galaxy in the Hydra I cluster; Arnaboldi 
et al. 2012) and/or an off-center appearance (e.g., the brightest cluster galaxy 
of Abell 1651; Gonz\'alez et al. 2005).
Outer envelopes that cannot be fitted well using a \ser\ profile often yield a 
larger \ser\ index in a single-component fit. Some authors have adopted more 
complex profiles to account for this additional structure (e.g., Graham et al.
1996; Gonz\'alez et al. 2005; Zibetti et al. 2005; Seigar et al. 2007). In 
general, however, ellipticals with extended envelopes are considered to be rare,
special cases.

Despite the abundant evidence that elliptical galaxies do have structural 
complexities on different scales, it is still customary to regard the bulk of 
the stellar surface brightness distribution---excluding the central 
``deviations''---in these systems as a single entity described by a single 
\ser\ profile.  In light of recent observational and theoretical developments on 
the formation and evolution of massive galaxies, we believe that it would be 
instructive to take a fresh look at this problem.  We take as our starting point
a collection of high-quality optical images available for a relatively large, 
well-defined sample of nearby, bright elliptical galaxies selected from the 
Carnegie-Irvine Galaxy Survey\footnote{{\tt http://cgs.obs.carnegiescience.edu}}
(CGS: Ho et al. 2011, Paper~I; Li et al. 2011, Paper~II).  We carefully analyze 
these data with a two-dimensional (2-D) image decomposition method, 
demonstrating that the majority of nearby ellipticals contain three well-defined 
subcomponents. The physical implications of these substructures are discussed in 
a companion paper by S. Huang et al. (in preparation).

\vskip 0.3cm
\begin{figure*}[t]
\centerline{\psfig{file=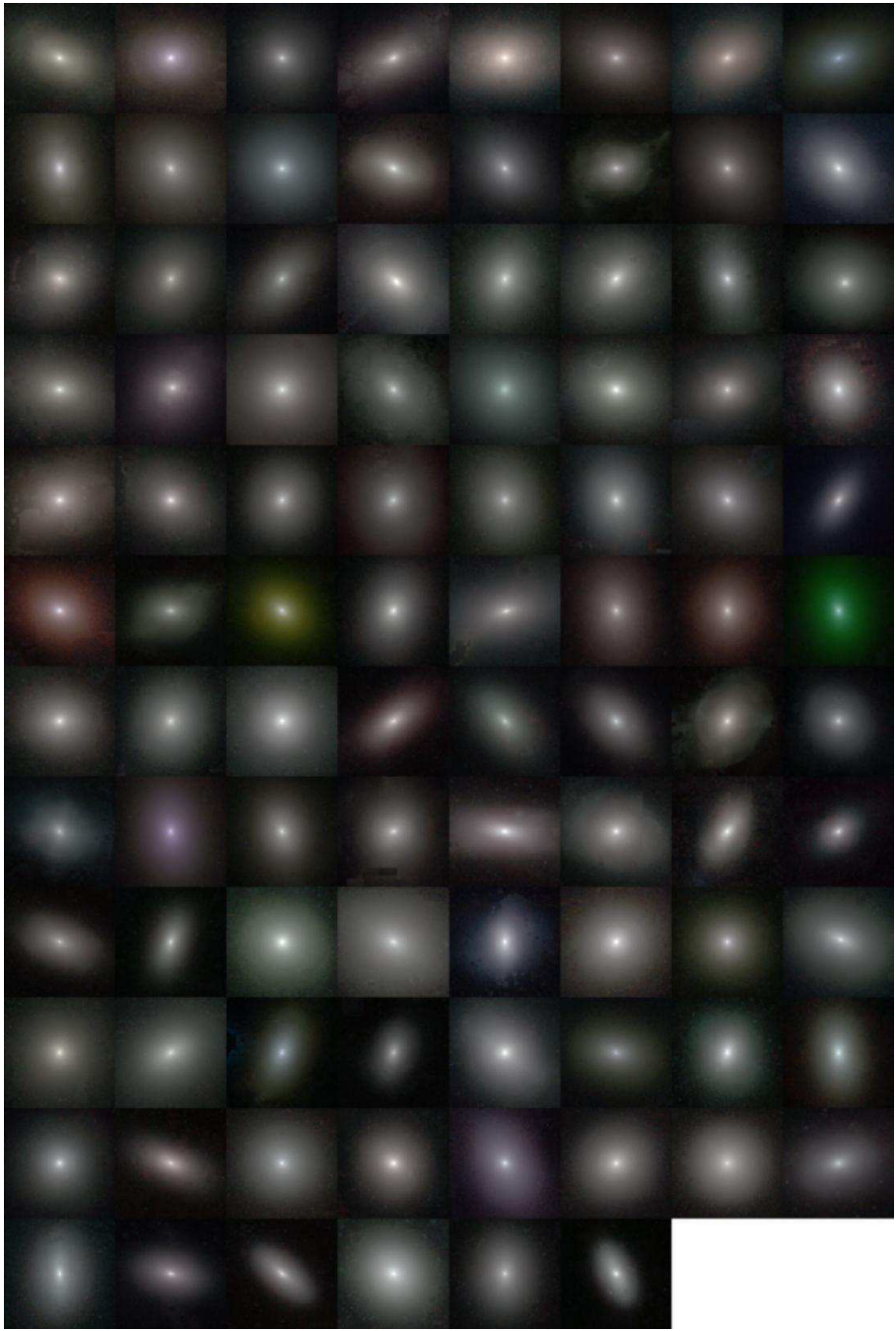,width=12.0cm}}
\figcaption[fig1.eps]{
Mosaic of three-color, star-cleaned images of the 94 elliptical galaxies in our 
sample. The dimensions of each panel vary, with each side corresponding to $1.5
D_{25}$; north is up and east is to the left.  From upper-left to bottom-right,
the galaxies are arranged in the order of decreasing $M_{{\it B}_T}$. The image
colors here are not directly related to the actual galaxy colors, and the 
discrepant colors are seen in a few objects having data of lower signal-to-noise
ratio.
\label{fig1}}
\end{figure*}
\vskip 0.3cm

This paper is organized as follows. Section~2 gives a brief overview of the 
elliptical galaxy sample, the observations, and basic data reduction procedures. 
In Section~3, we give a description of our 2-D image decomposition method, sky
background measurement, and other details related to the model fitting.  
Section~4 summarizes the main properties of the model parameters, the 
distributions of key structural parameters, and several important scaling
relations. The physical nature of the galaxy subcomponents is discussed in 
Section~5 within the framework of elliptical galaxy formation. Section~6 briefly
mentions interesting future directions, ending with a summary in Section~7.

As in other papers in the CGS series, we assume $H_0$ = 73~km~s$^{-1}$ 
Mpc$^{-1}$, ${\Omega}_m=0.27$, and ${\Omega}_{\Lambda}=0.73$.

\section{Sample and Data}

CGS is a long-term project to investigate the photometric and spectroscopic 
properties of nearby galaxies (Ho et al. 2011). The first part of the project is 
an optical photometric survey of galaxies that span the full range of the Hubble 
sequence in the local Universe.  The sample consists of a statistically
complete 

\vskip 2.5cm
\centerline{\psfig{file=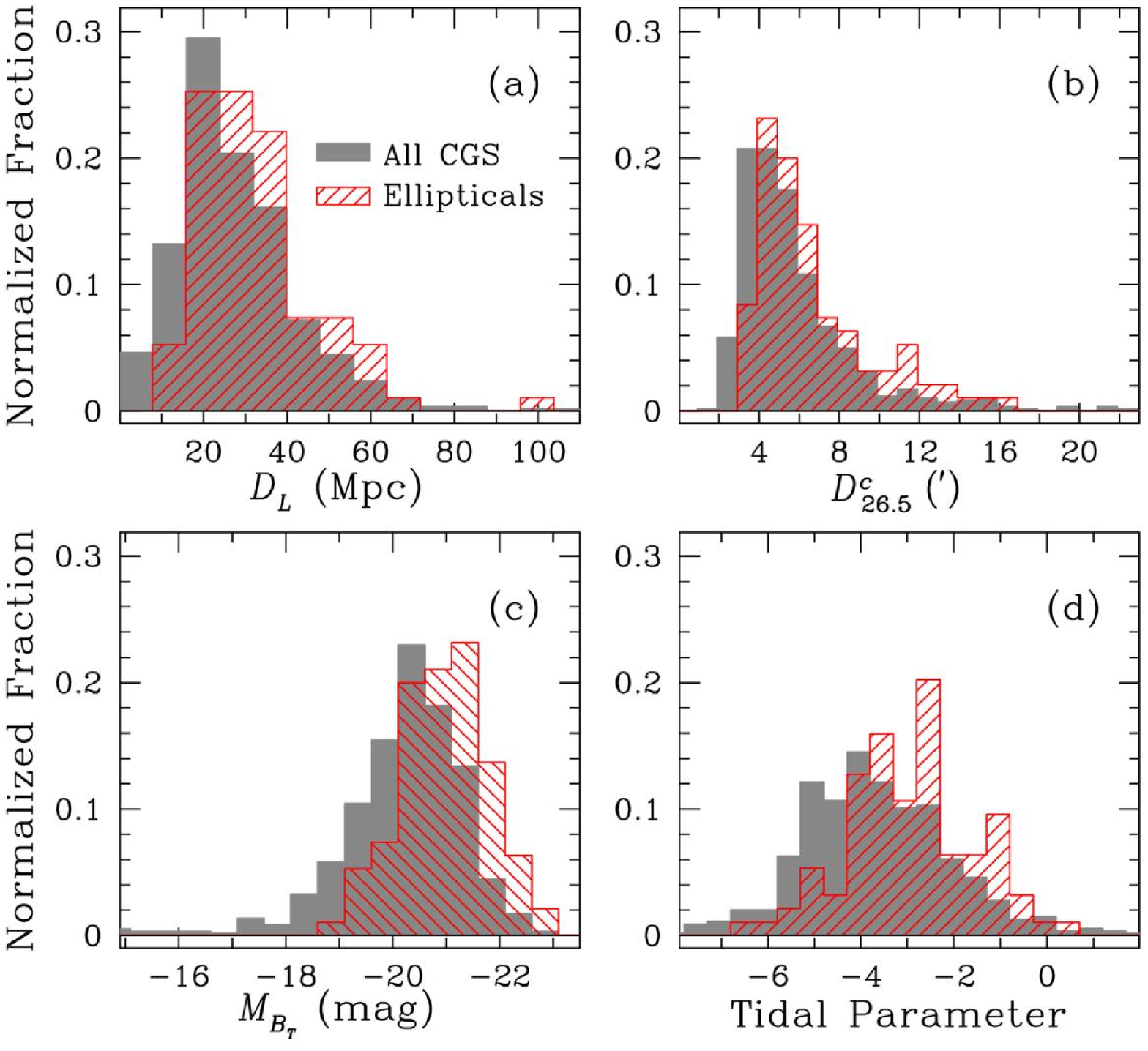,angle=270,width=8.9cm}}
\figcaption[fig2.eps]{
Normalized distributions of several properties for the 
whole CGS sample and the 94 elliptical galaxies sample in this work:
(a) luminosity distance; (b) corrected isophotal diameter corresponding 
to an $I$-band surface brightness of 26.5 \sb; (c) $B$-band absolute total 
magnitude, corrected for Galactic extinction; (d) tidal parameter. The detailed 
definition of the tidal parameter can be found in Paper~I. All the data are 
taken from Paper~I.
\label{fig2}}
\vskip 0.5cm

\noindent
sample of 605 galaxies drawn from the Third Reference Catalogue of Bright 
Galaxies (RC3; de~Vaucouleurs et al. 1991) and is defined by $B_{T} \leq 12.9$ 
mag and $\delta < 0$\deg. As most of the galaxies are nearby (median $D_L$ = 
24.9 Mpc), luminous (median $M_{B_T} = -20.2$ mag, corrected for Galactic 
extinction), and

\vskip 0.3cm
\centerline{\psfig{file=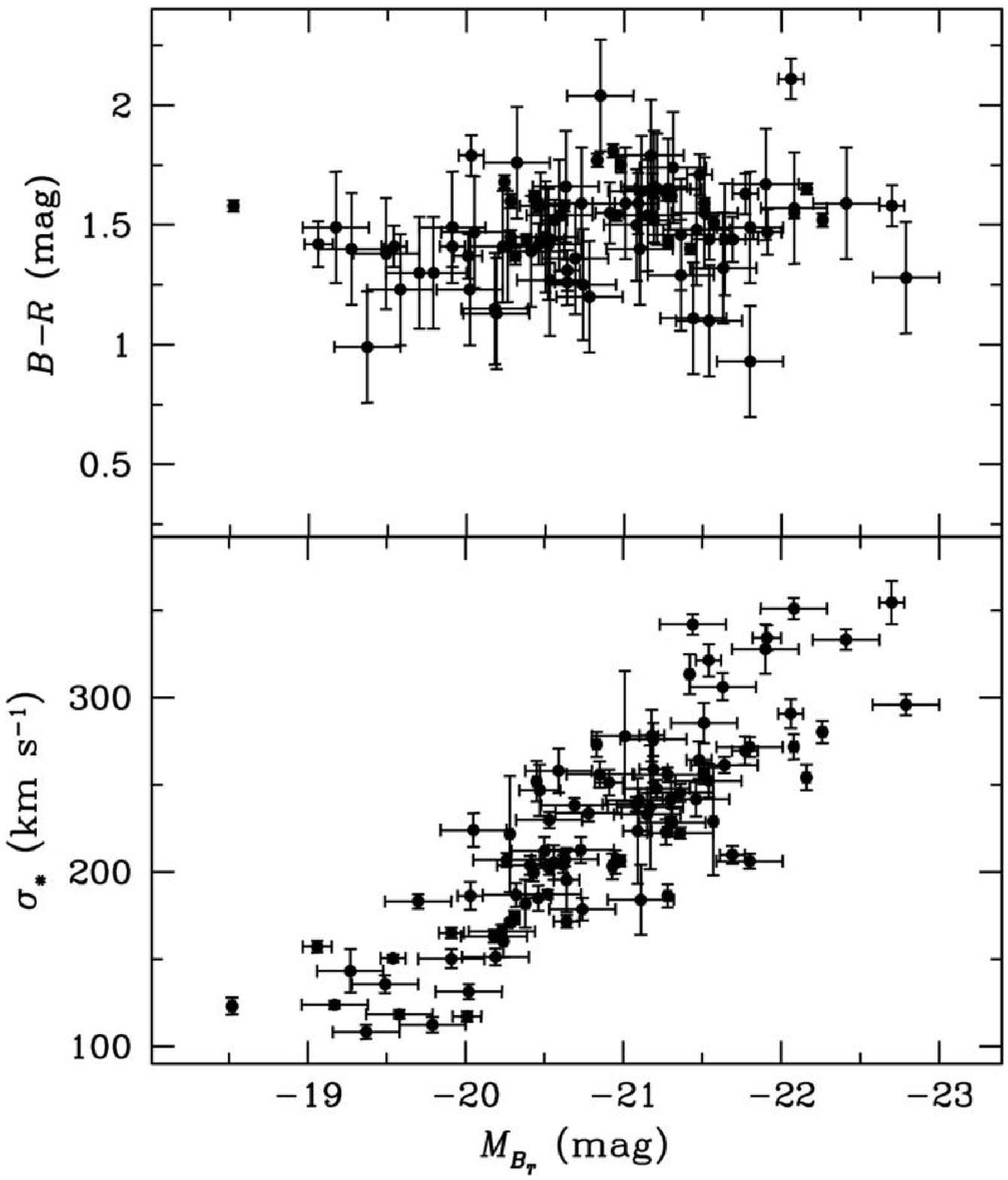,width=8.75cm}}
\figcaption[fig3.eps]{
The color-magnitude (top) and the Faber-Jackson (bottom) relation for the 
elliptical galaxies in our sample. The total $B$-band absolute magnitude is used
to represent the luminosity.
\label{fig3}}
\vskip 0.4cm

\vskip 0.4cm
\centerline{\psfig{file=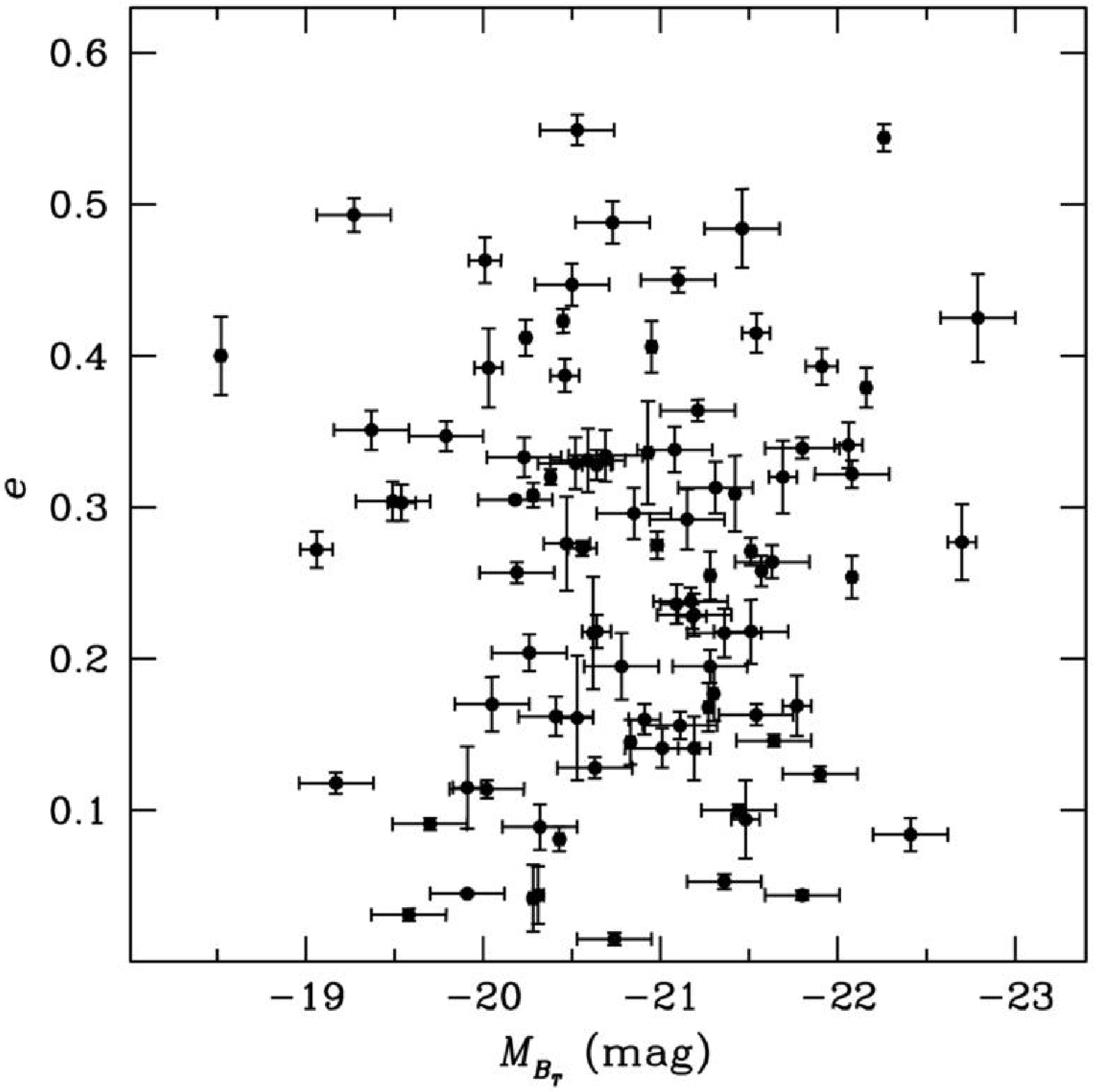,width=8.90cm}}
\figcaption[fig4.eps]{
The distribution of our sample of elliptical galaxies on the plane of total 
luminosity and average ellipticity as described in Paper~II. The uniform 
distribution on this plane indicates that our sample is representative of the 
elliptical galaxy population within this luminosity range.
\label{fig4}}
\vskip 0.8cm

\noindent
angularly large (median $B$-band isophotal diameter $D_{25}$ = 3\farcm3),
it is feasible to obtain high signal-to-noise ratio images with good spatial
resolution. Visually selecting a sample of ``pure'' ellipticals, even for nearby
galaxies, is challenging.  Ellipticals can be easily confused with face-on,
unbarred S0 galaxies.  We begin with the 103 galaxies in CGS originally 
classified as ellipticals in the RC3.  Nine galaxies were rejected for various
reasons: NGC~6758 does not have sufficient data; ESO~356-G004 (Fornax) is a 
spheroidal galaxy; ESO~495-G021 (Henize 2-10) is a very active star-bursting 
dwarf; and six (NGC~3309, 3311, 4105, 5090, 5193, 7176) are in merging or 
strongly interacting systems.  An additional object, NGC~2974, is severely 
contaminated by a nearby bright star, which makes image decomposition very 
unreliable; we do not rejected it from the final sample, but we exclude it from 
detailed image decomposition.  The final sample considered in this paper 
contains 94 galaxies deemed to be ellipticals in the RC3, whose morphological 
assignments were based on visual classification.  As will be discussed later, a 
few of the galaxies turn out to be misclassified S0s. We do not remove these a 
priori, as we are interested in knowing how well we can separate them using our 
image decomposition method, in light of recent kinematic studies (Emsellem 
et al. 2007; Cappellari et al. 2011a) that have blurred the morphological 
boundary between E and S0 galaxies.  In any case, the small number of 
potentially misclassified Es do not change any of our main statistical 
conclusions.

Although our final sample consists of relatively isolated ellipticals, seven 
of them (IC~3370, NGC~596, 1340, 1700, 2865, 3923, 5018) clearly show shell or 
tidal features in their $V$-band images. NGC~3923, in fact, is a prototypical
Type~I shell elliptical galaxy (Malin \& Carter 1980), and the others are 
known in the literature as post-starburst or ``young'' ellipticals. Using 
looser criteria, our visual examination of the images reveals another seven 
that may show signs of tidal or shell-like structures, bringing the total 
frequency of Es with such features to $\sim$15\% within CGS.  Given the modest 
surface brightness sensitivity and limited field of view (FoV) of our images, 
this percentage is, of course, only a lower limit.  Some of our galaxies overlap
with those in the OBEY survey\footnote{{\tt http://www.astro.yale.edu/obey/}} 
(Tal et al. 2009), whose images have a larger FoV and are $\sim$2 mag deeper in
$V$.  It is interesting to note that, after applying contrast enhancement 
techniques (e.g., adaptive histogram equalization; Peng et al. 2002), we can 
detect in the CGS $V$-band images most of the low-surface brightness features 
seen in the OBEY data.  We will not discuss these low-surface brightness 
features further in this paper.  We simply note that none of these low-level
features significantly impact our image decomposition of the light distribution 
in the main body of the galaxy.

Figure~1 presents a montage of the three-color, star-cleaned images of the 
final sample of 94 elliptical galaxies.  The galaxies are arranged in the 
order of decreasing absolute $B$-band absolute magnitude.  Some basic properties 
of the sample, compared to the parent CGS, are shown in Figure~2.  According to 
the environment information presented in Paper~I, most (80\%) of the E sample
lives in group-like environments; only six are in a cluster (Fornax), and 14 can 
be considered field galaxies.  Overall, the CGS objects form a representative 
sample of bright, local Es.  Figure~3 illustrates that the CGS Es populate a 
well-defined red sequence and define a relatively tight Faber-Jackson (1976) 
relation.  Our sample uniformly populates the luminosity-ellipticity plane 
(Figure~4), as does other well-studied, homogeneous samples of elliptical 
galaxies, such as those contained in the Atlas$^{\rm 3D}$ project (Cappellari 
et al. 2011b; see also de~Zeeuw et al.  2002). 

Papers~I and II give full details of the observations and data reductions, 
which will not be repeated here.   The overall quality of the CGS images is 
quite high, both in terms of resolution (pixel scale 0\farcs259; median seeing 
$\sim 1$\asec) and FoV (8\farcm9$\times$8\farcm9).  The surface brightness 
sensitivity of the images, limited principally by the accuracy of sky 
subtraction, reaches $\sim 27.5$, 26.9, 26.4, and 25.3 mag~arcsec$^{-2}$ in 
the $B$, $V$, $R$, and $I$ bands, respectively.  This work will only use the 
$V$-band images. Although the $V$ band is not the best tracer of the underlying
old stellar population, it suffers the least from the ``red halo'' effect on the 
point-spread function (PSF) (Appendix~A; also see Michard 2002 and Wu et al. 
2005).  Compared to the $R$ and $I$ bands, the $V$ band has an additional 
advantage of having lower sky background brightness.  This enables more accurate 
sky subtraction and allows us to probe to lower surface brightnesses.

\section{Two-dimensional Decomposition}

The vast majority of previous studies of the photometric structure of elliptical 
galaxies have relied on one-dimensional (1-D) analysis of their light
distribution.  Apart from small deviations near the center, the radial profile 
of ellipticals, especially when viewed under ground-based resolutions, is 
usually modeled as a single component\footnote{An exception are brightest 
cluster galaxies, which have long been recognized to contain an additional 
extended envelope (e.g., Schombert 1986).}, often fit with de~Vaucouleurs' 
(1948) law or a \ser\ (1968) function.  Although this approach is 
straightforward and generally suffices to describe the overall global structure, 
it is not very effective in isolating individual subcomponents, should they
exist, because it cannot easily take advantage of the geometric information
encoded in the 2-D image that can help to break the degeneracy between 
overlapping subcomponents.  In this study, we will employ 2-D image 
decomposition to study the detailed photometric structure of the CGS
ellipticals.  We relax the assumption that the main body of ellipticals is a 
single-component object.  We take advantage of our high-quality images and a 2-D
fitting technique to explore more complex, multi-component models.

Our analysis uses the most recent version of {\tt GALFIT}\footnote{{\tt
http://users.obs.carnegiescience.edu/peng/work/galfit}} (Peng et al. 2002, 
2010), a flexible fitting algorithm designed to perform 2-D decomposition of 
galaxy images.  \galfit\ can fit an arbitrary number and combination of
parametric functions, and version 3.0 has the additional capability of modeling 
non-axisymmetrical structures such as bars, spiral arms, and tidal features.  
In the case of elliptical galaxies treated in this study, we consider only the
\ser\ function.  We adopt this function not because of any physical motivation, 
but rather because it has enough flexibility to accommodate most of the main 
structural components in galaxies.  Being widely used in the literature, it also
provides a convenient basis for comparison with published results.

The \ser\ profile has the following form:

\begin{equation}
\mu(R) = \mu_{e}\exp\left[-\kappa\left(\left(\frac{R}{R_e}\right)^{1/n} - 1\right)\right].
\end{equation}

\noindent
It has seven free parameters: the central position $x_0$ and $y_0$, the
profile shape parameter $n$, effective radius $R_e$, surface brightness 
$\mu_e$ at $R_e$, axial ratio $b/a$, and position angle PA.  The 
parameter $\kappa$ is a variable that depends on $n$.

Galaxy isophotes are seldom perfect ellipsoids. Even in 1-D analysis, it is 
common practice to improve the fits of elliptical isophotes with the addition 
of perturbations in the form of Fourier modes (e.g., Jedrzejewski 1987). 
\galfit\ 3.0 now allows Fourier modes to be added to 2-D image fitting.  As 
shown in Peng et al. (2010), Fourier modes enable one to model a wide range of 
realistic features, including lopsidedness, boxy or disky shapes, and even 
more complicated structures often seen in actual galaxy images.  The Fourier 
modes in \galfit\ have many potential applications.  For this study, we will 
focus on just two: the $m = 1$ mode to quantify the overall degree of asymmetry
in the galaxy and the $m = 4$ mode to simulate boxy or disky isophotes.  We turn
on these modes only after we have first achieved an overall satisfactory model 
without them, for the purposes of further improving the residuals.

\subsection{Sky Background}

An accurate measurement of the sky background is crucial for our analysis, as 
we are interested in modeling structures covering a wide dynamic range in 
surface brightness, including faint features at large radii.  Although the FoV 
of CGS images is large enough to cover the full extent of most galaxies, some 
ellipticals have sufficiently large angular sizes ($D_{25} > 7$\amin) that it 
would be very difficult to measure accurate sky backgrounds.  An additional
complication, as explained in Papers~I and II, is that some of the CGS images
contain residual flat-field errors that imprint low-level, large-scale 
variations in the background.  Although the strategy outlined in Paper~II can 
recover the sky background to an accuracy of $\sim$0.6\% for relatively compact
galaxies, it is likely that the sky determination is more uncertain for the more
extended sources that fill an appreciable portion of the detector.  The sky 
level for the angularly large galaxies was estimated in a model-dependent manner
by fitting the outer portions of the light profile with a \ser\ function, and 
the uncertainty of the sky was estimated from a set of empirical calibrations 
determined from simulated images in which the galaxy occupies different relative 
fractions of the FoV.

Here, for the elliptical galaxy sample, we explore an alternative method to 
measure the sky background and its uncertainty, one that can be applied 
uniformly and consistently across the whole sample.  As described in greater 
detail in Appendix~B, we determine the sky from multi-component fits to the 
entire image, using as many \ser\ components as necessary to achieve a good 
model for the galaxy, while simultaneously solving for the sky as an additional 
component.  We adopt as the final sky level the average of the sky values from 
all the acceptable fits.  The uncertainty in the final sky value is estimated 
from the amplitude of the background fluctuations in the image, using an 
empirical relation calibrated from simulated data.  For a subset of the more
compact galaxies, we verify that this model-dependent estimate of the sky yields
results consistent with the more traditional model-independent, direct method of
measuring the sky background (Paper~I).  Appendix~C describes additional checks 
using a set of Sloan Digital Sky Survey (SDSS) images that do not suffer from
FoV limitations.

\subsection{Model Fitting Procedures}

With the sky value and its uncertainty in hand, we fit each galaxy with a series
of models, each consisting of one to four \ser\ components.  Unlike in the 
earlier phase of model fitting, wherein the purpose of the multiple components 
was to solve for the optimal sky value, here we use multiple components to 
explore the possibility that the galaxy contains real substructure.  Our 
approach is conservative: start simply and slowly, systematically increase the
complexity as the data permit.  We always begin with a single \ser\ component, 
and we add additional components, one at a time, after evaluating the residuals
and a number of other constraints, as explained in the next section.  Here we
briefly summarize the main steps of the fitting process.

We start with the exposure time-corrected $V$-band images, object masks, and 
empirical PSF images from Paper~I. The object masks were created from the 
segmentation images from Sextractor \footnote{
{\tt http://www.astromatic.net/software/sextractor}}. After removing the segment 
for the target elliptical galaxy, the sizes of the rest of the segments were 
increased according to their brightness to achieve better coverage of the 
objects that need to be masked. The PSF images were obtained using standard IRAF
procedures. Details of the methods for building object masks and PSF images are 
given in Paper~I. The sky value is fixed to the previously determined value, but
we allow for an additional tilt to account for a possible residual gradient in
the sky.  Normally this gradient, if present, is extremely small and does not
affect at all the final models.  

For the single-component model, the initial guesses for the total magnitude, 
effective radius, ellipticity, and position angle are taken from Paper~I;
the \ser\ index is set initially to $n$ = 2.5.  For the multi-component models,
reasonable choices for the initial parameters are made based on the best-fit 
values from the single-component trial, generally letting the inner component 
assume a larger effective surface brightness and a slightly higher \ser\ index.  
We require the different components to share the same central position.  The 
Fourier modes are initially turned off.

Next, we carefully inspect the residual image with the original object mask
overlaid.  In some cases we find that the original object mask needs to be 
enlarged to better cover the outermost regions of bright objects such as 
background galaxies, nearby companions, or saturated stars.  Occasionally we
discover previously unmasked stars directly superposed on the target of 
interest, often close to the galaxy center. Since these objects can introduce 
potential uncertainty into the multi-component model, we aggressively masked 
them out, and we repeat this process until the rest of the image is free of 
these artifacts.

The final set of models is generated with the updated, refined object mask 
and with the Fourier modes enabled for each component.  We restrict our 
attention to the $m=1$ mode to account for global asymmetry and the $m=4$ mode 
to simulate different isophote shapes.  Since the elliptical galaxies in this 
sample are already known to have little morphological peculiarity, the inclusion 
of the Fourier modes does not have an effect on the main results; they primarily 
serve to improve the residuals in the final models.

For each object we perform a series of tests to evaluate the reliability and 
robustness of the results, to explore the potential degeneracy among different 
model subcomponents, and to obtain a realistic estimate of the uncertainties 
of the model parameters.  Apart from the best-fitting sky value, we rerun the
models with the sky changed to plus and minus the estimated sky uncertainty.
We also test the sensitivity of the final models to our choice of input
parameters, by perturbing the initial parameters to dramatically different 
values.  In general, the models, even those with more than one component, are 
not very sensitive to the choice of initial parameters, so long as the initial 
guesses are not wildly off the mark.

The innermost regions of the galaxies pose special challenges.  Although our 
models are convolved with the nominal seeing of the observations (as formally
represented by the empirical PSF image), in practice the seeing profile is not 
known perfectly, and any residual errors in the seeing estimate can lead to 
significant mismatch between the model and the data.  Furthermore, we know from 
high-resolution studies (e.g., Lauer et al. 1995; Ravindranath et al. 2001) that
elliptical galaxies often contain substructure within the central few hundred 
parsecs.  Even for nearby galaxies, this inner substructure is not well resolved
with our typical seeing, and failure to account for it in our models could bias 
our global fits.  To evaluate the impact of seeing and unresolved inner 
substructure, during the model selection process we compare fits with and 
without a central mask applied.  After extensive testing (Appendix~D), we 
conclude that a circular central mask with a radius equal to twice the full 
width at half-maximum (FWHM) of the PSF is the most appropriate choice for most
of the observations. These central masks are used for two purposes in this work.
First, for each model, an extra run is performed using exactly the same 
parameters but with the central mask applied. These tests evaluate the impact 
from the seeing and the robustness of the central component. Second, to compare
with the approach taken by K09 in constructing their 1-D single-\ser~models, we 
also apply a central mask to our 2-D single-component models. Although the 
method used to decide the size of the central mask is different between K09 and
this paper, we can show that the single-component models in both studies behave
consistently on important scaling relations. 

\vskip 0.3cm
\begin{figure*}[t]
\centerline{\psfig{file=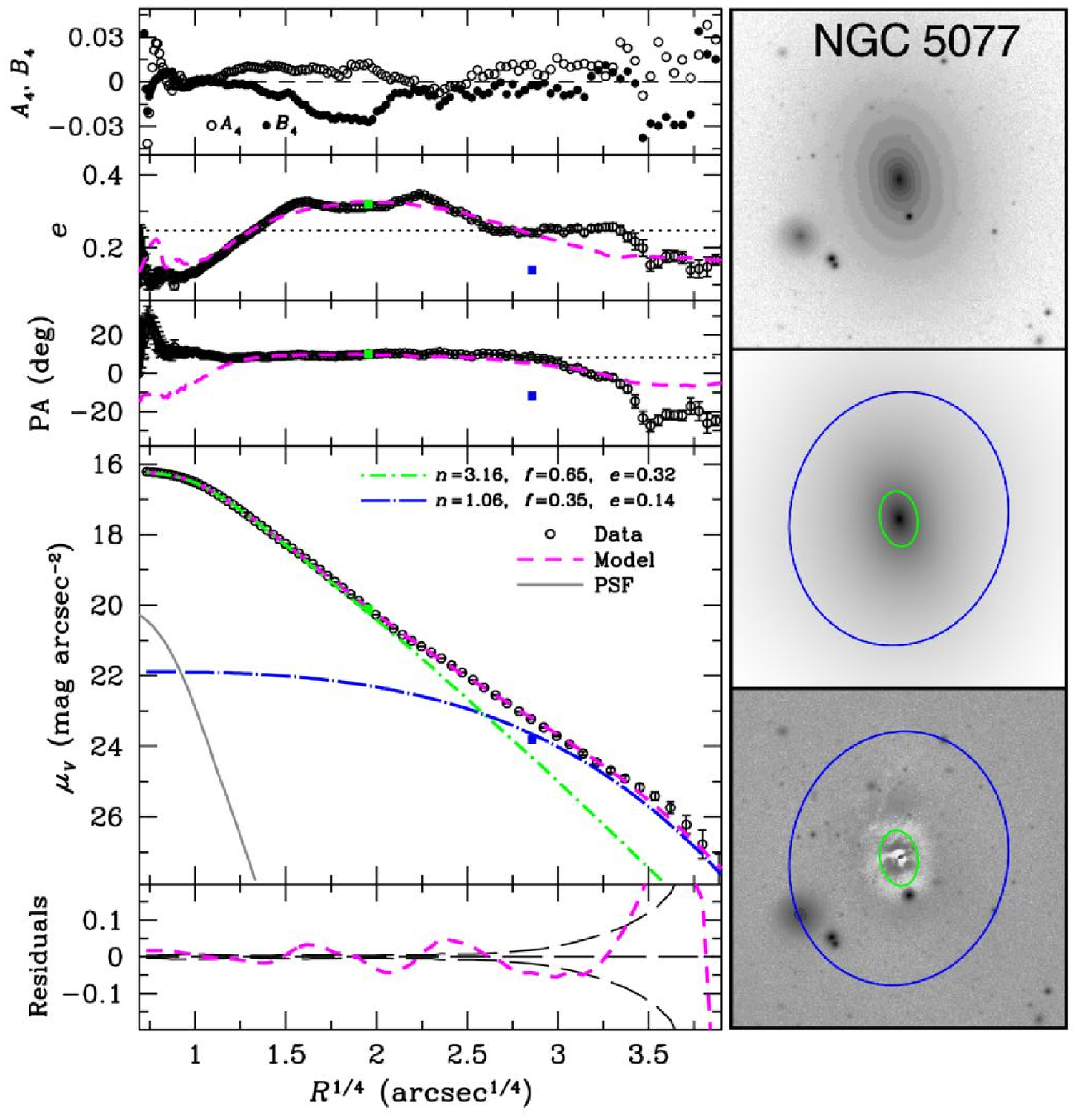,width=16.5cm,angle=270}}
\figcaption[fig5.eps]{
The best-fit two-component model of NGC~5077.  The right panels 
display, from top to bottom, images of the original data, the best-fit model, 
and the residuals; each panel has a dimension of $2 R_{50}$ and is centered on 
the nucleus.  The left panels compare the 1-D profiles of the data and the 
model; from top to bottom, we show the isophotal shape parameters $A_4$ and 
$B_4$, $e$, PA, $\mu$, and the residuals between the data and the model.  The 
observed $\mu$ profile is extracted with $e$ and PA fixed to the average value 
of the galaxy (Paper~II), which is indicated by the dotted line in the $e$ and 
PA panels. Here, the deviations of the isophotes from perfect ellipses are 
parameterized by the forth harmonics of the intensity distribution 
($A_4$, $B_4$) as described in Jedrzejewski (1987). Individual subcomponents, 
each extracted using the geometric parameters generated by {\tt GALFIT}, are 
plotted with different line types and colors; the PSF is plotted with arbitrary 
amplitude.  Colored squares on the $e$, PA, and $\mu$ plots mark the
ellipticity, position angle, and effective surface brightness of each component.  
The same color scheme is used in the greyscale images, where the ellipses trace 
$R_{50}$, $e$, and PA of each component.  In the residual plot, the three black
dashed lines indicate the position of zero residuals and the level of 
photometric error.
\label{fig5}}
\end{figure*}
\vskip 0.3cm

\begin{figure*}[t]
\centerline{\psfig{file=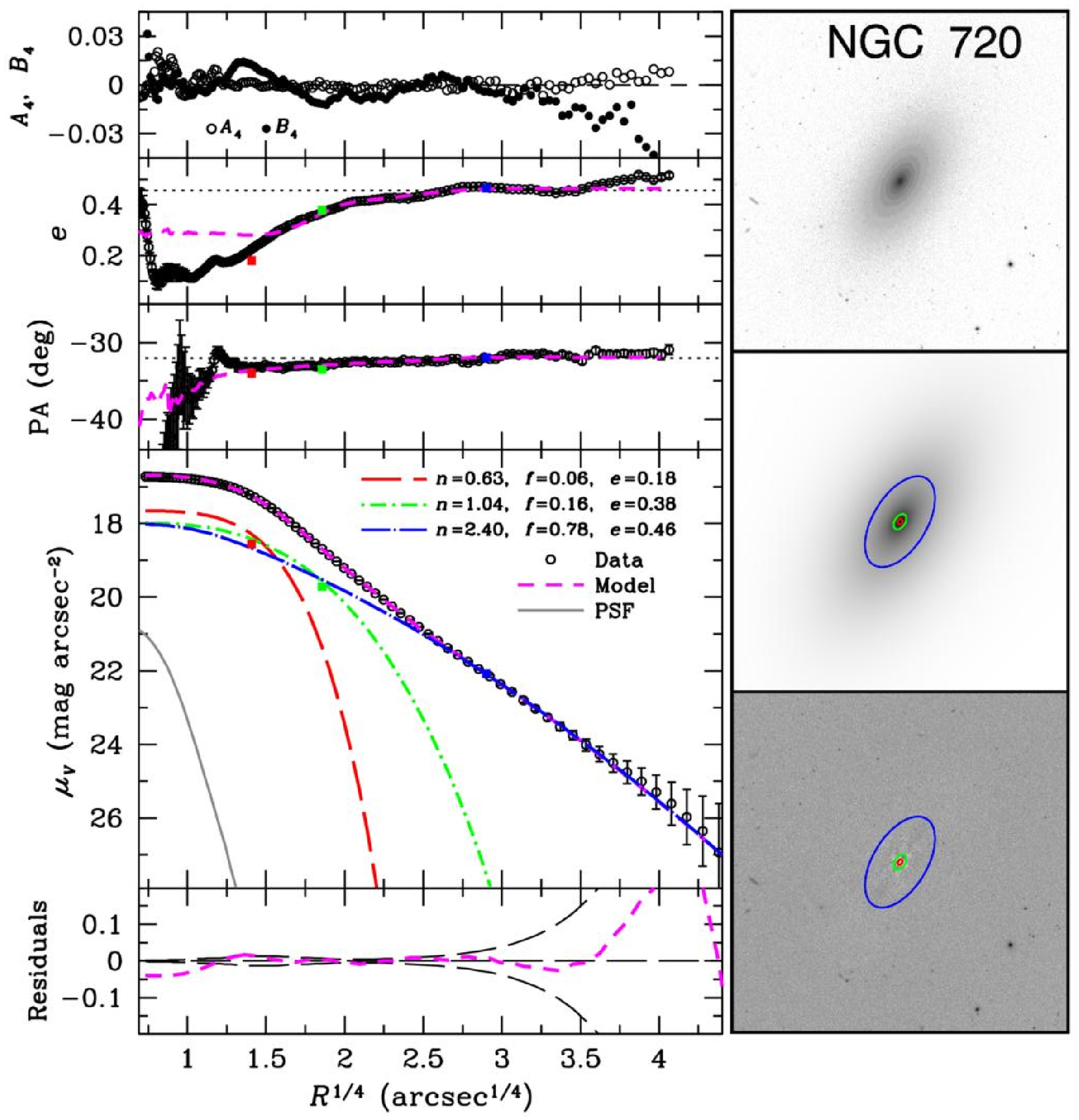,width=17.0cm,angle=270}}
\figcaption[fig6.eps]{
Best-fit three-component model for NGC~720.
In this and subsequent figures, the inner, middle, and outer component are
plotted in red, green, and blue, respectively. All other conventions are 
the same as in Figure~5.
\label{fig6}}
\end{figure*}
\vskip 0.3cm

\begin{figure*}[t]
\centerline{\psfig{file=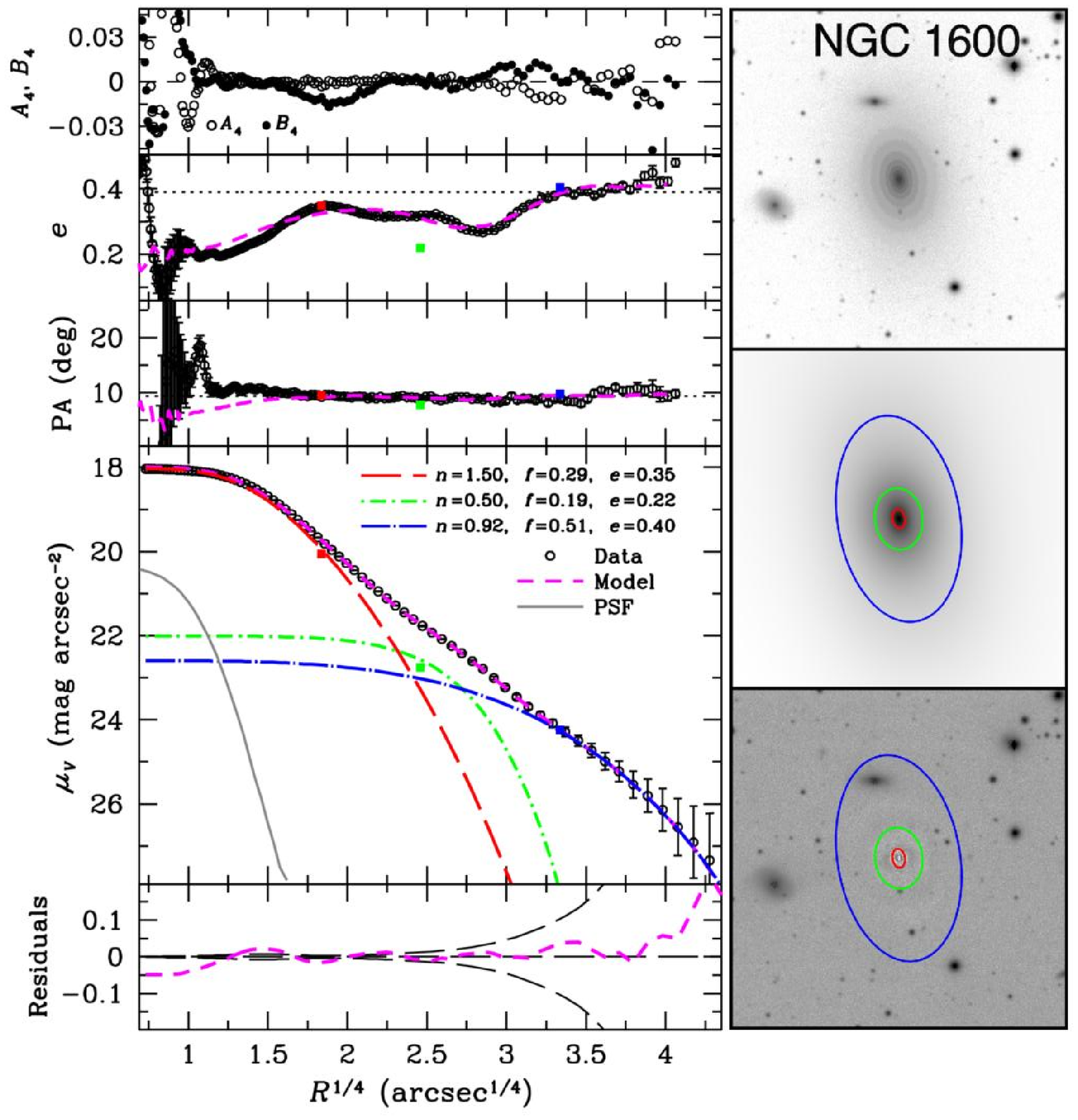,width=17cm,angle=270}}
\figcaption[fig7.eps]{
Best-fit three-component model for NGC~1600.  See Figure~5 for 
details.
\label{fig7}}
\end{figure*}
\vskip 0.3cm

\subsection{Sample Fits}

Using procedures described in the above section, a series of models with a
different number of components was generated for each galaxy. 
Therefore, the ``best'' model needs to be carefully selected for further 
analysis. For this purpose, a reasonable definition of ``best'' model is 
proposed. Also, we design a consistent method to compare these models both 
qualitatively and quantitatively. This crucial definition and method will be 
discussed in detail in the next section. But, first, we 
show a few examples of best-fit models to illustrate the range of typical 
structures encountered in our sample. The full set of 
fits for the 94 CGS ellipticals is presented in Appendix~E, and a summary of 
the final, best-fit parameters is given in Table~1.

NGC~5077, a member of a small group, is an elliptical in pair with NGC~5079.
The best model, summarized in Figure~5, consists of a dominant 
inner ($R_e = 2.69$ kpc) component with a high \ser\ index ($n = 3.2$) and an 
extended ($R_e = 12.3$ kpc), diffuse outer component with $n = 1.1$.  Both the 
ellipticity and the position angle are reasonably well reproduced, except for 
the innermost region where the color index maps (Paper~II) suggest that there 
are dust patches.  Although the 2-D residuals are not very clean, the existence 
of extended dust structures prevents us from exploring more complex models.
Only five galaxies in our sample have two-component models as their best fit. 
Indeed, the residual patterns of all the two-component models can
be improved by introducing additional components, but various complications 
(e.g., strong contamination from nearby stars or poorly resolved substructure 
near the nucleus) preclude us from pursuing more complex models. 

NGC 720 is a well-studied elliptical galaxy known to have a central disky 
structure (Goudfrooij et al. 1994; Rembold et al. 2002). Its ellipticity 
gradually increases toward larger radius. Our best-fit model (Figure~6) 
contains three components: a compact inner component with $R_e = 0.46$ kpc, $n 
= 0.63$, and luminosity fraction $f = 0.06$; an intermediate-scale disk-like 
component with $R_e = 1.38$ kpc, $n = 1.0$, and $f = 0.16$; and a dominant 
outer component with $R_e = 8.18$ kpc, $n = 2.4$, and $f = 0.78$. Some 
residuals can still be seen in the innermost region, but the model in 
general reproduces well the overall structure of the galaxy, especially its
ellipticity profile.

Figure~7 gives another example of an elliptical (NGC~1600) that requires a 
multi-component model to describe the full radial range captured in our 
images.  Three components give a reasonably good fit, 
especially when we take into consideration the rather complex ellipticity 
profile, which cannot be recovered with either one or two components alone.  
The constancy of the PA with radius precludes the possibility that the 
ellipticity changes are induced by isophotal twists.  Interestingly, none of 
the three components has a \ser\ index higher than 1.5; both the middle and 
outer components formally have $n < 1$.  Our tests indicate that the 
three-component structure is robust with respect to changes in the sky 
background, although the exact value of the \ser\ index for the outer 
component can vary from $n = 0.7$ to $2.1$.

The above two cases (NGC 720 and NGC 1600) are good examples of the most 
common configurations encountered in our survey. Roughly 75\% of the sample 
can be described by a qualitatively similar three-component model. Taking into 
consideration that many of the galaxies modeled by two components may, in 
fact, also be better fit with three components, the general picture that 
emerges from our analysis is that the structure of most nearby ellipticals 
consists of three photometrically distinct components.

It is often extremely difficult to recognize large-scale subcomponents in the 
1-D azimuthally averaged surface brightness profiles of ellipticals.  Different 
subcomponents, if present, usually have very low relative contrast and blend 
nearly seamlessly with one another.  Sometimes neither the 1-D nor the 2-D 
residuals offer much help to discriminate between models, because different 
models often produce only minor differences in the residuals.  However, this
ambiguity can be resolved if the subcomponents have different shapes, even if
the differences are 

\vskip 1.5cm
\begin{figure*}[tb]
\centerline{\psfig{file=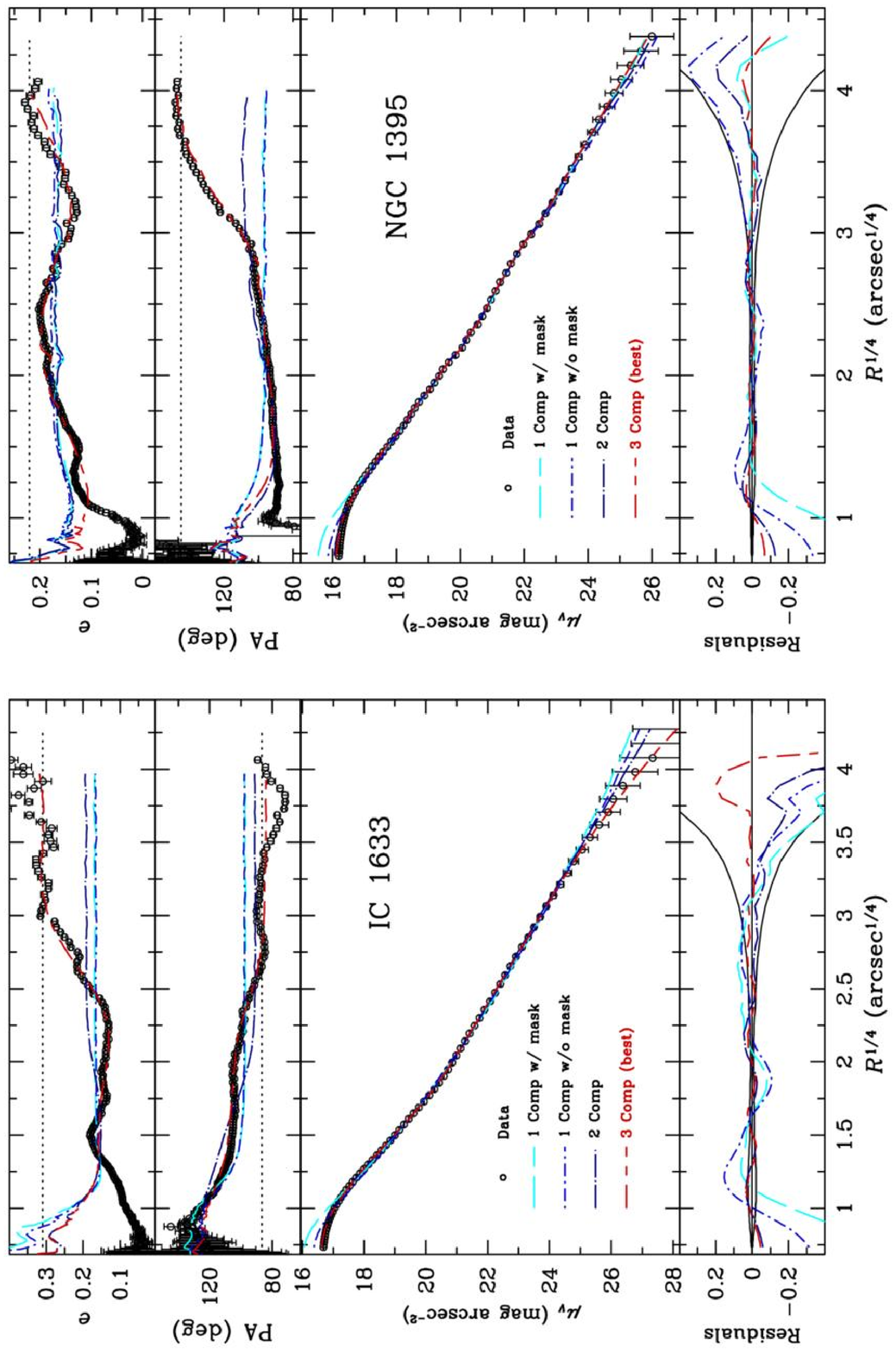,width=19.0cm,angle=270}}
\figcaption[fig8.eps]{
Comparison of different models for IC~1633 and NGC~1395. We show a 
single-component \ser\ model, with and without a central mask, followed by a
two-component model and the best-fitting three-component model. The
single-component models clearly give inferior fits.  While the models with both
two and three components yield much improved residuals, only the 
three-component model can self-consistently match the $\mu$, $e$, and PA 
profiles. It is clear that the three-component model does a superior job of
recovering not only the $\mu$ profile but also the geometric constraints 
provided by the $e$ and PA profiles.
\label{fig8}}
\end{figure*}
\vskip 0.9cm

\noindent
subtle. When evaluating the viability of different models, we find that the 
information furnished by the radial variation of ellipticity and position angle
provides very valuable constraints to decide which is the best solution.  This 
is illustrated in Figure~8 for IC~1663 and NGC~1395, where we compare the 1-D
information for different models.  Even though the 1-D residuals of the two- and 
three-component fits are quite similar, clearly only the three-component models 
can \emph{simultaneously} follow the surface brightness, ellipticity, and 
position angle profiles.  This not only demonstrates the improvement provided by 
the three-component models but illustrates the distinct advantages of 2-D image
decomposition compared to traditional 1-D profile fitting.  Of course, one can 
always attempt to fit multiple components in 1-D, and, in principle, one can
perhaps even obtain better residuals.  However, it will be very hard to prove
the necessity of any extra component or to understand its nature.  2-D analysis 
takes maximum advantage of the full information contained in a galaxy image.  
Any further, unnecessary compression of information should be avoided. 

Interestingly, roughly 20\% of the galaxies require even more complex models 
to explain their structure. The 18 galaxies with four-component models come in 
three flavors.  

\vskip 0.5cm

\begin{itemize}

\item{
Real substructures: 
The most common category consists of galaxies that contain additional 
substructure such as a small edge-on disk (e.g., NGC~7029) or an outer ring 
(e.g., IC~2006).  Figure~9 illustrates the case of NGC 7029, a galaxy 
classified as S0 by Sandage \& Tammann (1981) but as E6 in the RC3.  The 1-D 
information in Figure~9 and the residual images in Figure~10 clearly 
demonstrate the impressive improvement made by the four-component model, which 
includes a moderately compact ($R_e = 2.31$ kpc), edge-on ($e\equiv 1-q=0.81$), 
disk-like structure described by a very low \ser\ index ($n$ = 0.47).  Apart 
from the embedded disk, the global spheroidal structure of the rest of 
the galaxy resembles that typically seen for the bulk of the sample,
namely consisting of an inner, middle, and outer component.  NGC~7029, having a
structure intermediate between an idealized elliptical and S0, illustrates the 
ambiguity that can arise. From the standpoint of the four-component structure, 
the galaxy may be regarded as an extreme case of an S0 with a small disk 
($f = 0.02$) and a large spheroid. At the same time, it can be viewed as an 
elliptical with an embedded disk, perhaps acquired from a recent merger event. 
Either 
way, this example showcases the power of our 2-D, multi-component decomposition 
technique. The disk component in NGC~7029 is so subtle and inconspicuous that it 
would be very difficult, if not impossible, to isolate it with any other method.
}

\begin{figure*}[t]
\centerline{\psfig{file=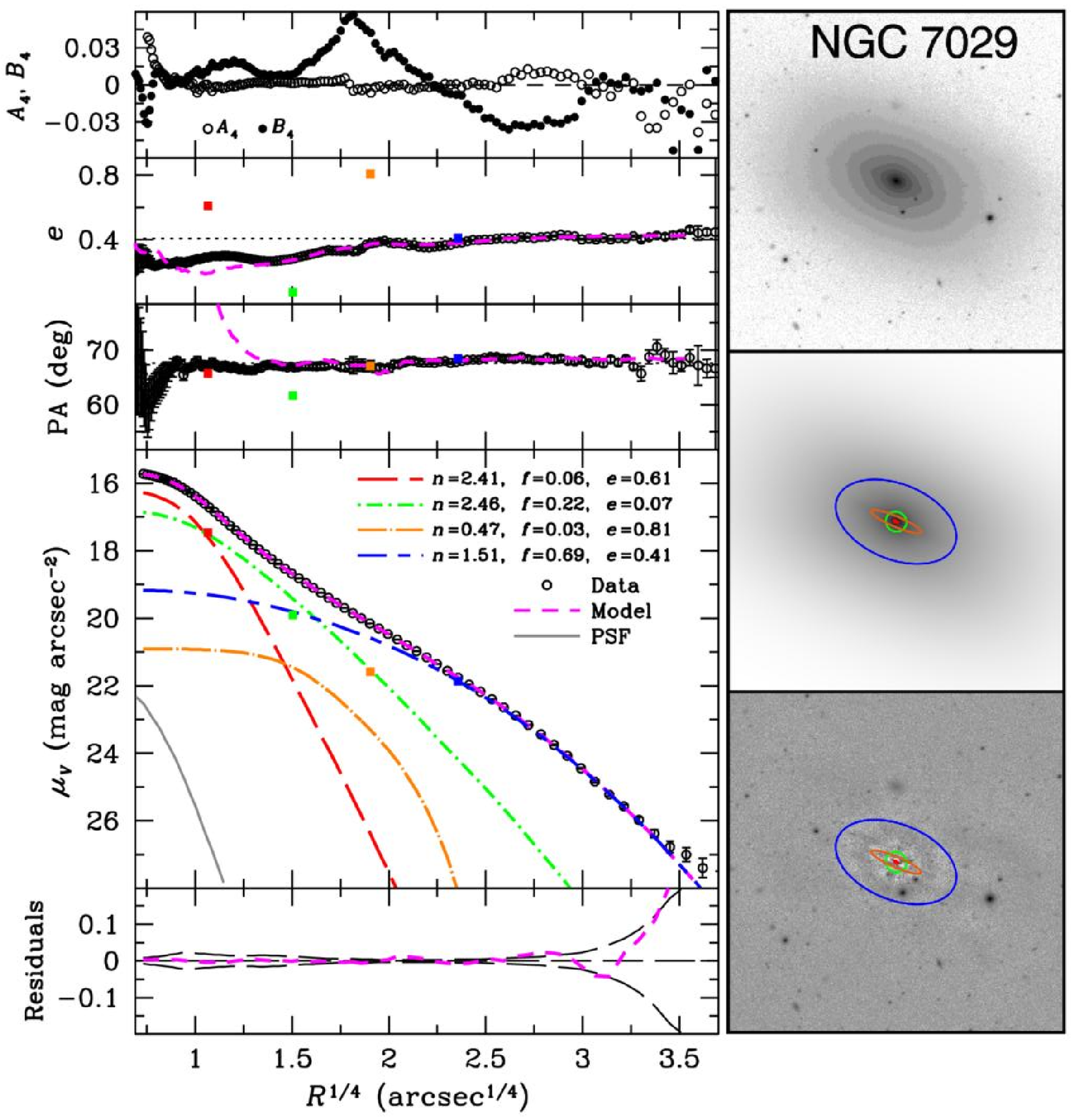,width=14.0cm,angle=270}}
\figcaption[fig9.eps]{
Best-fit four-component model for NGC~7029.  See Figure~5 for details.  The 
best fit requires a component (shown in orange dot-long dashed line) with high 
ellipticity.  The radial profiles of surface brightness and geometric 
parameters do not provide strong constraints for the high-ellipticity component 
because it contains only a small fraction (3\%) of the total light, but, 
as illustrated further in Figure~10, it can be clearly seen in the 
2-D analysis.  This highlights the importance of the 2-D approach.
\label{fig9}}
\end{figure*}
\vskip 0.3cm

\begin{figure*}[t]
\centerline{\psfig{file=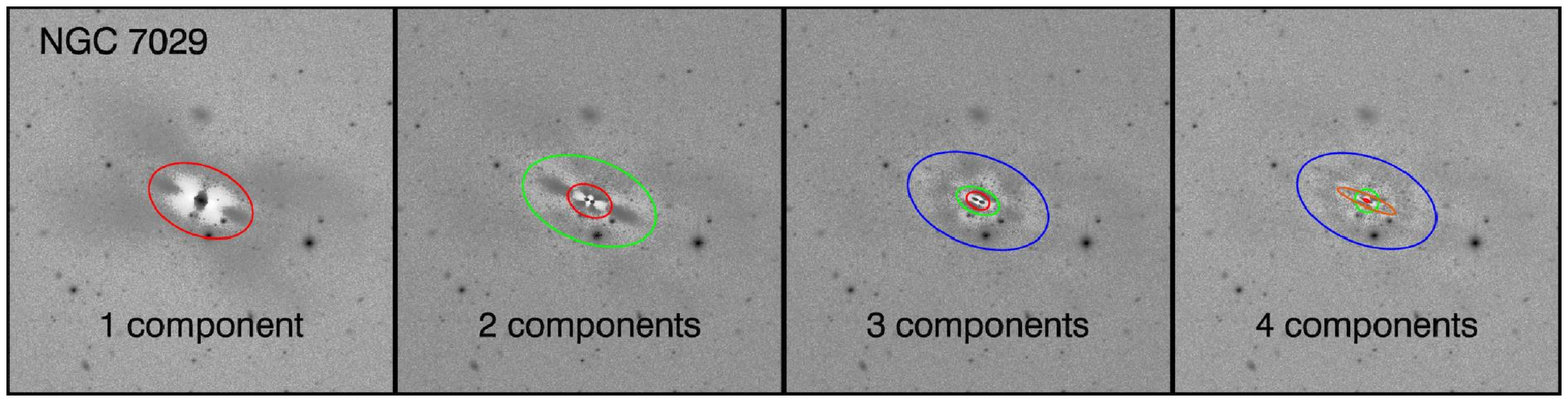,width=17.0cm}}
\figcaption[fig10.eps]{
Comparison of residual images from different models for NGC~7029. 
Only the four-component model correctly reproduces the inner, high-ellipticity 
structure.
\label{fig10}}
\end{figure*}
\vskip 0.3cm

\begin{figure*}[t]
\centerline{\psfig{file=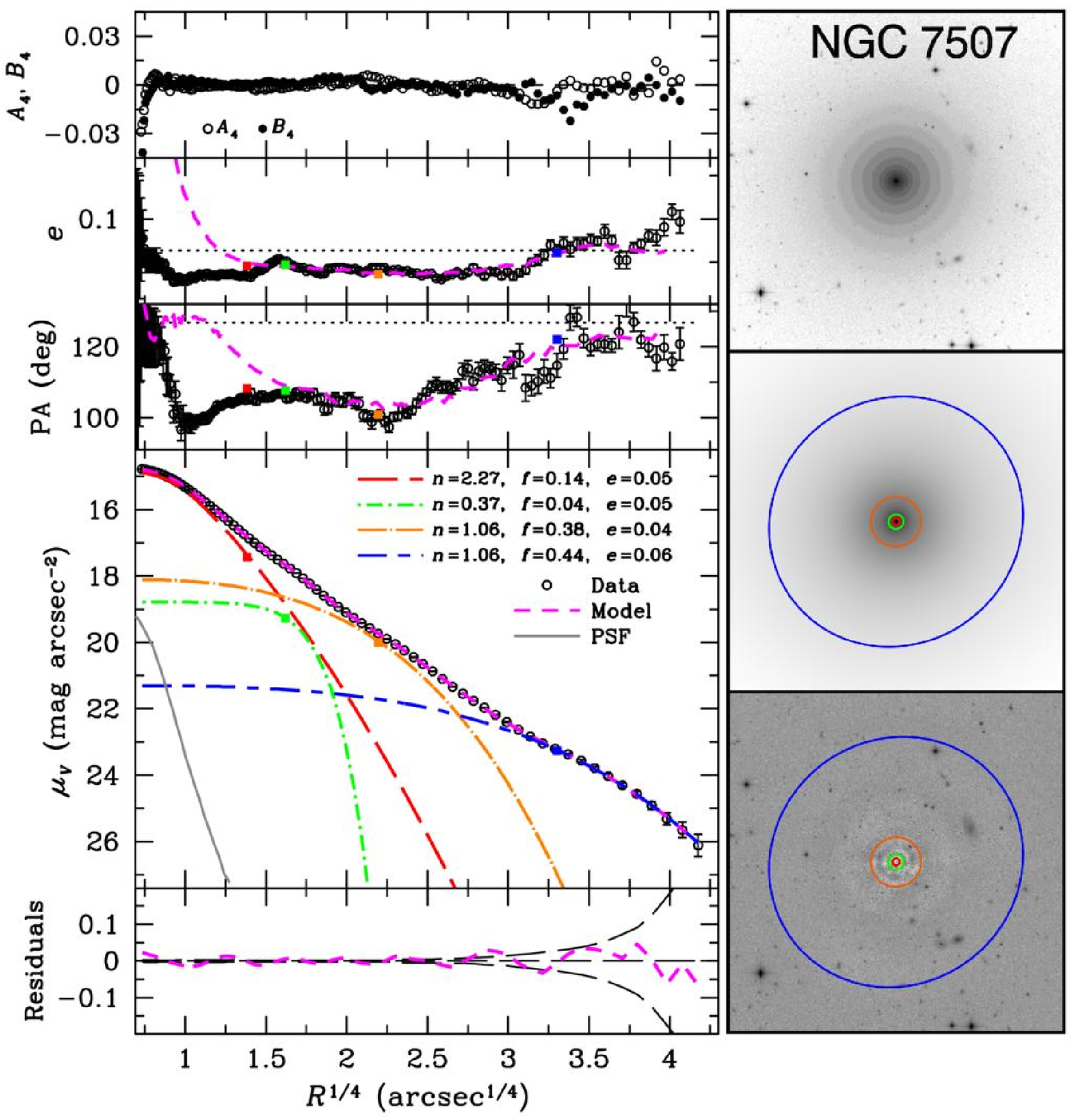,width=14.0cm,angle=270}}
\figcaption[fig11.eps]{
Best-fit four-component model for NGC~7507.  See Figure~5 for 
details.  The geometric information is less useful to constrain the fit in 
this case because of the low inclination angle of the galaxy. This model 
includes a very small component (4\% of total luminosity; green dot-short 
dashed line), which does not have an obvious physical explanation,
but is necessary to produce a good model, as illustrated further in 
Figure~12.
\label{fig11}}
\end{figure*}
\vskip 0.3cm

\begin{figure*}[t]
\centerline{\psfig{file=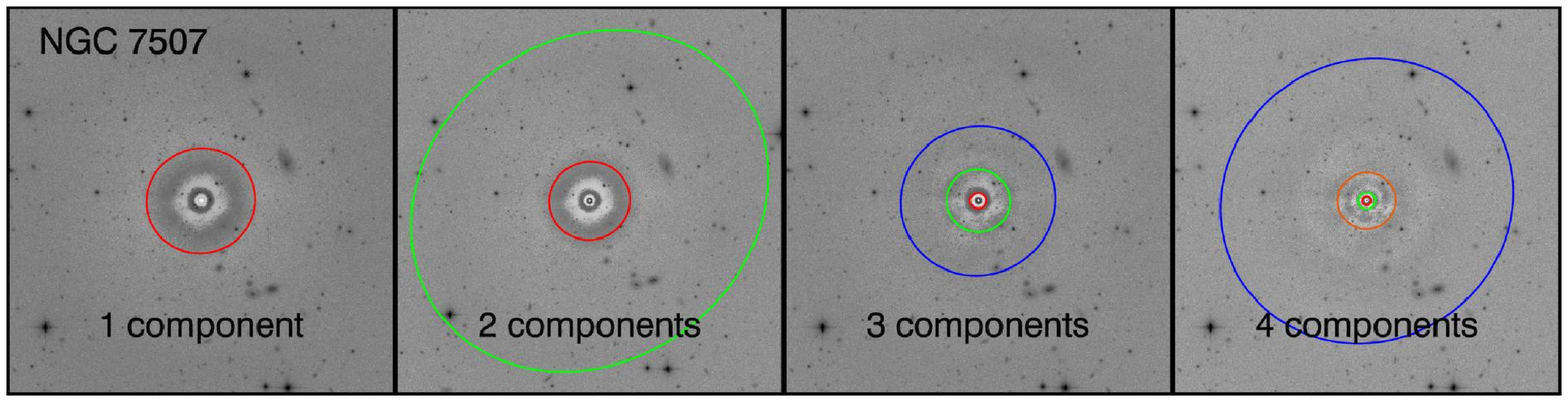,width=17.0cm}}
\figcaption[fig12.eps]{
Comparison of residual images from different models for NGC~7507. 
The four-component model clearly gives the best fit.
\label{fig12}}
\end{figure*}
\vskip 0.3cm

\begin{figure*}[t]
\centerline{\psfig{file=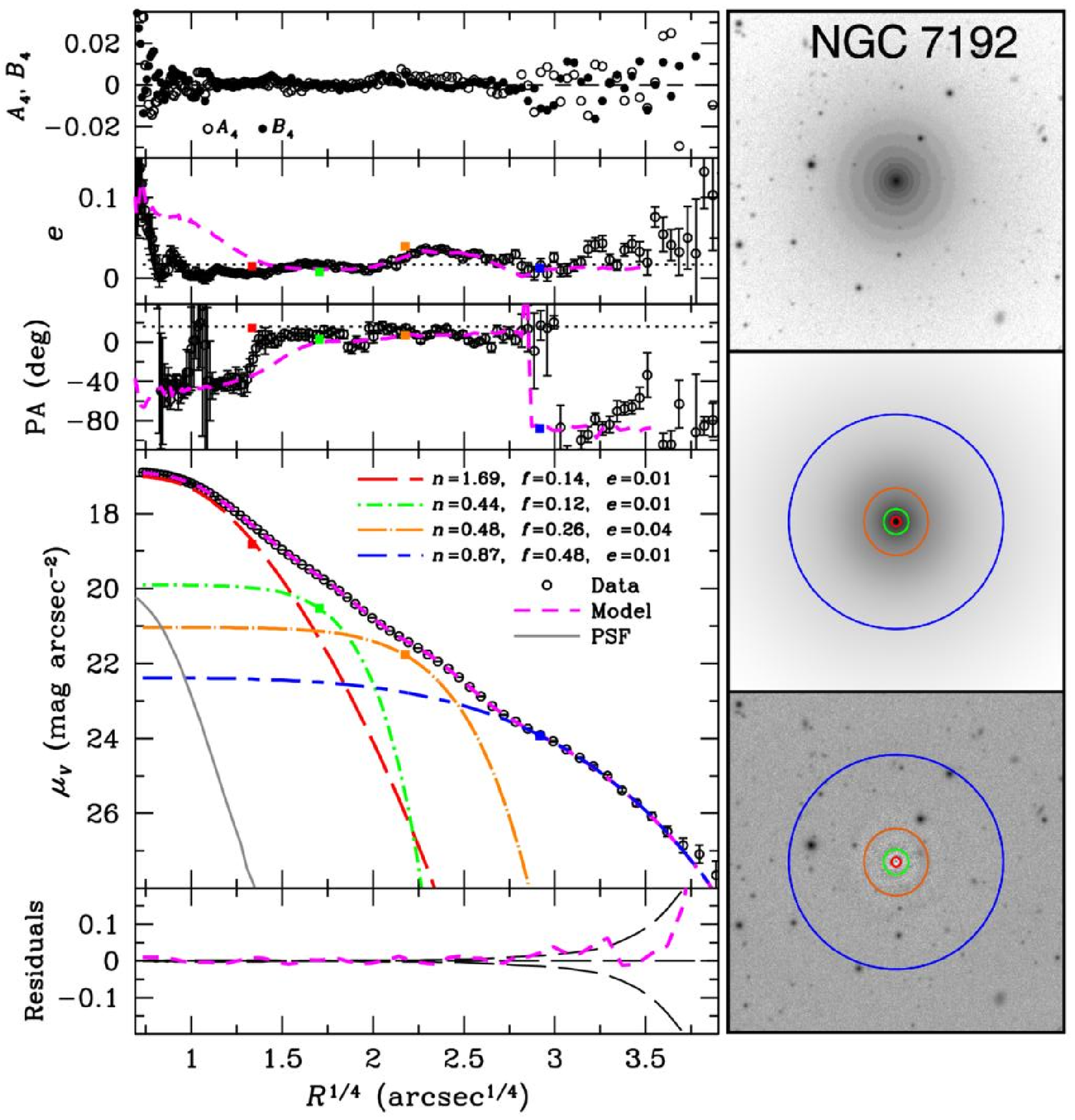,width=14.0cm,angle=270}}
\figcaption[fig13.eps]{
Best-fit four-component model for NGC~7192.  See Figure~5 for details. 
Although it is also a face-on system, the near-exponential ($n = 0.87 \simeq 
1$) extended component is consistent with it being a disk, and thus this  
galaxy may be a misclassified S0.  Additional models are shown in Figure~14.
\label{fig13}}
\end{figure*}
\vskip 0.3cm

\begin{figure*}[t]
\centerline{\psfig{file=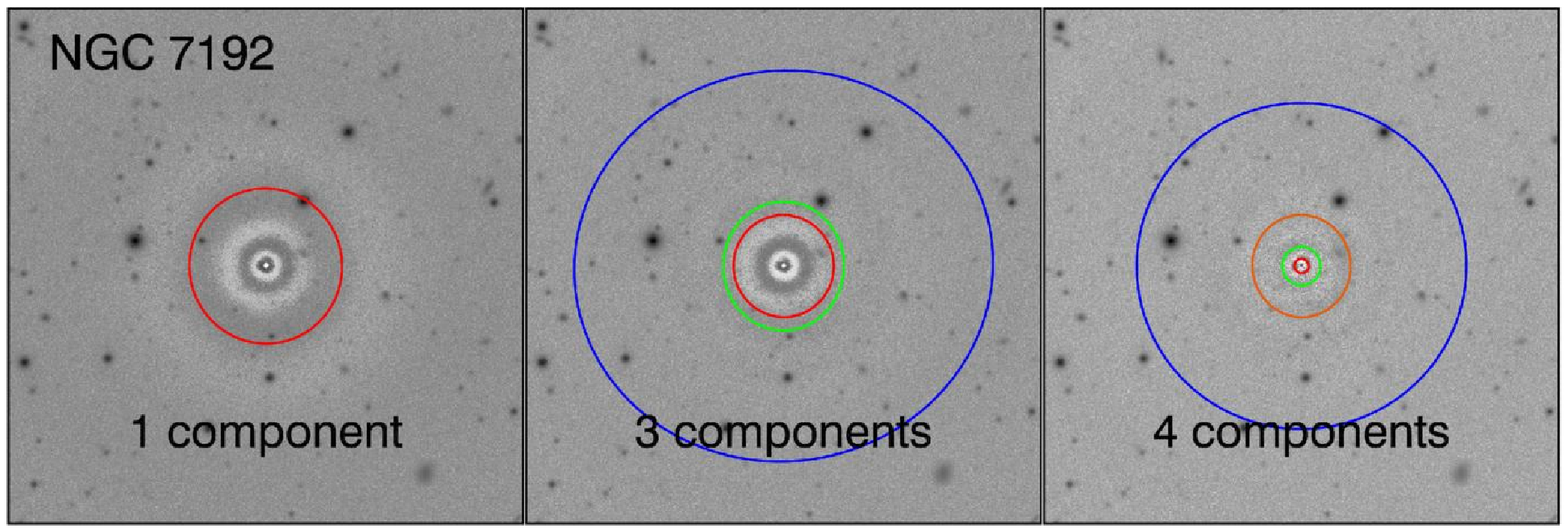,width=13.5cm}}
\figcaption[fig14.eps]{
Comparison of residual images from different models for NGC~7192. 
From left to right, residual images are shown for models with one, three, and 
four \ser\ components; the two-component model is unreasonable and is excluded.
The four-component model clearly gives the best fit.
\label{fig14}}
\end{figure*}
\vskip 0.3cm

\item{
Inner complications due to dust or unresolved structures:
Another class of four-component models describes systems that require a minor 
component to account for central dust structures or other nuclear features. 
This does not necessarily mean that all four components are physically 
separate and meaningful. For example, a dust lane present in the central 
region could induce a ``break'' in the inner luminosity distribution. Without 
taking this effect into account, the model could produce an unreasonable 
component with extremely high \ser\ index. Sometimes introducing another small 
component can greatly help alleviate such an effect, even if such a component
is not a physically separate entity.  Figure~11 
shows the case of NGC~7507, an almost perfectly face-on elliptical that 
is considered a prototypical E0 galaxy. The structure of this galaxy, including 
the mild change of position angle, is very well reproduced by a four-component 
fit.  Apart from an extra, very small component in the inner region, the rest 
of the galaxy is well fit by a standard three-component structure.   The size 
of this small component ($R_e = 0.75$ kpc) is very similar to the component 
that dominates the inner region ($R_e = 0.4$ kpc), and its \ser\ index is 
quite low ($n = 0.37$).  Although this component has a very low luminosity 
fraction ($f = 0.04$), it is essential for constructing a reasonably good 
model for this galaxy. By comparing the residual images of models with 
different numbers of components (Figure~12), it is easy to see that the 
four-component model fares the best.  As for the nature of this small 
component, it should be viewed as part of the inner structure since it is 
likely to associated with the ``extended region of dust absorption on the 
southwest side of the nucleus'' (Goudfrooij et al. 1994).
}

\begin{figure*}[t]
\centerline{\psfig{file=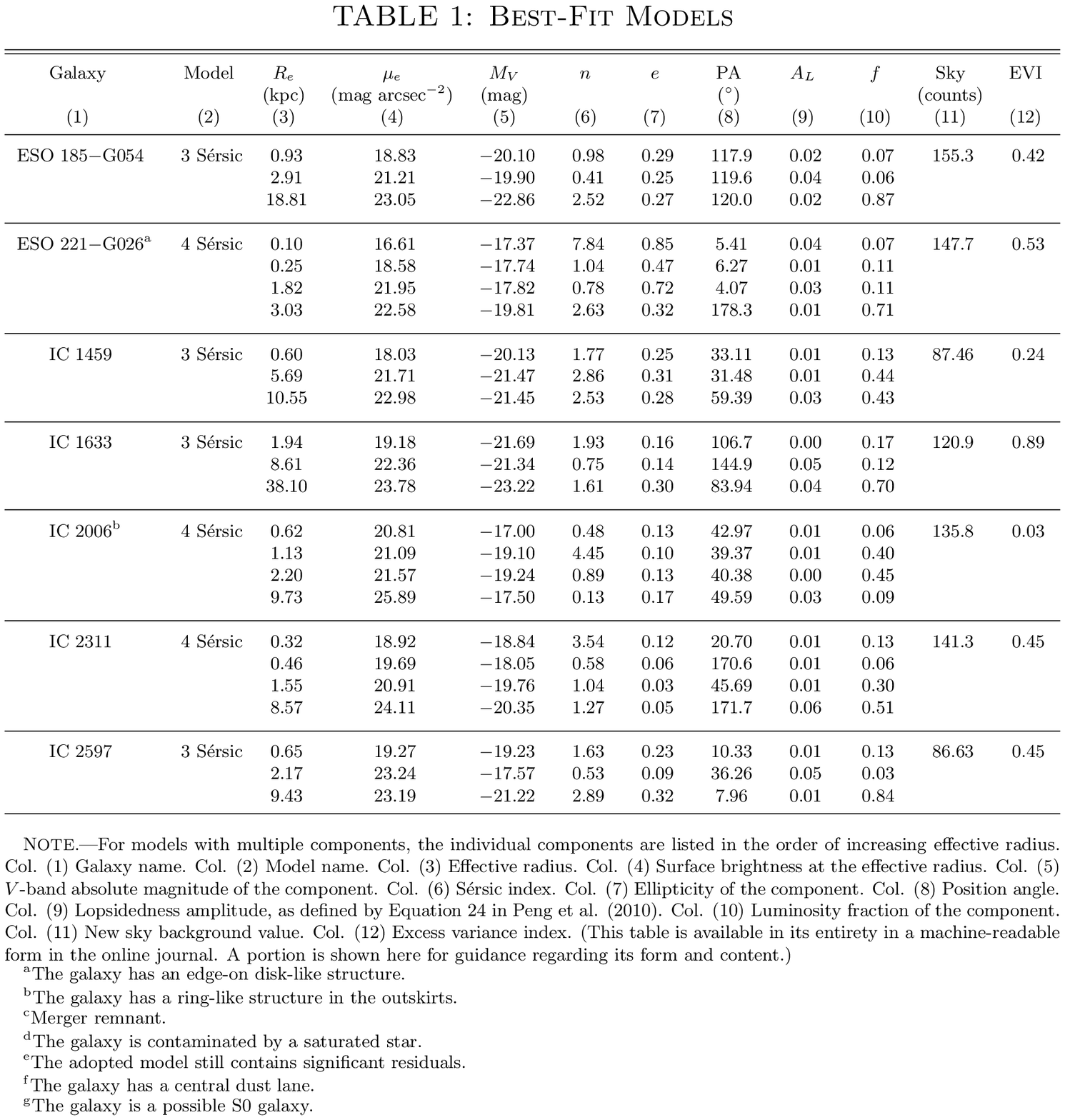,width=18.5cm,angle=0}}
\end{figure*}

\item{
Misclassified S0s: 
The third category of galaxies that requires an extra fourth component to 
achieve a satisfactory fit comprise misclassified S0s. For these galaxies, the 
evidence for their S0 nature comes from either the existence of a bar-like 
structure in the residual image of the two- or three-component model (e.g., 
NGC~584) or the presence of disk-like properties in the outermost component, 
as in the case of NGC 7192 (Figures 13 and 14).  This almost face-on 
galaxy is classified as a cD galaxy in the RC3 on account of the contrast 
between the inner region and the outer extended feature, but it was classified 
as an S0 in the Carnegie Atlas of Galaxies (Sandage \& Bedke 1994).  Our 
four-component model recognizes this galaxy as an S0 from its multiple 
``lens'' structure. The innermost bulge component has the highest \ser\ index 
($n = 1.7$) and accounts for about 14\% of the total luminosity. The disk, 
comprising 48\% of the total luminosity, is the outermost component with $n 
\approx 0.9$.  The ``lens'' structures are the two intermediate-scale components 
with \ser\ indices $n \approx 0.4-0.5$, which altogether account for 38\% of the
total light.  Our four-component model is qualitatively similar to the $K$-band 
decomposition presented by Laurikainen et al. (2010), although the detailed 
parameters of the subcomponents are different.  Unlike Laurikainen et al.,
we do not restrict the model parameters to fixed values (e.g., setting $n = 1$ 
for an exponential disk). 
}
\end{itemize}

Apart from the misclassified S0s, the additional component invoked for the 
other two types of four-component fits comprises only an insignificant 
fraction of the light ($\le 5$\%).  Nevertheless, it can be extremely helpful 
in terms of improving the global model of the galaxy.  Setting aside 
this extra perturbation, the other main components have properties similar to, 
and can be easily identified with, the standard three components that normally 
suffice to describe the overall structure of the sample.  In other words, 
these more complicated four-component models do not violate our main thesis: 
that nearby ellipticals contain three main structural components.

\subsection{Definition of a ``Good'' Model}

There is no unique or rigorous definition of a ``good'' model
in image decomposition.  Whether a fit is ``good'' or ``good enough'' depends
on the fitting method as well as on the scientific application.
In 1-D fitting, the residual profile is often used to judge the quality of the 
fit. Yet, we know that the 1-D method cannot fully take into account geometric 
constraints, and sometimes it cannot resolve the degeneracy between different 
components. Both of these limitations are especially problematic for 
elliptical galaxies, which tend to show only mild or gradual changes in 
surface brightness and geometric parameters.  As a case in point, K09 
successfully fit a single \ser\ function to the 1-D global surface brightness 
distribution of the giant elliptical galaxy M87 (NGC~4486; see their 
Figure~10) after judiciously choosing a radial range that excludes the inner 
and outer portions of the profile.  From the standpoint of the residual 
profile, their fit is perfectly acceptable within the specified radial 
range used in the fit.  However, does this mean that this is the correct model 
for M87?  Two clues suggest otherwise.  First, as can be seen from Figure~10 
of K09, the ellipticity of the outer regions of M87 is significantly higher 
than that of its inner region; the position angle varies with radius too, 
albeit less drastically.  Although the radial variation of $e$ and PA can 
conceivably be explained by isophotal twisting of a single component, an 
alternative (and according to the results of this paper, more likely) 
interpretation is that it is a manifestation of the superposition 
of more than one physical component.  Second, the single-component fit 
yields a very high \ser\ index of 11.8. Although there is no strong physical 
justification to restrict the \ser\ index below any particular value, such a 
high index may be driven by the fact that M87 has an extended envelope.

For 2-D image decomposition, the quality of the model is judged usually in one 
of three ways. The most direct one is to examine the residual image, after 
appropriate scaling and enhancement. This method is straightforward and 
efficiently identifies obvious problems with the fit, since a ``bad'' model 
often leaves strong residual features. However, when degeneracy between
components exists in the model, or when two models only differ at a moderate 
level, the residual image alone does not offer enough discrimination.  Another 
common approach uses empirical parameters to quantify the goodness-of-fit. The 
general idea is to compare the relative absolute flux 
left in the residuals, either to the photometric uncertainties or to the 
luminosity of the galaxy inside a certain aperture. This method can efficiently 
weed out problematic models for a large sample, or it can be used to compare 
the average quality of different models within a given sample. However, 
since it uses integrated information, it is 
not effective at diagnosing the details of the model or for comparing various 
models when the differences are subtle. 
Lastly, it is common to extract the 1-D surface brightness profile from a 
2-D model and compare it with that derived from the original data. This method 
is not as intuitive as inspecting the residual image directly, but it is 
relatively sensitive to mild differences between models, as shown in Figures~8. 

Different scientific applications set different standards and require 
different strategies for model comparison. While single-component 2-D fitting 
can efficiently extract global parameters for large samples of distant 
galaxies (e.g., Hoyos et al. 2011), whose resolution and image depth often do 
not warrant more complex structural analysis, here we are facing the opposite 
regime.  Our sample is specifically designed to yield maximum information on 
the internal structure of nearby, bright galaxies, and we must strive for a 
higher standard of model acceptability and devise a more comprehensive 
strategy for model comparison.  For our purposes, we define the ``best'' model 
as one that contains a \emph{minimum number of components with reasonable, 
robust parameters that describes visibly distinct structure}.  Here the word 
``visibly distinct'' has two layers of meaning. First, both the surface 
brightness distribution and the systematic radial change of geometric 
information should be recovered down to a certain level, which means that the 
level of the residuals should be consistent with the stochastic RMS error. 
Second, the amount of information that can be recovered is also 
directly related to the data quality (e.g., spatial resolution, seeing,
signal-to-noise ratio). Therefore, a selected ``best'' model should only be 
considered as the model that recovers ``enough'' photometric information on a 
certain image. While we do not visually identify components by eye, we should 
mention here that often the individual components correspond to noticeable 
breaks in the surface brightness profile or distinct features on 2-D images. 

Increasing the number of components adds more degrees of freedom.  It is 
always possible to reduce the residuals or $\chi^2$ by making the model 
more complex.  This is dangerous and can lead to over-interpretation of the 
data.  As previously stated, our approach errs on the side of caution: we 
restrict our models to the fewest components possible and slowly increase in 
complexity only as absolutely justified by the data.  We further demand that
the fitted components have ``reasonable'' parameters, ones that are 
well-constrained by prior empirical knowledge or by the data themselves. For 
example, \ser\ indices with extremely high (e.g., $n$ \gax\ 10) or low (e.g., 
$n$ \lax\ 0.1) values should be regarded with suspicion, and effective radii 
larger than the size of the image are untrustworthy.  If the fit returns a 
component with high ellipticity, a flattened structure should be directly 
identifiable on the image or in the residuals.  In all cases, the model 
parameters should be robust with respect to potential systematic errors 
introduced by the seeing or the sky background. In short, while there is no 
fully objective or completely automatable procedure to select the 
best-fit model, we exercise due diligence and uniform methodology that is based 
on reasonable expectations for the structure of galaxy sub-components to examine
the models case-by-case, using \emph{all} the constraints furnished by the data, 
both in their 1-D and 2-D representation.

\subsection{Model Selection and Sample Statistics}

To quantify the relative goodness-of-fit for the various models and to compare 
them in a statistical manner, we compute the integrated absolute value 
of the 1-D residuals inside $V$-band $R_{50}$ and $R_{80}$, the radii enclosing 
50\% and 80\% of the total light (Paper~I).  Figure~15 shows that the residuals 
decrease markedly in going from single- to multiple-component models.  The 
improvement from two to three components is less 

\vskip 0.3cm
\centerline{\psfig{file=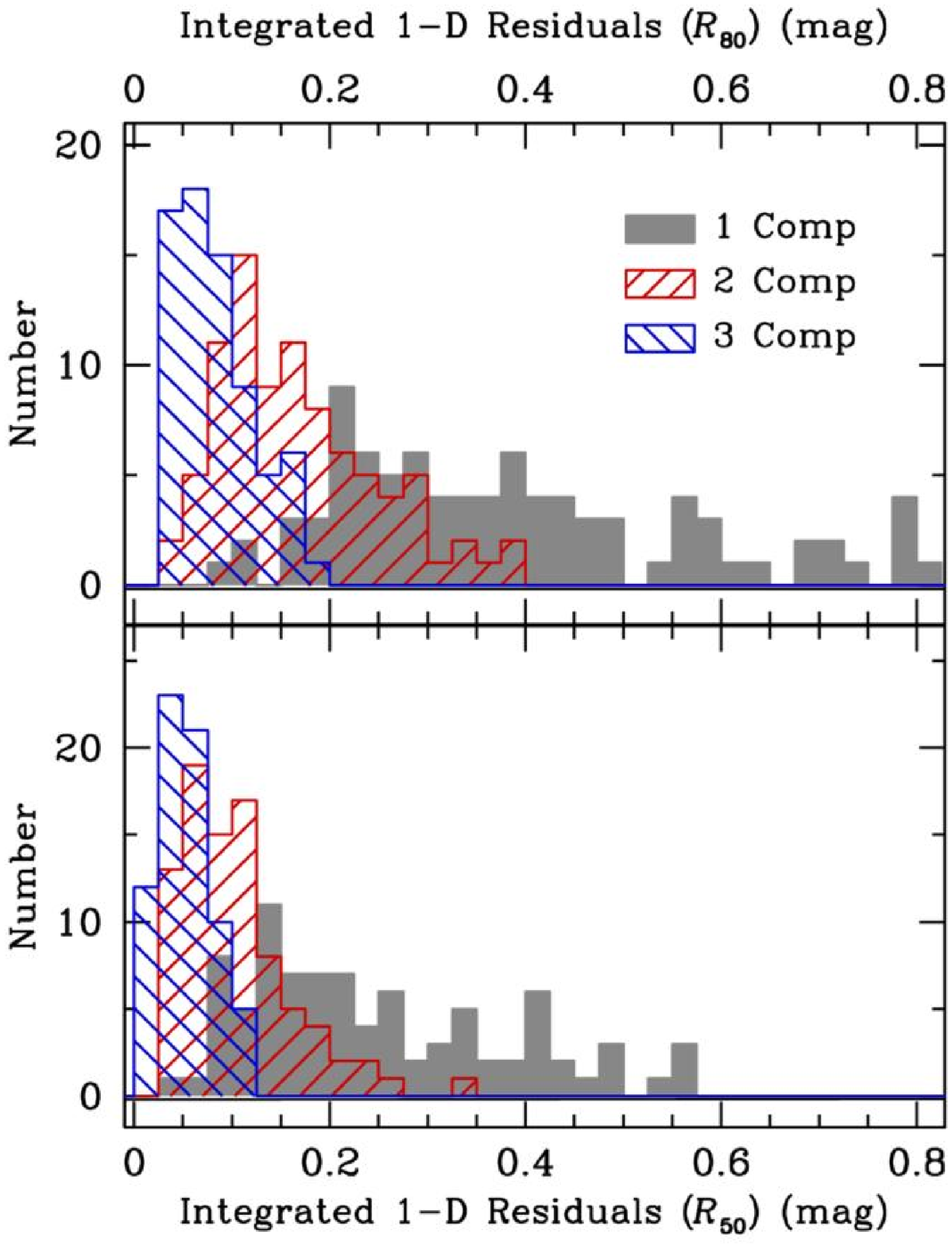,width=8.90cm}}
\figcaption[fig15.eps]{
Distribution of integrated 1-D residuals within (top) $R_{80}$ and 
(bottom) $R_{50}$.  These parameters are used to diagnose the goodness-of-fit 
of models with different number of components.  The models include 
one (without central mask), two, and three \ser\ components.
\label{fig15}}
\vskip 0.3cm

\noindent
dramatic, but it is clear that 
three components give systematically smaller residuals than two components, and 
that the tail of high-residual objects disappears.  These trends are present for 
the integrated residuals within both $R_{50}$ and $R_{80}$.  Following Hoyos 
et al. (2011), we also define the excess variance index,

\begin{equation}
{\rm EVI} = \frac{1}{3} \, \left(\frac{{\sigma_{\rm res}}^2}{\langle{\sigma_{\rm img}}^2\rangle}-1\right),
\end{equation}

\noindent
where $\sigma_{\rm res}$ is the root mean square of the residuals within 
$R_{80}$ 

\noindent
(Hoyos et al. use a somewhat different aperture, but this difference 
is immaterial) and ${\sigma_{\rm img}}$ is the root mean square of the Poisson 
noise in the same area.  In a perfect model,  ${\rm EVI} = 0.0$; if ${\rm EVI} 
\gg 1$, the model deviates significantly from the actual data.  Figure~16 
shows that, as expected, EVI generally correlates with the integrated 1-D 
residuals.  According to Hoyos et al. (2011), ${\rm EVI} > 0.95$ is probably 
indicative of an inadequate model.   We cannot strictly apply the same 
numerical criterion to CGS, whose data are very different from those 
considered by Hoyos et al., but we can easily see that EVI decreases 
dramatically as the number of components increases from one to three.  
Impressively, most galaxies that require a three-component fit have 
${\rm EVI} < 1$; by contrast, most of the single-component fits return 
${\rm EVI} \gg 1$.  In no instance is a single-component fit 
ever superior to a more complex model according to the EVI. This may sound like 
a trivial statement since the inclusion of more components adds more degrees 
of freedom to the fit, hence will result in lower residual level. But we
remind the reader that our 

\vskip 0.3cm
\centerline{\psfig{file=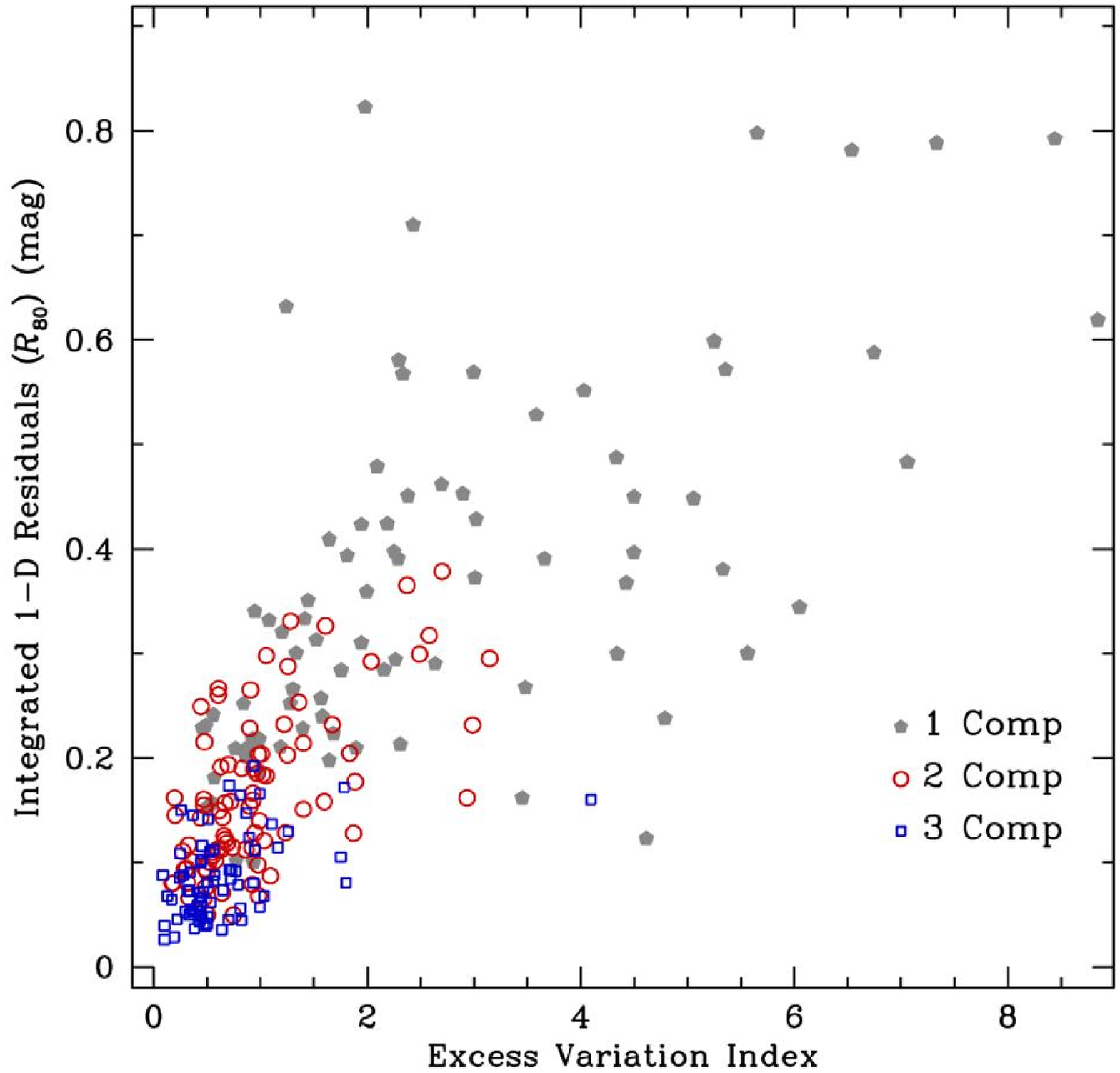,width=8.75cm}}
\figcaption[fig16.eps]{
The relation between the integrated 1-D residuals within $R_{80}$
and the excess variation index (EVI) defined by Equation 2, another parameter 
that indicates the goodness-of-fit of the models.  The models include one 
(without central mask), two, and three \ser\ components.
\label{fig16}}
\vskip 0.3cm

\noindent
standard for the ``best'' model and the 
method to select such model rely only partly on the residual level. We 
require the best model to be one that recovers as much information as 
possible using the minimum number of components. Therefore, the above statement 
should be understood as ``no single-component model provides enough information 
to recover the 2-D luminosity distribution of ellipticals in this sample.''
Even without further analysis, we can conclude from this that nearby ellipticals 
almost always have global structures more complicated than can be described by a 
single \ser\ function.  The majority require multiple photometric components.

Among the sample of 94 CGS ellipticals, 70 can be robustly fit 
with a three-component \ser\ model.\footnote{A possible edge-on structure is 
present in NGC~3585, but it is very subtle and difficult to isolate; since the 
standard three-component model already yields a reasonable fit with low 
residuals, we decided not to pursue a more complicated fit.  The merger 
remnants NGC~1700 and NGC~2865 can be fit with more complex models to account 
for their irregular, low-surface brightness tidal features, but we prefer to 
be conservative and restrict the final models to the standard three
components.} And, using different types of tests, including ones with 
dramatically different initial parameters, these models also prove to be 
quite stable. Five galaxies (IC~4742, NGC~1404, 1453, 5018, 5077) were modeled 
with two components, but their fits suffered from a variety of complications 
(e.g., nearby bright objects, central dust features, strong tidal features,
etc.) that preclude us from obtaining a more refined model. The remaining 18
galaxies require four components. Among these, three (ESO~221-G026, IC~2006, 
NGC~7029) need a fourth component to account for minor structures such as an 
edge-on disk or an outer ring; seven (IC~2311, NGC~3078, 4786, 5796, 6876, 
7145, 7196, 7507) contain an extra central, compact component of unspecified 
nature, but most likely attributable to unresolved nuclear dust features;
eight (IC~4797, NGC~584, 3904, 4033, 4697, 6673, 7144, 7192) we deem to be 
misclassified S0s.

The final decompositions for the sample are presented in Appendix~E, where
we show an atlas designed to help visualize the 1-D and 2-D information of
each best-fit model.  Table~1 lists the parameters for the final models.
Table~2 summarizes the statistical properties of the parameters for 
the subcomponents of the 70 galaxies fitted with a three-component model.

\vskip 0.3cm
\centerline{\psfig{file=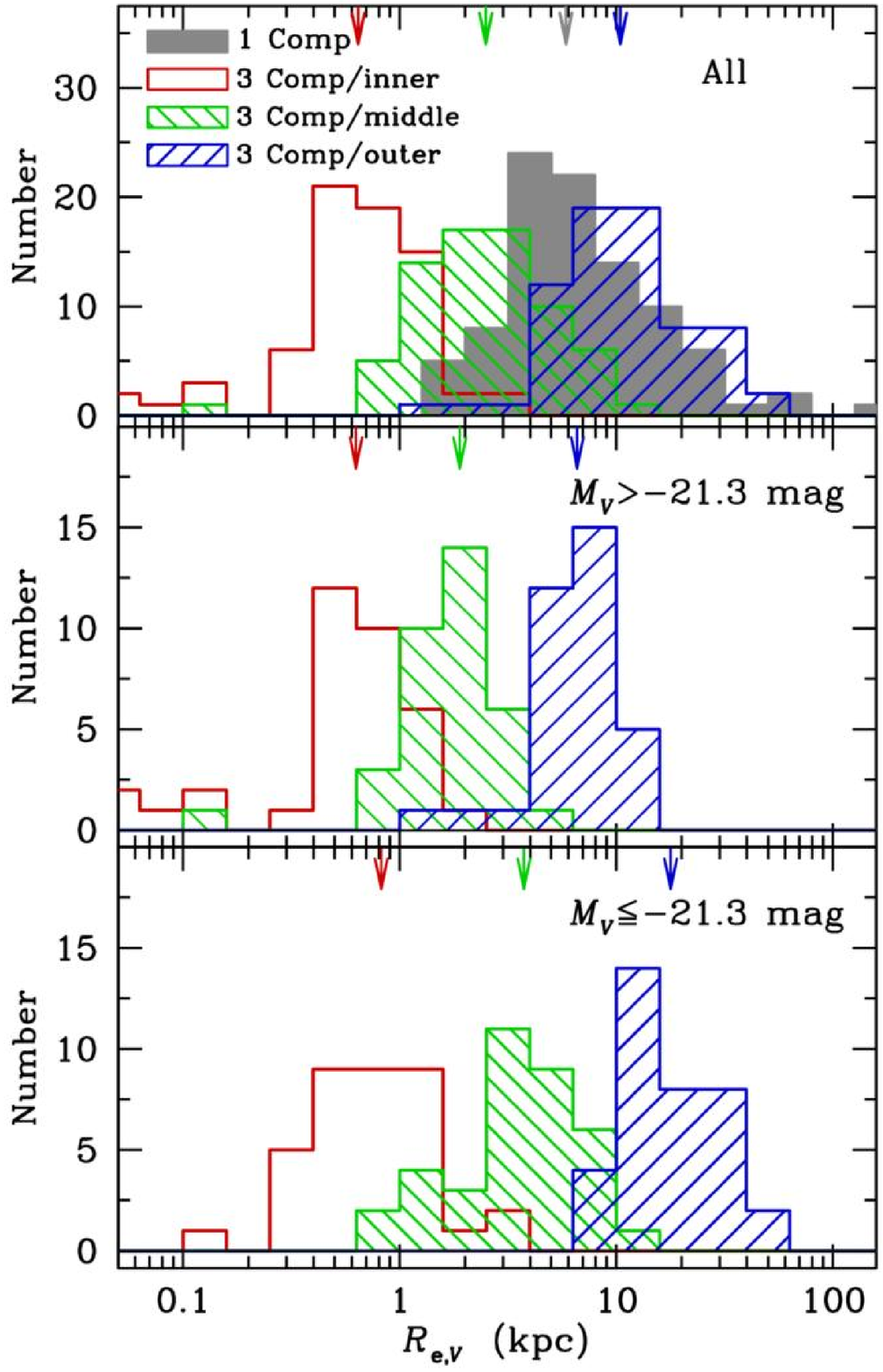,width=8.90cm}}
\figcaption[fig17.eps]{
Distribution of effective radius in the $V$ band.  The top panel 
compares the individual subcomponents of the three-component model with the 
single-component model.  The middle and bottom panels separate the sample into 
the low-luminosity and high-luminosity groups, respectively, defined to be 
those above and below the median value of $M_V = -21.3$ mag.  The median 
values of $R_{e,V}$ for the various components are indicated by the arrows.
\label{fig17}}
\vskip 0.45cm

\subsection{Uncertainties}

It is nontrivial to calculate errors for the parameters of our multi-component
fits.  Lessons learned in the context of single-component fitting of galaxy 
images (e.g., Guo et al. 2009; Yoon et al. 2011) do not easily translate to 
our case.  Because our galaxies are very bright and highly resolved, the
error budget is limited neither by signal-to-noise ratio nor by the 
PSF, which affects only the very central region.  Instead, the dominant source
of uncertainty comes from errors in sky background level estimation.  We 
approach this problem empirically, by manually adjusting the sky pedestal within 
the range of values constrained by our sky measurement procedure (Section~3.1
and Appendix~B) and systematically re-running the fits.   For a given parameter, 
the largest absolute variation normalized to the average value from the 
different outputs (e.g., $|\Delta {R_e}|/{R_e}$), gives its relative 
uncertainty.  On average, the relative uncertainties are quite small.  The 
parameters most vulnerable to sky uncertainty are the effective radius and the 
\ser\ index. For the inner component, the mean uncertainty for $R_e$ is 
$\sim$8\% and for $n$ $\sim$17\%.  They get worse for the other two 
components, increasing to 18\% and 35\%, respectively, for the middle 
component and 25\% and 39\% for the outer component.  These uncertainti-  

\vskip 0.8cm
\centerline{\psfig{file=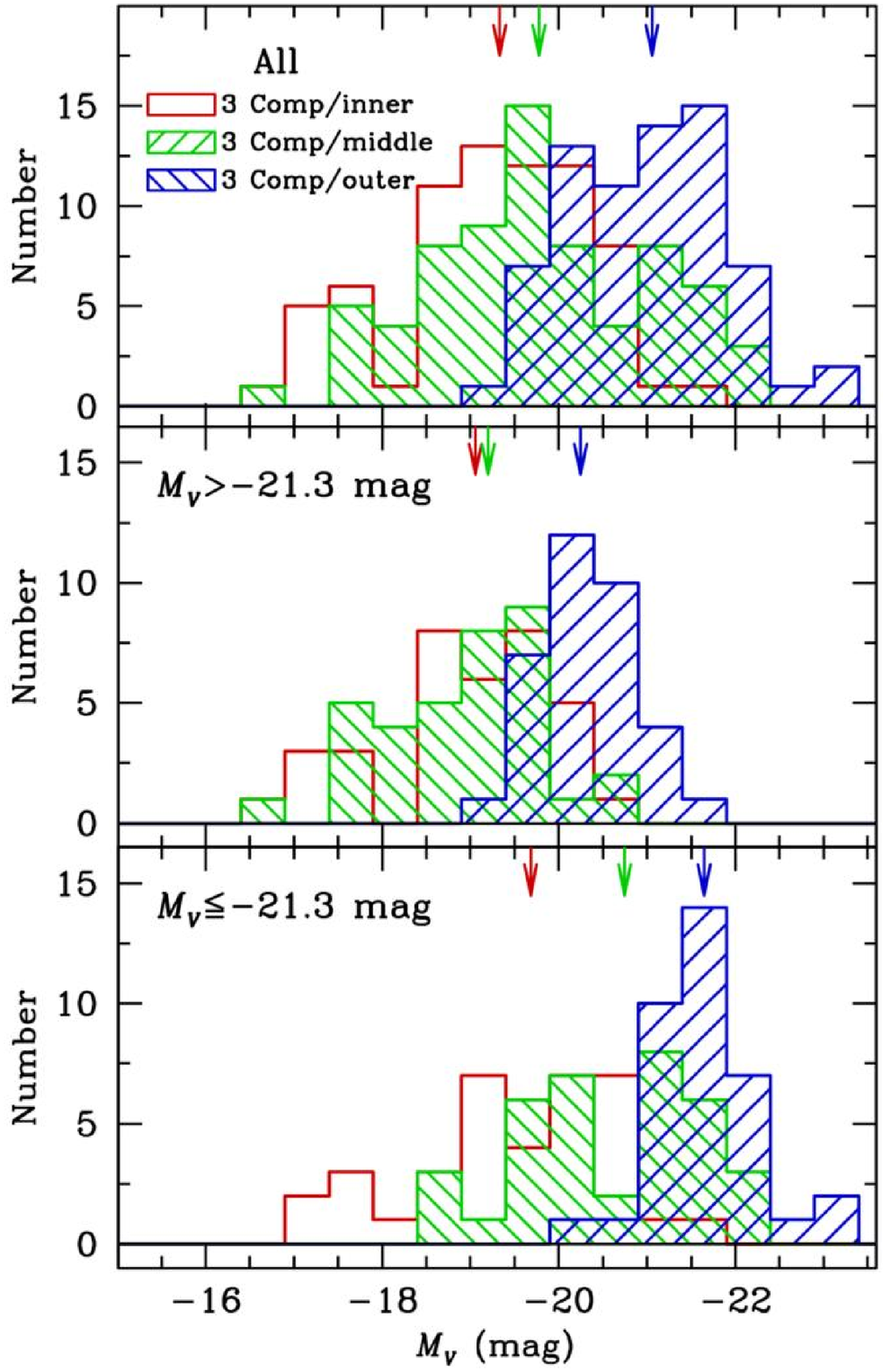,width=8.9cm}}
\figcaption[fig18.eps]{
$V$-band absolute magnitude of the subcomponents for the 
three-component models, corrected for Galactic extinction.  The three panels 
show the distribution for the entire sample and then separately for the 
low-luminosity and high-luminosity groups.
\label{fig18}}
\vskip 0.9cm

\noindent
es apply to galaxies for which ${\it R}_{\rm img}/{\it R}_{50}$ \gax\ 5; they can 
be significantly larger for more extended sources. Notwithstanding these 
complications, the main result of this paper --- that most nearby 
elliptical galaxies have a three-component structure --- is robust.


\section{Results}

The central result that emerges from this work is that the global photometric 
structure of nearby elliptical galaxies can be described most generally by 
three \ser\ components. Ranking them by physical size, we designate them the 
inner, middle, and outer components.  This section (see also Table~2) 
summarizes the statistical properties of key parameters for these these 
subcomponents and demonstrates that they follow well-behaved photometric 
scaling relations.  To put these results in context with the literature, we 
also compare them with results from more traditional single-component fits.  
As many properties of ellipticals are known to depend systematically on 
luminosity or stellar mass, we discuss the sample collectively as well as 
separately for a high-luminosity and a low-luminosity subsample, where the 
division between the two is taken to be $M_V = -21.3$ mag, the median $V$-band 
total luminosity. In a companion paper, we explore the implications of these 
results for  

\vskip 0.8cm
\centerline{\psfig{file=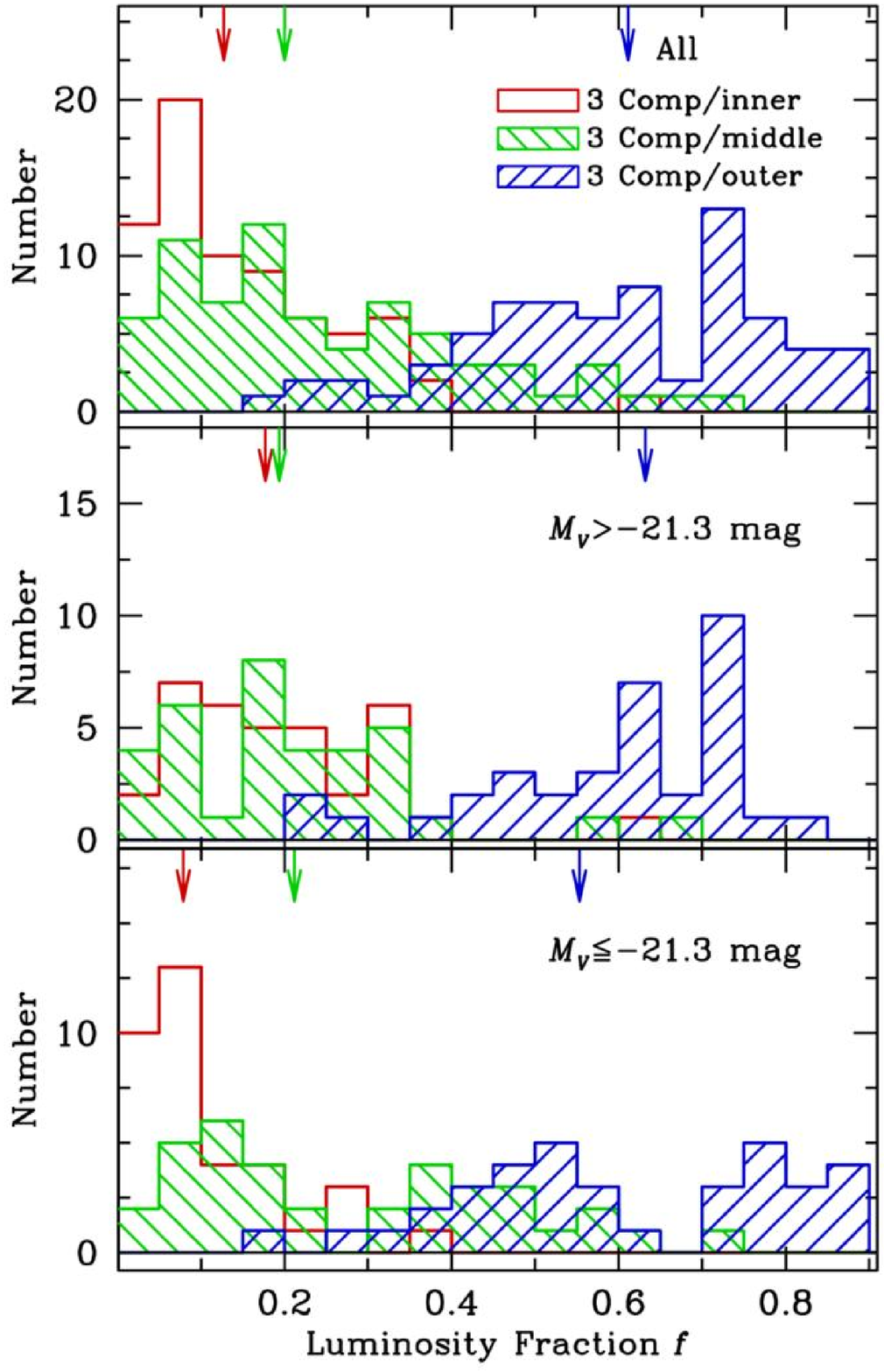,width=8.90cm}}
\figcaption[fig19.eps]{
Luminosity fraction of the subcomponents for the three-component 
models.  The three panels show the distribution for the entire sample and then 
separately for the low-luminosity and high-luminosity groups.
\label{fig19}}
\vskip 0.85cm

\noindent
the current paradigm of elliptical galaxy formation.  

\subsection{Distribution of Key Parameters}

The size and size ratios of the substructures offer useful insights into the 
size evolution of ellipticals, currently a topic of much research (e.g., Naab 
et al. 2009; van~Dokkum et al. 2010).  Figure~17 shows the distribution of 
effective radii.  For a traditional single-component \ser\ fit, the median 
$R_e$ is 5.8 kpc; this value is insensitive to whether the fit is done with or 
without a central mask. Applying our new fitting methodology, the three 
subcomponents become well separated in size, having median $R_e$ = 0.6, 2.5, 
and 10.5 kpc for the inner, middle, and outer components, respectively.  On 
average, high-luminosity ellipticals have larger effective radii for all three 
components. The outer component of the high-luminosity sample has a typical 
$R_e \approx 17.8$ kpc, 5 times larger than the middle component ($R_e = 3.5$ 
kpc); by contrast, in the low-luminosity group the outer component is only 
3 times as large as the middle component ($\sim 6.6$ vs. 1.9 kpc).  

In both groups the inner component is a compact, sub-kpc structure.  It has 
relatively low luminosity ($M_V \approx -19.3$ mag; Figure~18) and accounts 
for only 8\%--20\% of the light (Figure~19).   Most of the light resides in 
the middle and outer components, with the latter responsible for $\sim$60\% of 
the light in 

\vskip 0.2cm
\centerline{\psfig{file=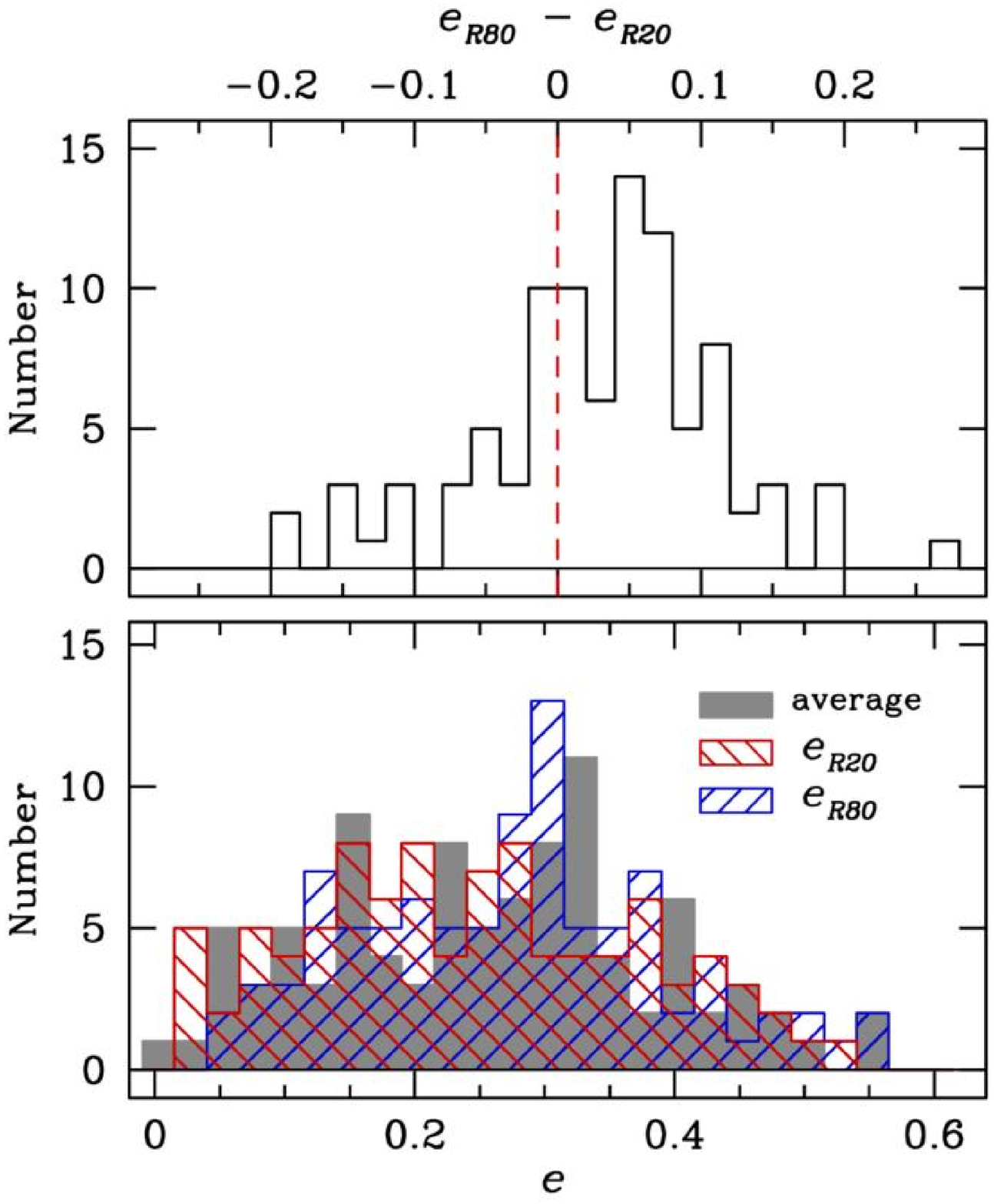,width=8.75cm}}
\figcaption[fig20.eps]{
Bottom: Distribution of the ellipticity averaged over the whole 
galaxy and as measured at two radii, $R_{20}$ and $R_{80}$.  Top: 
Distribution of the ellipticity difference between $R_{80}$ and $R_{20}$.
The red dashed line marks zero. The outer regions are generally more 
flattened than the inner regions.
\label{fig20}}
\vskip 0.9cm

\noindent
both luminosity groups and the former for $\sim 20\%$.  
Thus, the outer, most extended component dominates the luminosity---and 
presumably the mass---of nearby ellipticals.  

One of the most unexpected outcomes from our analysis is that the ellipticity
of the isophotes systematically increases toward large radii.  This is a
robust, model-independent result, which can be seen even from a straightforward
comparison of the 1-D ellipticity profile at $R_{20}$ and $R_{80}$.  Figure~20
shows that ${\it e}_{R80}- {\it e}_{R20}$ is skewed slightly ($\Delta e
< 0.1$), but definitely, toward positive values: the outer regions are
generally more flattened than the inner regions. This phenomenon is shown
more explicitly in Figure~21 for the 70 ellipticals decomposed with three 
components. As before, the outer component is, on average, more 
flattened than either the inner or middle components. When the sample is 
separated into the two luminosity bins, we see that the effect occurs almost 
exclusively within the high-luminosity subsample (panel b), not in the 
low-luminosity objects (panel c). The ellipticity distribution of the total 
sample exhibits another interesting feature: the middle component contains a 
high-$e$ tail. Closer inspection reveals that most of these highly flattened 
substructures reside in the low-luminosity sample, and they can be traced to 
small-scale, sometimes dusty disk-like structures embedded within the main body 
of the galaxy.

\begin{figure*}[t]
\centerline{\psfig{file=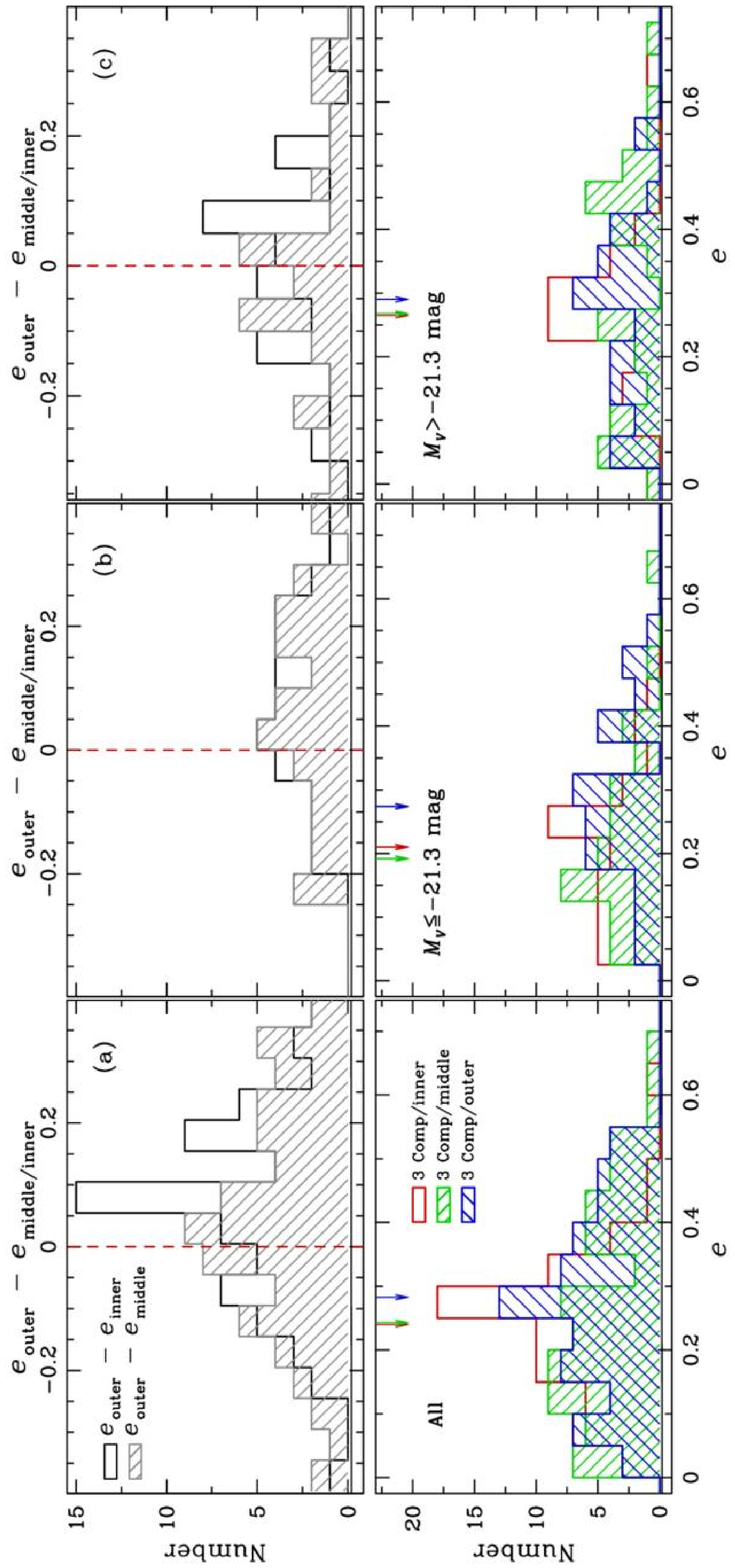,width=18.4cm,angle=270}}
\figcaption[fig21.eps]{
Ellipticity of the galaxies fit with a three-component model, 
grouped into (a) the entire sample, (b) high-luminosity galaxies with $M_V 
\leq -21.3$ mag, and (c) lower luminosity galaxies with $M_V > -21.3$ mag.  
For each group, the bottom panel gives the $e$ distribution of each of the 
subcomponents, with the respective median value indicated by arrows, and the 
top panel shows the difference between $e$ for the outer and middle/inner 
components.  Note that it is the high-luminosity objects that have a flattened 
outer component.
\label{fig21}}
\end{figure*}

Ever since the seminal work of Caon et al. (1993), elliptical galaxies have 
been increasingly modeled using the \ser\ function.  Apart from some 
investigations of brightest cluster galaxies (e.g., Gonz\'alez et al. 2005;
Donzelli et al. 2011), most studies describe the surface brightness 
distribution of the main body of ellipticals (after excluding the innermost 
and outermost regions; e.g., K09) using a single \ser\ index.  For relatively 
luminous ellipticals, the resulting \ser\ indices are large, typically $n 
\approx 3-6$, but occasionally as high as 10 or even greater.   Indeed, a 
high \ser\ index is often synonymous with early-type galaxies in general, 
and with ellipticals in particular.  In this work, we demonstrate that the 
overall structure of ellipticals, in fact, is best described by not one, but 
three photometric subcomponents.  It is thus not surprising that the \ser\ 
indices of the individual subcomponents turn out to be substantially smaller 
than the values for the single-component fits (Figure~22).  The inner, 
middle, and outer components all have rather similar \ser\ 
indices; their formal median values are $n = 2.0, 1.2$, and 1.6, respectively.
Very few fits yield $n = 4$.  On the other hand, a sizable fraction of the
components have $n$ \lax\ 1.  This is particularly noticeable for the 
middle component of the low-luminosity sample. In combination with the 
tendency for this component to appear flattened (Figure~21), it supports 
the notion that low-luminosity ellipticals contain embedded disks. 

We further note that the \ser\ index of the inner component for the 
high-luminosity subsample ($n = 1.5$) is significantly lower than that of 
the low-luminosity subsample ($n = 2.6$).  If we follow Lauer et al. (2007) 
and approximate the inner slope of the profile as a simple power law, 
$\mu(R) \propto R^{-\gamma\prime}$, the 
corresponding power-law index measured within the inner 3\asec\ is 
$\gamma\prime = -0.2$ for the high-luminosity 
sources and $\gamma\prime = -0.6$ for the low-luminosity sources.   This is 
remarkably close to the canonical values of the average inner slopes for 
``core'' and ``power-law'' (or ``extra-light'') ellipticals obtained from 
\emph{HST} observations (e.g., Lauer et al. 2007).  Although the resolution 
of the CGS images cannot compete with that of \emph{HST}, it is reassuring 
to know that our detailed decomposition of the ground-based data can 
essentially recover the gross properties of the nuclear regions.

Sky background errors greatly affect the \ser\ index.  Given 

\vskip -0.4cm
\centerline{\psfig{file=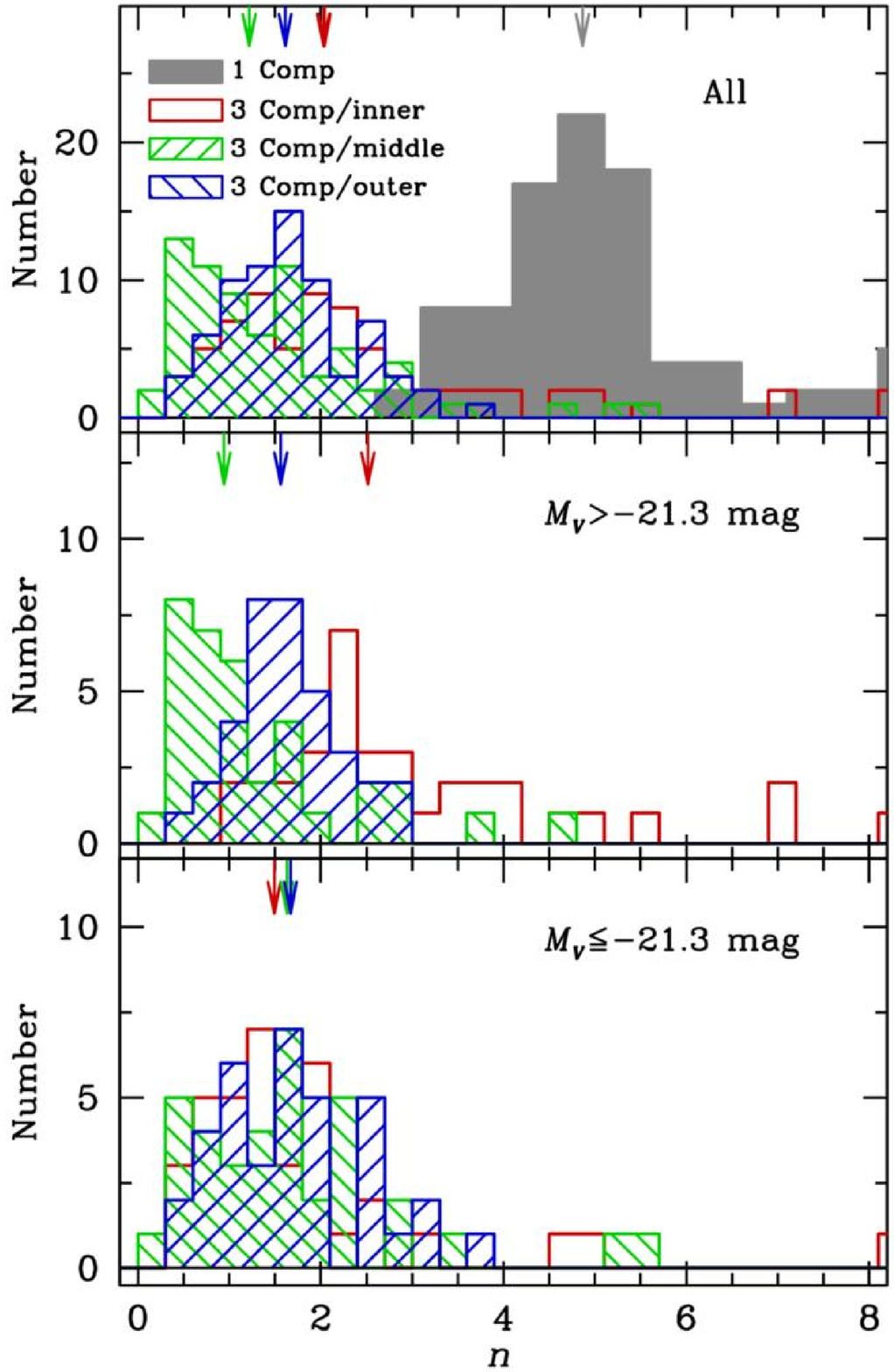,width=8.60cm}}
\figcaption[fig22.eps]{
Distribution of \ser\ index $n$ for the entire sample and then 
separately for the low-luminosity and high-luminosity groups.  The top 
panel also includes the distribution of $n$ for the single-component models.
\label{fig22}}

\vskip 0.3cm
\centerline{\psfig{file=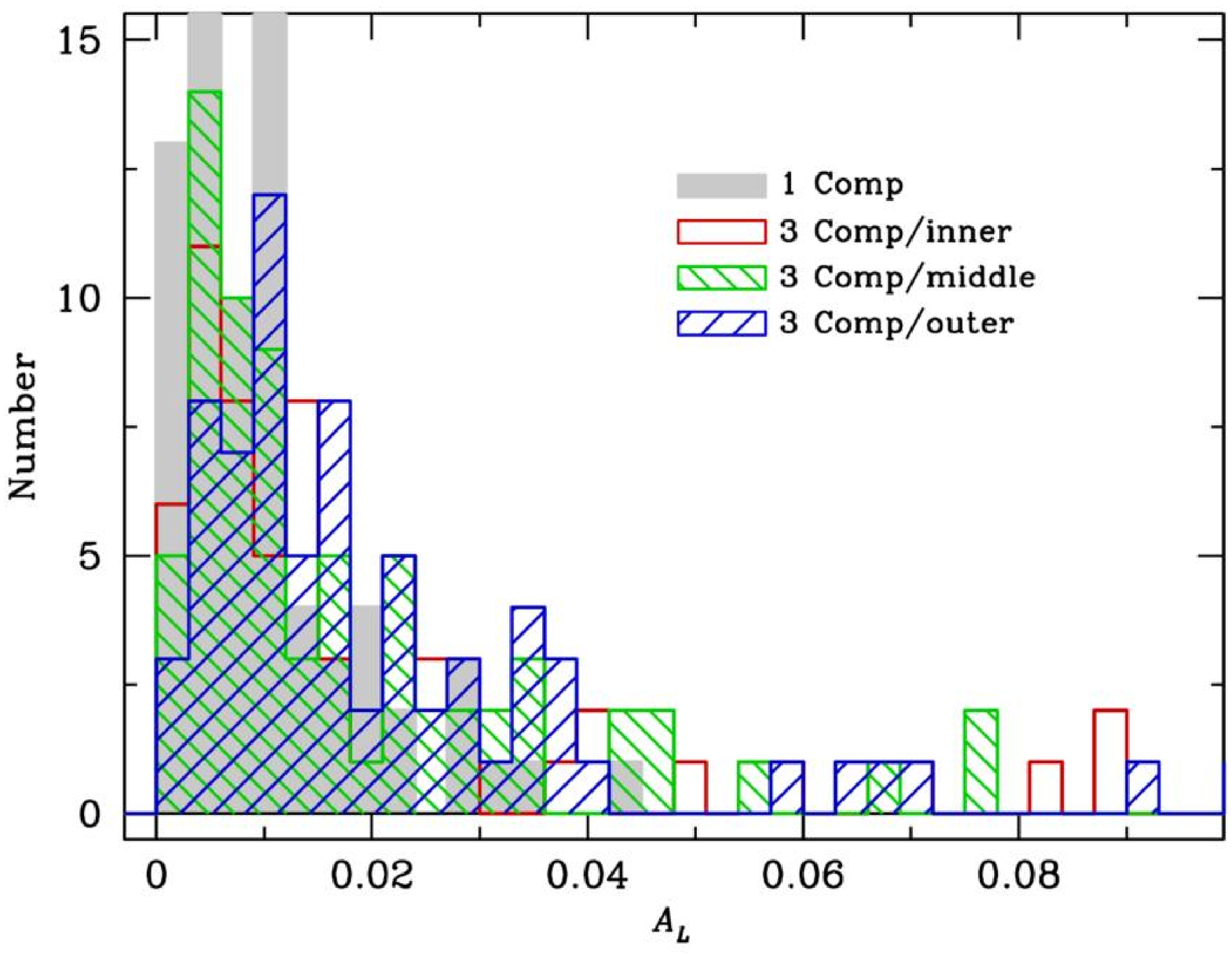,width=9.00cm}}
\figcaption[fig23.eps]{
Distribution of the global asymmetry index, $A_L$, defined as the 
absolute value of the amplitude of the $m=1$ Fourier mode, shown 
separately for the subcomponents of the three-component fits and 
the single-component fits.
\label{fig23}}
\vskip 0.55cm

\noindent 
the complications 
involved in our sky determination (Section~3.1 and Appendix~A), we are wary of 
over-interpreting the results on the \ser\ indices.  To test the robustness 
of our conclusions, we re-examined the statistics for the subset of 48 objects 
with image sizes larger than $6\times R_{50}$, for which we have very high 
confidence in their sky estimates.  Reassuringly, none of the main conclusions
change qualitatively.

Lastly, we look at the output from the Fourier analysis.  Apart from some 
applications limited to quasar host galaxies (e.g., Kim et al. 2008), to our 
knowledge the Fourier mode capability of \galfit\ 3.0 has yet to be applied 
extensively to study galaxy structure.  On the one hand, it is reassuring that 
the Fourier modes do not strongly affect the results of our final models; on 
the other hand, we also find that the Fourier modes of each component are 
affected by different sources of uncertainties. For example, even after 
improving the object masks, nearby saturated stars and small galaxies can 
still greatly affect the amplitude of the Fourier modes.  Dust features are
also problematic.  Given this complexity, we first restrict our attention to the 
$m=1$ mode.  As the lowest order mode, it should be more reliable, and its 
interpretation (global lopsidedness) is least ambiguous. Given the overall 
symmetry of our elliptical galaxies, we expect---and find---that the amplitude 
of the $m=1$ Fourier mode for each component to be very small.  
We also find that the $m=1$ Fourier mode is actually not very sensitive to 
low-surface brightness features found in the outskirts of ellipticals.  Faint 
tidal features contribute little to the global lopsidedness.  Properly fitting 
these features would require higher-order modes or additional model components, 
which is beyond the scope of this work.

Figure~23 shows the distribution of the parameter $A_L$, defined as the 
absolute value of the amplitude of the $m=1$ Fourier mode (Equation 24 of 
Peng et al. 2010), calculated for both single- and multi-component models. 
Not surprisingly, the individual subcomponents attain systematically higher 
values of $A_L$ than the global, single-component fit, because they are more 
sensitive to local perturbations.  However, we find that not all high-
$A_L$ components actually correspond to strong lopsidedness; instead, in some
cases they seem to arise from different sources of contamination.  

\vskip 0.1cm
\centerline{\psfig{file=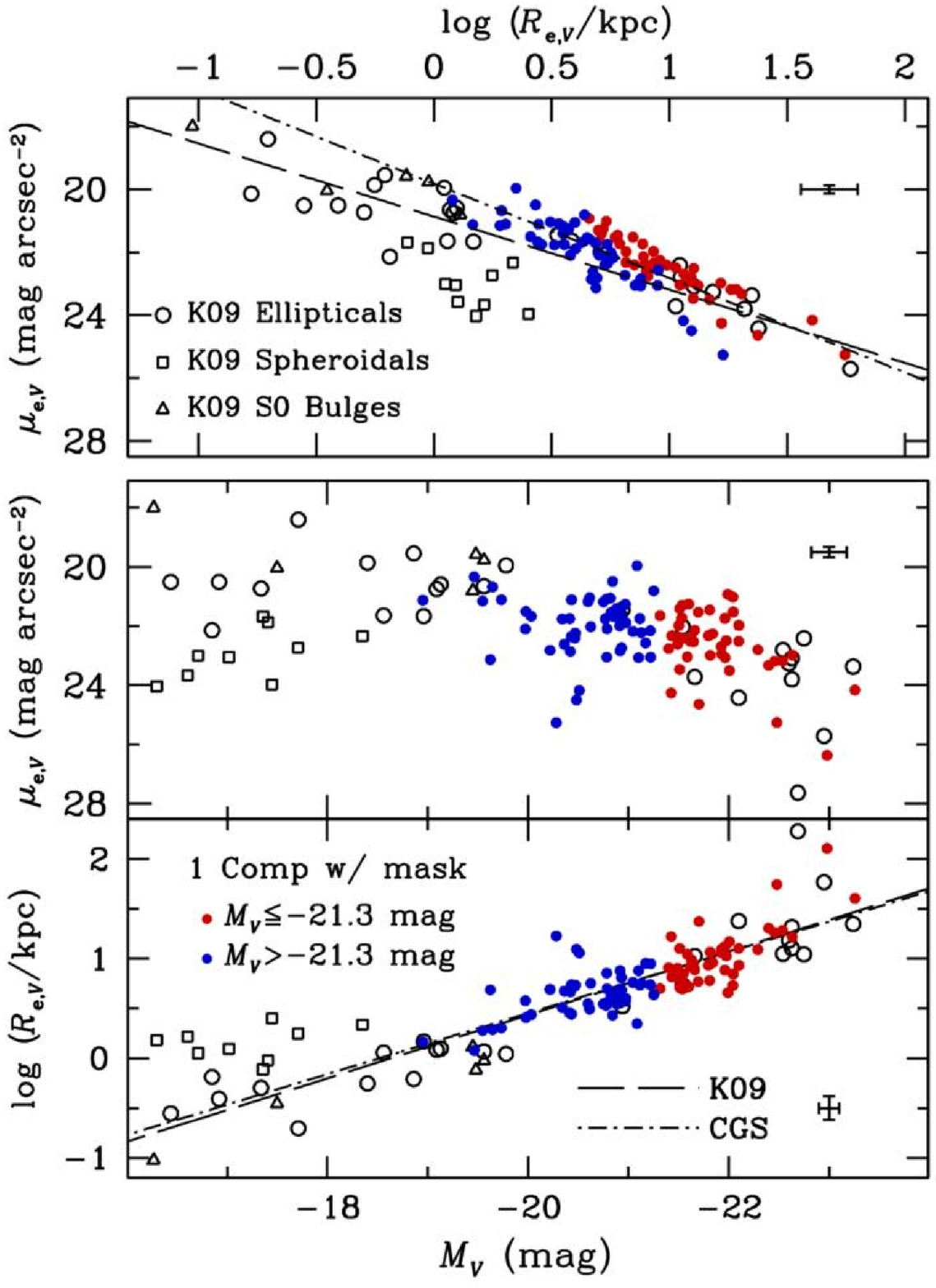,width=8.9cm}}
\figcaption[fig24.eps]{
Correlations among luminosity, effective radius, and effective 
surface brightness for the 94 CGS ellipticals, compared to the sample of 
early-type (elliptical, spheroidal, and S0) galaxies from K09.  For 
consistency with K09's approach, we fit our galaxies using a single-component 
\ser\ model with a central mask applied.  We divide our sample into 
high-luminosity ($M_V \leq -21.3$ mag; red points) and low-luminosity 
($M_V > -21.3$ mag; blue points) groups. Representative uncertainties are 
given in the panels.  The lines on the $\mu_{e}-R_e$ and $M_V-R_e$ plots are 
linear least-squares fits to the K09 and CGS samples.
\label{fig24}}
\vskip 0.4cm

\subsection{Photometric Scaling Relations}

Galaxies follow a number of scaling relations, whose slope and scatter provide 
insights to diagnose their formation and evolution.  Among these relations, 
the ones that involve purely photometric parameters have been used most 
extensively as tools for understanding the structure of different types of 
galaxies.  The mutual correlations among luminosity, effective radius ($R_e$), 
and surface brightness at the effective radius (${\it \mu}_e$) have been most 
frequently used to reveal the physical connections within the family of hot
stellar systems.  K09 (and references therein) extensively apply this tool to
understand the relationships between ellipticals, bulges, and spheroidals, and
Kormendy \& Bender (2012) extend this analysis to the subcomponents within S0 
galaxies.  In this section, we use these very same tools to explore the physical
nature of the three subcomponents in ellipticals.

We begin by verifying that the CGS ellipticals comply with the global 
photometric scaling relations obeyed by other samples of well-studied 
early-type galaxies.  This comparison is made in Figure~24, where we use the 
sample of K09 as reference. For consistency with K09's approach, we show 
single-component fits with a central mask, although the results without a 
central mask are very similar. Our objects do not extend to the lowest 
luminosities covered by K09's sample, but over most 

\vskip 0.3cm
\begin{figure*}[t]
\centerline{\psfig{file=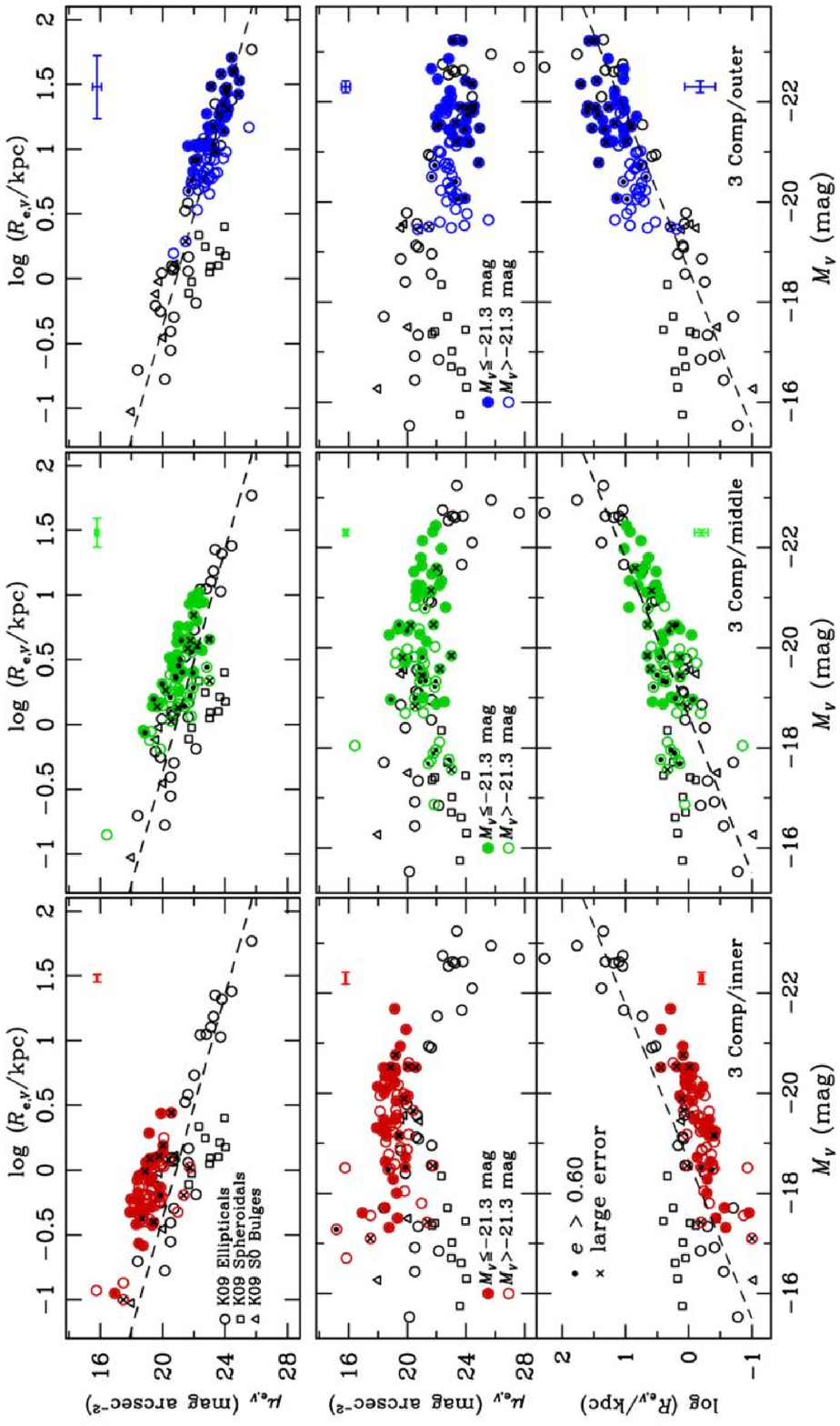,width=18.8cm,angle=270}}
\figcaption[fig25.eps]{
Correlations among luminosity, effective radius, and effective
surface brightness for each of the subcomponent in the three-component 
models (left: inner; center: middle; right: outer).  The typical uncertainties 
for these properties are illustrated using the error bar on the right side 
of each plot; points with the largest uncertainties are marked with a black 
cross.  Components with $e > 0.6$ are further highlighted with a small black 
dot.  The CGS sample is divided into two luminosity groups, denoted by the 
solid ($M_V \leq -21.3$ mag) and open ($M_V > -21.3$ mag) colored points.  As 
in Figure~24, data from K09 are overplot for reference; the dashed lines are 
linear least-squares fits to the K09 ellipticals.
\label{fig25}}
\end{figure*}
\vskip 0.3cm

\noindent
of the luminosity range in 
which they do overlap the two samples follow very similar, tight correlations in 
the $M_V-R_e$ and $R_e-\mu_e$ plots; their formal linear least-squares fits 
(dashed and dot-dashed lines) are consistent. The distributions of the two
samples also show good agreement in the $M_V-\mu_e$ plane.  It is worth noting
that in these plots the high- and low-luminosity subsamples merge seamlessly 
without any systematic differences in slope or scatter.

Figure~25 repeats this sequence of plots, but now showing each of the 
three subcomponents separately.  The linear least-squares fits for the 
K09 sample are shown for reference in the $M_V-R_e$ and $R_e-\mu_e$ plots.
Unlike K09, we do not circularize the $R_e$ values, but this difference is 
only significant for components with high ellipticity; we highlight the points 
with $e>0.6$, as well as those with relative uncertainties larger than 30\%.
The purpose of this exercise is two-fold.  First, we wish to establish whether
the individual subcomponents identified in our fits follow any trends in these 
photometric scaling relations.  If, for example, the subcomponents have no 
physical significance but are merely artifacts of decomposition, we would not 
expect them to display any coherent, let alone tight, relations.  Second, if 
well-behaved patterns do indeed emerge, we hope to use them to gain insights 
into the physical nature of the subcomponents, in much the same way as 
these relations have helped to guide our understanding of the physical 
processes governing various types of galaxies (Kormendy 1985; Bender et al.
1992; Burstein et al. 1997; K09) and major substructures within 
them (e.g., Kormendy \& Bender 2012).  We summarize the key points to note:

\begin{enumerate}

\item The three major subcomponents \emph{do} follow well-defined loci 
  in the $M_V-R_e$, $R_e-\mu_e$, and $M_V-\mu_e$ planes.

\item The inner component occupies a region of parameter space similar to that 
  populated by the lowest luminosity ellipticals and the classical bulges of S0 
  galaxies.  The correlations for the high- and low-luminosity subsamples 
  are equally tight. It is especially tight for the luminosity-size relation, 
  where the slope seems to be flatter than that of the global relation. 

\item The middle components of the high-luminosity sources define tight 
  relations in the $M_V-R_e$ and $M_V-\mu_e$ planes that closely track the 
  global ones.  By contrast, low-luminosity sources contain middle components 
  that overlap surprisingly closely with the locus of spheroidal galaxies.  The 
  same trend is present, albeit more subtly, in the $R_e-\mu_e$ plot; at a 
  fixed $R_e$, the low-luminosity group is displaced toward lower surface 
  brightnesses, in the general direction of spheroidals.  Recall that the 
  middle component of low-luminosity ellipticals exhibits attributes 
  (low \ser\ indices, high ellipticity, occasionally dust features) reminiscent
  of disks (Section~4.1).

\item The outer component behaves very similarly for both luminosity groups.
  The most noticeable (but predictable) feature is their offset relative to 
  the global luminosity-size relation: at fixed $V$-band luminosity, the outer 
  components have larger $R_e$ than predicted from the correlation based on 
  global properties. The scatter is also larger, and the slope is marginally 
  flatter.  A formal linear least-squares fit for the entire sample results in 
  $M_V ({\rm mag}) = (-3.76\pm 0.65) \log R_e ({\rm kpc})-(0.23\pm 0.03)$, to be 
  compared with $M_V ({\rm mag}) = (-6.14\pm 0.48) \log R_e ({\rm kpc})-
  (0.33\pm 0.03)$ for the 
  combined collection of ellipticals and S0 bulges from K09. 
  As for the $R_e - \mu_e$ relation, the outer component is found
  to follow a very similar trend, albeit with somewhat larger scatter, as the 
  one defined by the global properties of elliptical galaxies.  In this context, 
  it is worth mentioning that the even more extended intra-cluster light 
  component of galaxy clusters also exhibits a correlation between effective 
  radius and mean surface brightness within that radius; the slope is similar
  but the scatter is larger than the relation of the brightest cluster members
  (Zaritsky et al. 2006).  Although the outer component of ellipticals 
  identified in this work is in general different from the intra-cluster light, 
  it is tempting to speculate, based on the similarity of their scaling
  relations, that they might share a common origin.
  It is also interesting to note that the distribution of the outer 
  component in the three planes shares some similarities with the trends for 
  the \emph{disks} of S0 galaxies, as presented by Kormendy \& Bender (2012; 
  see their Figure~17).  While we do not suggest that the outer component of 
  ellipticals are strictly analogous to S0 disks, it is intriguing that their 
  isophotes do tend to be somewhat flattened (Section~4.1).
  
\end{enumerate}

Apart from the primary mutual relations involving luminosity, size, and 
surface brightness, some studies have explored additional correlations 
involving the \ser\ index. Caon et al. (1993) and Trujillo et al. (2001), for 
example, discuss the general tendency for the \ser\ index to rise with 
increasing effective radius and decreasing effective surface brightness.  As 
previous investigations were based on global \ser\ indices measured from 
single-component fits, it is worth revisiting this issue using the newly 
derived indices based on our subcomponent decomposition, which we have shown 
to be significantly different from those obtained from traditional fits.  While 
our single-component fits do indeed recover the inverse correlation between 
\ser\ $n$ and $\mu_e$ (Figure~26, grey points), it is clear that, with the 
possible exception of the inner component of the low-luminosity subsample, the 
correlation completely disappears for the separate subcomponents. The 
uncertainties of the individual \ser\ indices are considerable (typical error 
bars shown on the lower-right corner of plot), especially for the outer 
component, but they cannot account for the huge observed scatter.

The redistribution of the \ser\ indices from large (e.g., $n$ \gax\ 4) to 
small values can be understood as a simple consequence of the behavior of 
the \ser\ function.  A \ser\ profile with a high $n$ has both a centrally 
concentrated core \emph{and} an extended envelope.   In this study, we 
promote the notion that the light distribution of ellipticals, particular the 
most luminous ones, separates into three portions that occupy small, 
intermediate, and extended scales.  Whereas the global profiles of these 
systems were once characterized by a single \ser\ function with large $n$, 
in our new description the compact and extended subcomponents each naturally 
accommodates a lower value of $n$.  Galaxies with the highest luminosity and 
mass (and, via the 

\vskip 0.8cm
\centerline{\psfig{file=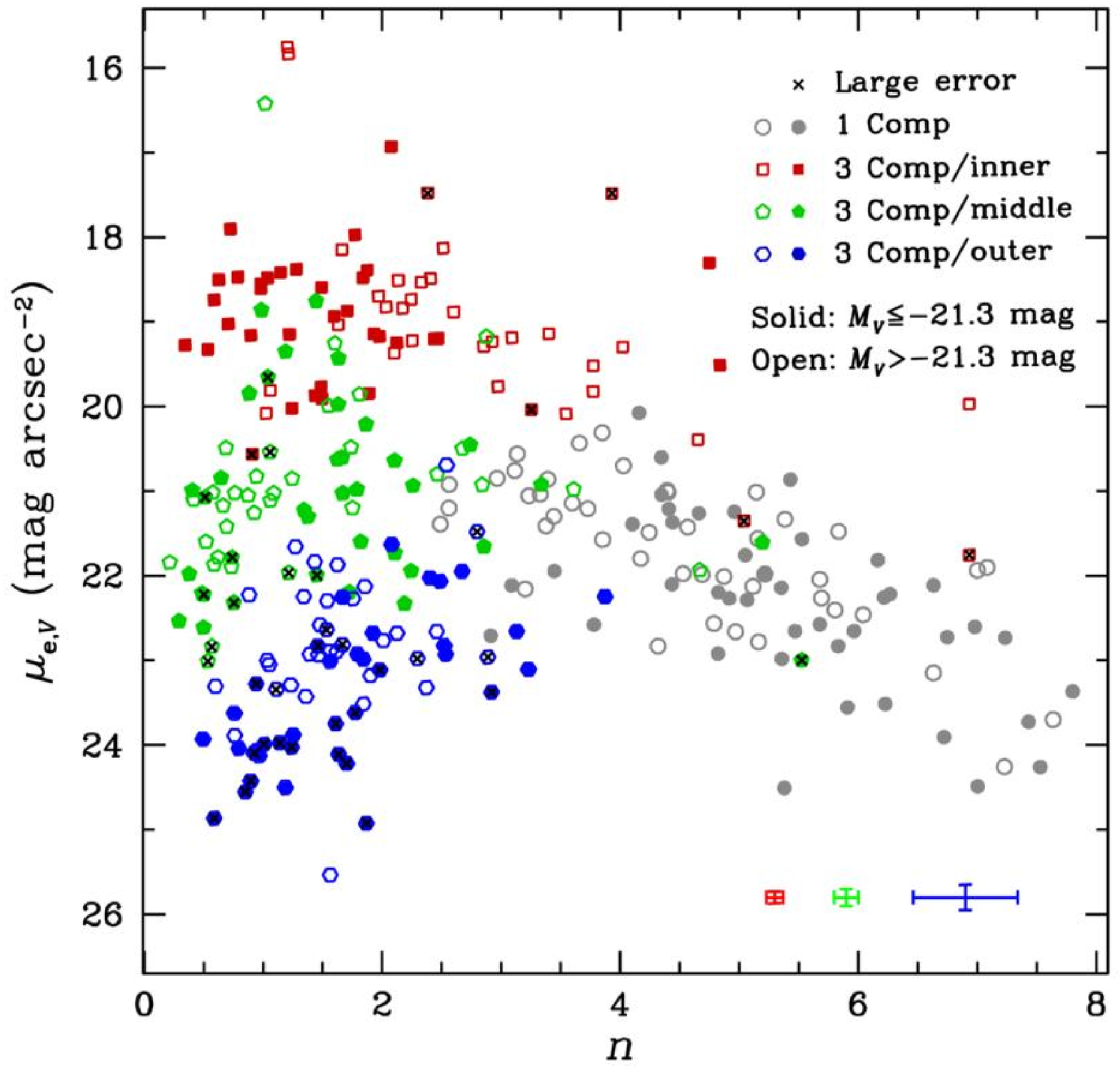,width=8.95cm}}
\figcaption[fig26.eps]{
Correlation between \ser\ index and the surface brightness at the 
effective radius.  We show the results for single-component fits and each of 
the subcomponents of the three-component models.  Points with large 
uncertainties are marked with a black cross, and the two luminosity groups are 
denoted by the solid ($M_V \leq -21.3$ mag) and open ($M_V > -21.3$ mag) symbols.
Typical error bars are given on the bottom-right corner of the plot.
\label{fig26}}
\vskip 0.8cm

\noindent
above-described scaling relations, the largest $R_e$ and 
lowest $\mu_e$) traditionally have had the largest values of \ser\ $n$ because 
they are the systems that have managed to build up an extended envelope.  
In the following paper of this series, this viewpoint will be explored in the 
context of the many recent efforts to quantify the evolution of the size 
and global \ser\ index of massive early-type galaxies.  

\subsection{The Stellar Mass-Size Relation}

The stellar mass-size relation offers another important diagnostic tool to 
investigate the formation and assembly of galaxies, both at low and high 
redshifts.  In the local Universe, it is well known that early-type galaxies 
follow a steeper stellar mass-size relation than late-type galaxies (Shen et 
al. 2003; Guo et al. 2009).  This difference points to potentially different 
formation physics and evolutionary pathways.  We wish to examine the stellar 
mass-size relation in light of our three-component decomposition of ellipticals.

For this analysis we need stellar masses.  While accurate stellar masses are 
difficult to estimate for galaxies, the relatively massive, local elliptical 
galaxies in the luminosity range of our sample have roughly uniformly old 
stellar population (e.g., Kuntschner 2000), which, to a first approximation,
can be assumed to have a single stellar mass-to-light ratio ($M/L$).  This 
assumption enables us to estimate reasonable stellar masses using empirical 
relations between broad-band optical colors and stellar $M/L$.  Different colors 
and choice of models, which depend on the stellar population library, initial 
mass function, and star formation history, can lead to systematically different 
masses.  As there is no clear evidence as to which combination is superior, we 
take the median of six mass estimations using the $B-V$ and $B-R$ relations from
Bell \& de~Jong (2001), Bell et al. (2003), and Zibetti \& Ferguson (2009).  
The standard deviation of these six estimates is taken to be the uncertainty,

\vskip 0.3cm
\centerline{\psfig{file=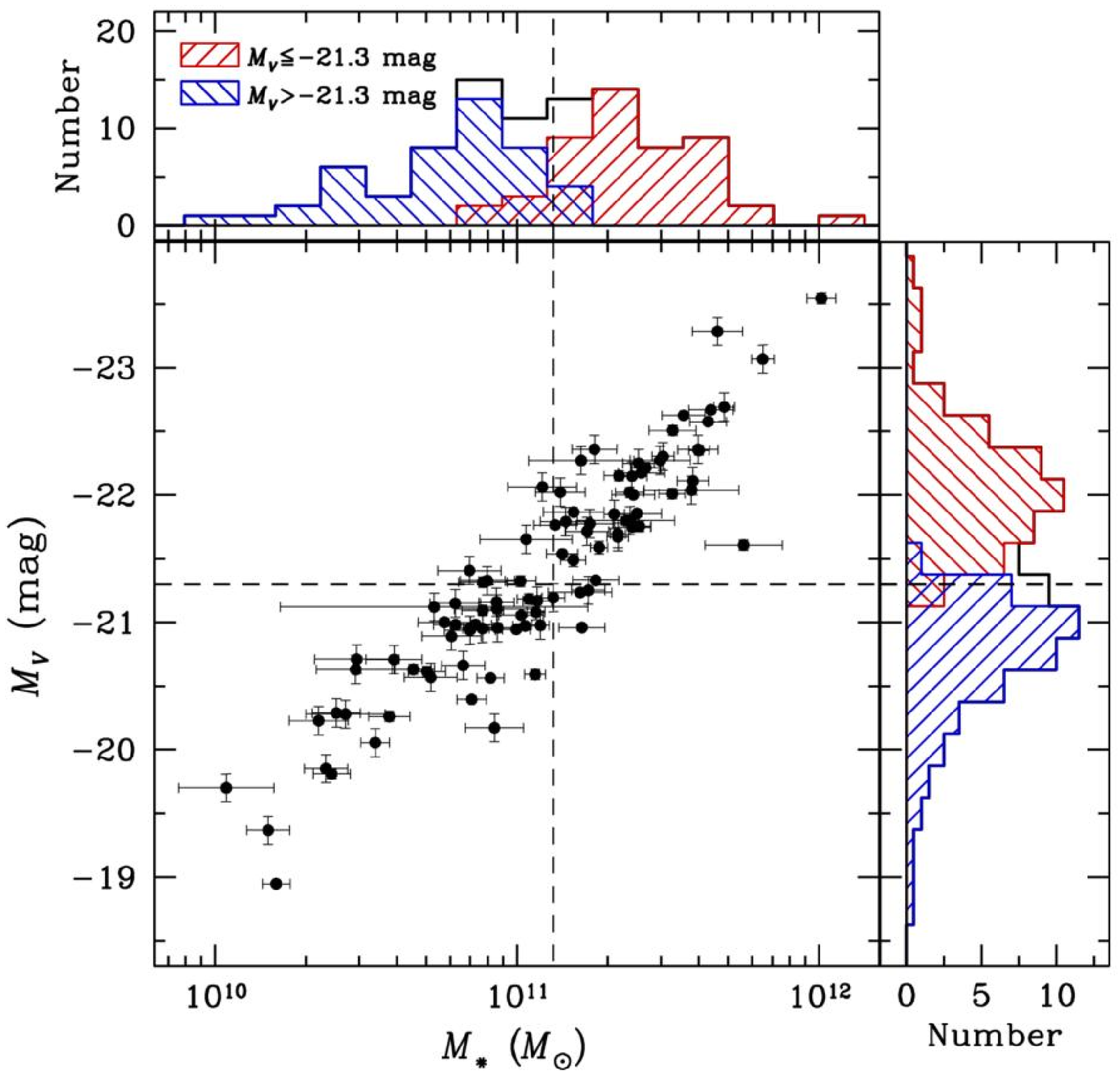,width=8.75cm}}
\figcaption[fig27.eps]{
Distribution total $V$-band absolute magnitude versus stellar 
mass, estimated from the optical colors as described in Section 4.3.  The 
final adopted mass for each galaxy is the median of the estimates
based on different colors and mass-to-light ratio relations, and the 
uncertainty is taken to be the scatter among these different estimates.
The median value of $M_V = -21.3$ mag corresponds to a median mass of $M_* = 
1.3\times10^{11}$ \solmass.  Throughout the paper, we define the 
high-luminosity and low-luminosity groups according to the median luminosity 
of the sample.
\label{fig27}}
\vskip 0.3cm

\noindent
which on average is 0.11 dex for $\log M_*$.   To estimate stellar masses for 
the subcomponents, whose colors are not yet robustly known, we need to make an
additional assumption that they have the same $M/L$ as the global value. 
Considering that the observed optical color gradients of the ellipticals are 
quite weak (Papers I and II), this is probably not an unreasonable assumption. 
In any case, the errors incurred from assuming a constant $M/L$ most likely do 
not exceed those associated with the image decomposition.

Figure~27 plots the final stellar masses versus $V$-band total absolute 
magnitude. The high- and low-luminosity subsamples are well separated in 
stellar mass. The median $V$-band luminosity ($M_V = -21.3$ mag) used to 
separate the two luminosity subsamples translates to $M_*= 1.3\times10^{11} 
\,M_{\odot}$.  In Figure~28 and Figure~29, the stellar mass-size and stellar
mass-luminosity fraction relations are shown. Our single-component fits recover 
a stellar mass-size relation that agrees with the relation defined by early-type
galaxies in Guo et al. (2009). For this comparison, we use the fits without a 
central mask; the fits that include a central mask yield a relation with 
somewhat smaller scatter but that still agrees well with that of Guo et al. 
(2009).

The following trends emerge for the individual subcomponents:

\begin{enumerate}

\item The inner components of the high-luminosity ellipticals delineate a 
  surprisingly tight sequence in the $M_* - R_e$ plane.  It runs approximately 
  parallel to, but is significantly offset below, that of the global relation 
  for early-type galaxies.  Aside from the somewhat larger scatter, the 
  same pattern can be seen for the lower-luminosity systems.  At a fixed stellar 
  mass, the inner component achieves much higher densities than the global 
  average.  

\item The behavior of the middle component is significantly different for the
  two luminosity groups.  For the high-luminosity subsample, the middle 
  component, with the exception of a few low-mass outliers, defines a 
  tight $M_* - R_e$ correlation that lies essentially on top of the best-fit 
  relation for early-type galaxies from Guo et al. (2009).  By contrast, the 
  points for the low-luminosity subsample show not only much more 
  scatter but also a shallower slope, one that roughly resembles that 
  delineated by the late-type SDSS galaxies (Guo et al. 2009).  In view of 
  the disky nature of the middle component in low-luminosity ellipticals 
  (Section 4.1), perhaps this similarity is to be expected. 

\item The outer component in both luminosity groups occupies a distinct 
  sequence offset above the global relation: at fixed mass, the most extended 
  envelope is significantly more diffuse than the global average.  The scatter 
  is also notably larger, perhaps reflecting the stochastic nature in which 
  this component was assembled or a more protracted history of formation.

\end{enumerate}

\vskip 0.3cm
\centerline{\psfig{file=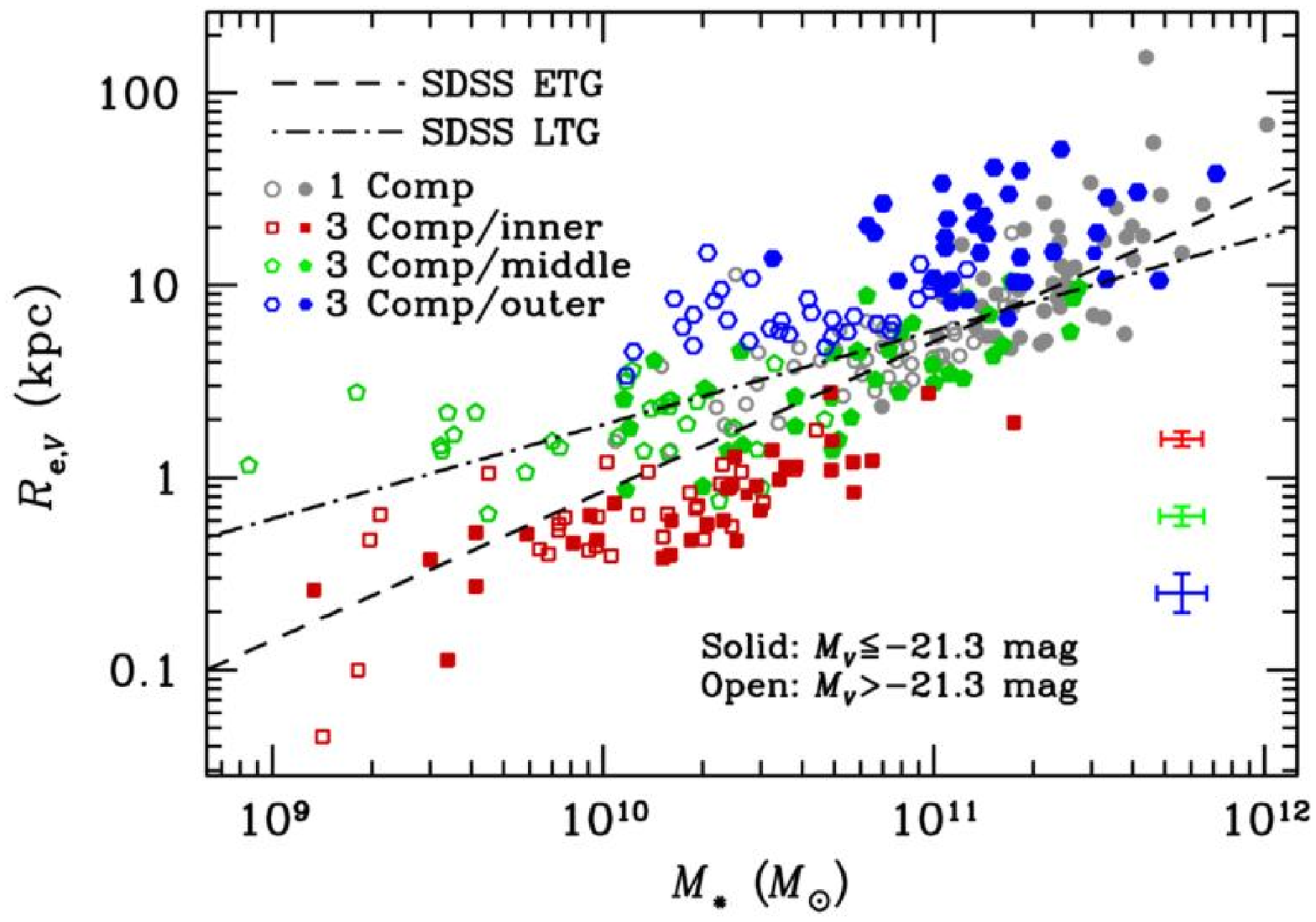,width=9.00cm}}
\figcaption[fig28.eps]{
Dependence of the size of the different subcomponents on the total 
stellar mass of the galaxy.  For reference, we also show the single-component 
models.  The two luminosity groups are denoted by the solid ($M_V \leq -21.3$ 
mag) and open ($M_V > -21.3$ mag) symbols.  The lines show the stellar mass-size 
relation for early-type galaxies (ETG) and late-type galaxies (LTG) derived 
from the SDSS analysis of Guo et al. (2009).
Typical error bars are given on the bottom-right corner of the plot.
\label{fig28}}
\vskip 0.3cm

\vskip 0.3cm
\centerline{\psfig{file=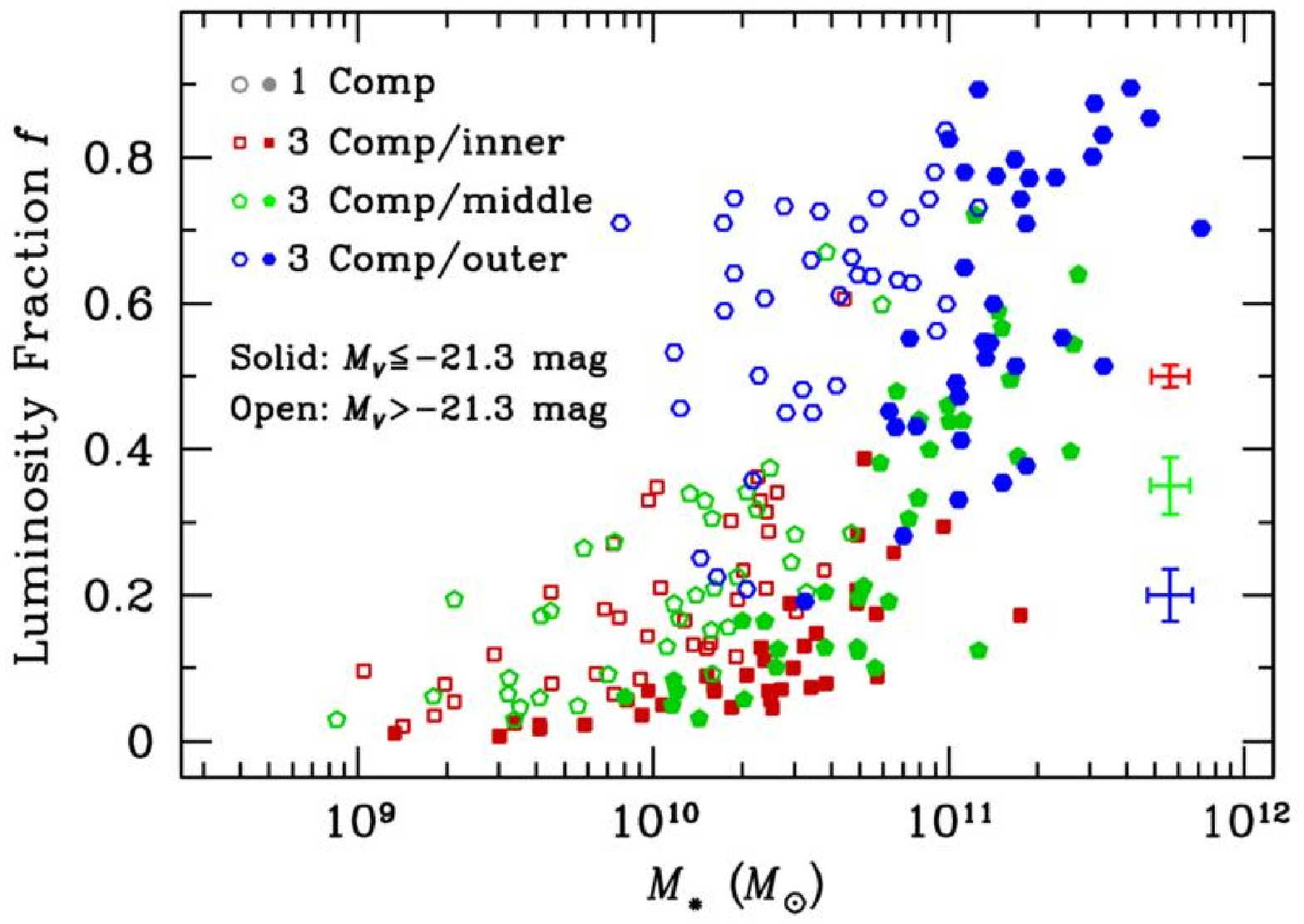,width=9.00cm}}
\figcaption[fig29.eps]{
Dependence of the luminosity fraction of the different
subcomponents on the total stellar mass of the galaxy.
Typical error bars are given on the bottom-right corner of the plot.
\label{fig29}}
\vskip 0.3cm

\begin{figure*}[t]
\centerline{\psfig{file=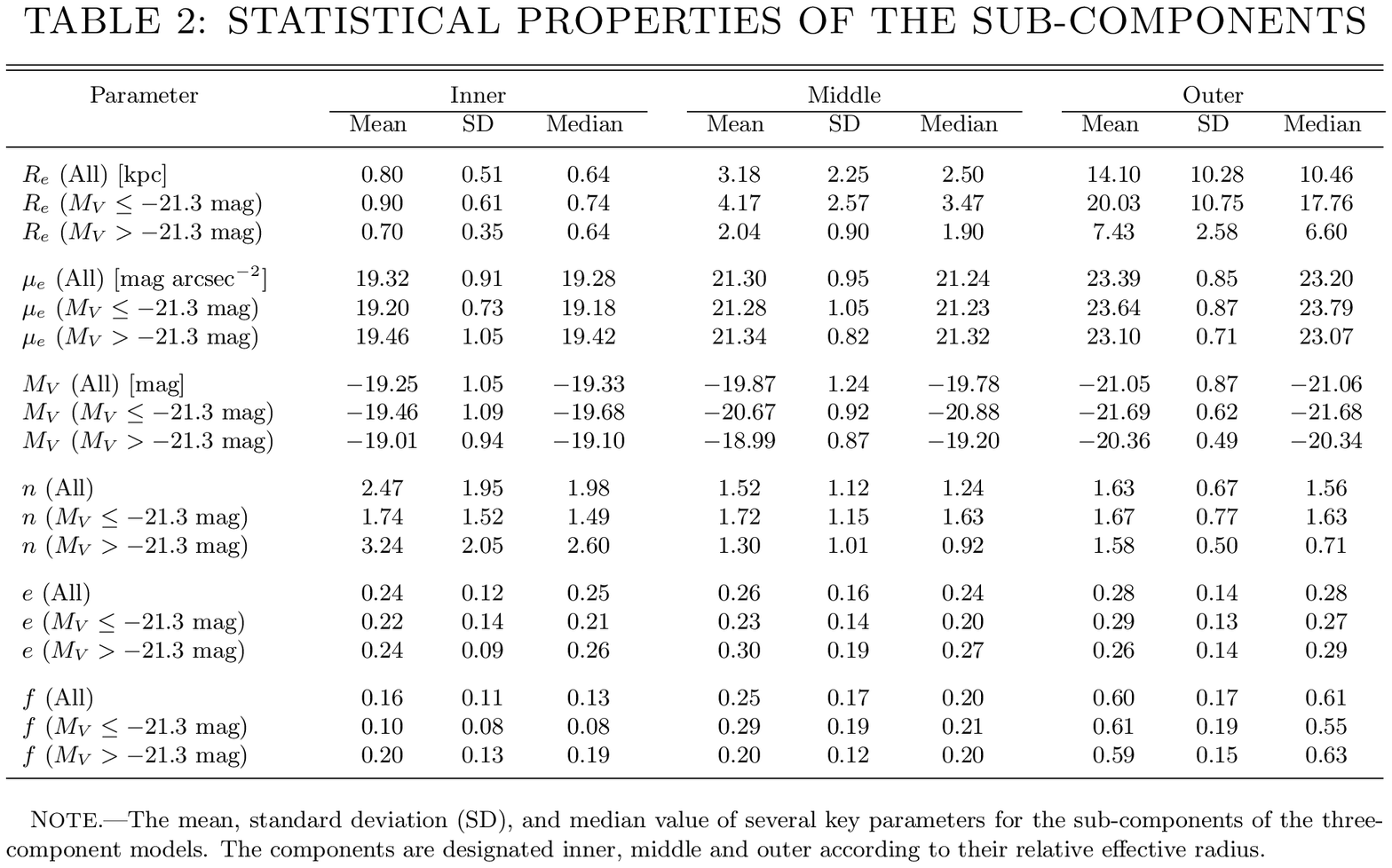,width=18.0cm,angle=0}}
\end{figure*}


\section{The Physical Nature of the Subcomponents}

A large body of the current literature on elliptical galaxies is built on the 
notion that they have a highly concentrated global structure that can be 
well described by a \ser\ profile with a high index.  Historically, this 
impression arose from the common practice of fitting the surface brightness 
distribution using a single component, almost always using azimuthally 
averaged 1-D light profiles.  Abandoning the traditional 1-D fits in favor of 
the more powerful 2-D approach and taking advantage of the improved 
capabilities afforded by the latest version of \galfit, we demonstrate in this 
work that nearby ellipticals, on scales of a few hundred pc to tens of kpc,
can be decomposed into three distinct components on small (\lax 1 kpc), 
intermediate ($\sim$3 kpc), and large ($\sim$10 kpc) scales.  We show that 
these subcomponents obey well-defined scaling relations on the photometric 
fundamental plane and in the stellar mass-size relation.  Below we argue that 
they provide ``fossil records'' that help decode the formation and 
evolutionary history of massive galaxies.

We are by no means the first to recognize that elliptical galaxies contain 
internal complexity or that their global profile cannot be modeled entirely 
using a single analytic function.  The inner regions of ellipticals have long 
been known to depart from the inward extrapolation of the outer global profile 
(Kormendy 1985; Lauer 1985).  Ever since the advent of high-resolution 
observations from the \emph{HST}, these central ``deviations'' have been 
commonly characterized using either a double power-law function (Ferrarese et 
al. 1994) or a ``Nuker'' law (Lauer et al.  1995).  Low-luminosity ellipticals 
typically have steeper central power-law cusps\footnote{K09 advocate changing 
the nomenclature from ``power-law'' to ``extra-light.''} than the shallower 
cores found in high-luminosity systems (Faber et al. 1997; Ravindranath et al. 
2001; Lauer et al. 2007).  Regardless of the exact details, the canonical 
paradigm depicts ellipticals largely as a single-component system, most 
frequently described by a \ser\ function over most of its radial extent.  
Departures from this basic form, if they exist (Graham et al. 2003), only 
enter as perturbations near the center (Trujillo et al. 2004; Ferrarese et al. 
2006; K09).  At the other extreme in scale, studies of brightest cluster 
galaxies have shown that their outer regions often possess extended emission 
in excess of the outward extrapolation of an $r^{1/4}$ law (Schombert 1986) or 
a \ser\ function (Graham et al. 1996; Gonz\'alez et al. 2005;
Donzelli et al. 2011).  However, the prevailing perception has been that these 
extended envelopes are confined to the rarest, most massive ellipticals in 
clusters.  

Here we propose that \emph{most}, perhaps all, massive ellipticals possess 
three principal subcomponents that are physically distinctive.  
Ordinary, run-of-the-mill ellipticals generically contain a compact core, an 
intermediate-scale main body, and an extended outer envelope.  The challenge 
now is to see whether a natural theoretical framework can be found to explain 
this substructure.

\subsection{Inner Component: Evidence of Dissipation}

The innermost compact component identified in our decomposition is directly 
analogous to the central structure revealed in \emph{HST} images.  Despite 
the inferior PSF of our ground-based images, the resolution of the CGS data
suffices to isolate the central component as a photometrically distinct 
entity, even if its detailed shape may be poorly determined.   At \emph{HST}
resolution the slopes of the central profile correlate with galaxy luminosity.
Lower luminosity galaxies possess steep ``extra-light'' or ``power-law'' cusps,
more luminous systems have flatter ``cores,'' and relatively few cases fall 
in between (Gebhardt et al. 1996; Faber et al. 1997; Ravindranath et al. 2001; 
Lauer et al. 2007; K09)\footnote{This issue remains contentious (Ferrarese et 
al. 2006; Glass et al. 2011).}.  Our decomposition, interestingly, recovers 
a qualitatively similar trend.  From Table~2, we find that low-luminosity 
subsample has a median \ser\ index of 2.6, which corresponds to a steeper 
inner profile than $n = 1.5$, the median \ser\ index for the high-luminosity 
members.  The surface brightness distribution of ellipticals at \emph{HST} 
resolution is usually parameterized in terms of a power-law slope.  To compare 
our results with those from previous \emph{HST}-based studies, we generated 
\ser\ profiles with the above median indices and, following Lauer et al. 
(2007),  measured the central logarithmic slopes $\gamma\prime$ within the 
inner 3\asec. For the high-luminosity subsample, the median (mean) 
$\gamma\prime = 0.16$ (0.24), whereas the corresponding value for the 
low-luminosity sources is $\gamma\prime = 0.42$ (0.58).  While the exact value 
of $\gamma\prime$ depends weakly on the chosen fitting region, the conclusion 
that more luminous ellipticals have flatter, less cuspy inner profiles than 
lower luminosity ellipticals is robust.  Indeed, our CGS-derived measurements 
of $\gamma\prime$ are surprisingly close to the typical values for the ``core'' 
and ``cusp'' subsamples from Lauer et al. (2007), although we are not 
confident enough in our ground-based measurements to judge whether the inner 
slopes are truly dichotomous with respect to the galaxy luminosity. It is 
noteworthy that the parameter correlations for the inner component on the 
photometric fundamental plane show very similar behavior (Figure~25) for 
the two luminosity subsamples.  At the same time, the stellar mass-size 
relation of the inner component for lower luminosity ellipticals does exhibit 
slightly larger scatter at the low-mass end (Figure~28). 

A crucial distinction between our results and those of previous studies is 
that we do not necessarily regard the central component as a continuous 
extension or modification of an exterior component.  In our decomposition the 
central substructure is simply a separate component that naturally emerges 
from the fits.  

In a series of works based on hydrodynamical simulations of gas-rich 
galaxy-galaxy mergers, Hopkins et al. (2009a,b,c) presented a two-component 
model to unify the global and central structure of elliptical galaxies.  In 
their picture, the extra-light or power-law component is the relic of a 
dissipational starburst resulting from a gas-rich merger event.  Subsequent 
mergers of these now gas-poor, extra-light ellipticals preserve this central 
dense, compact component, until the newly formed binary black hole scours the 
central density profile into an even flatter, lower density core.  Hopkins et 
al. (2009a,c) fit the 1-D profiles of their simulated galaxies and the 
observational data of a sample of both ``core'' and ``cusp'' elliptical 
galaxies using an inner exponential plus a single, outer \ser\ component.  
Their choice to use an exponential profile for the central component is partly 
due to the limited resolution of their simulations and partly motivated by the 
starburst nature of their model.  The results of our study show that this 
assumption is too restrictive and unnecessary.  Within our low-luminosity 
subsample, the best-fitting \ser\ indices have an average value of $n$ = 3.2, 
a standard deviation of 2.1, and a median value of 2.6.  Forcing the central 
component to a \ser\ index that is lower than it actually should be will bias 
the outer \ser\ component to a higher index.  This 
contributes, in part, to the systematically higher \ser\ indices, compared to 
our values, that Hopkins et al. (2009a) obtained for the main body of the 
galaxy.  The second, more dominant reason why our \ser\ indices are much lower 
is that we have split the main body of the galaxy into two pieces, in our 
nomenclature dubbed the middle and outer components, which individually have 
lower \ser\ indices (median $n = 0.9-1.6$) than when summed into a single
entity.  Apart from this difference, our inner component strongly 
resembles the extra-light component in the Hopkins et al. (2009a) models.  In 
both cases, for instance, the size scale is \lax\ 1 kpc, and the luminosity 
(mass) fraction is $\sim 10\%$.  The inner component of the high-luminosity 
subsample in our study has about a factor of 2 lower luminosity fraction 
(median $f = 0.08$) than in the low-luminosity subsample 
(median $f = 0.19$), qualitatively consistent with the expectation from 
Hopkins et al. (2009c; their Figure 21), but in neither group does the 
luminosity fraction drop with increasing galaxy mass (see Figure~29). 

\subsection{Middle Component: Embedded Disks and Compact Nuggets}

The physical nature of the middle component depends on the galaxy luminosity.  
Although some overlap inevitably arises from our semi-arbitrary division of 
the sample into two luminosity groups, the middle component in low-luminosity 
sources is more heterogeneous and qualitatively different from that in the 
high-luminosity subsample.  Its distribution within the photometric scaling 
relations (Figure~25) shares qualitative and quantitative similarities with 
the loci populated by spheroidals (K09; their Figure~37) and the disks of 
late-type spirals (Kormendy \& Bender 2012; their Figure~20), and its stellar 
mass-size relation (Figure~28) broadly resembles that of late-type (i.e. 
disk-dominated) galaxies.  Additional evidence for the disk-like nature of the 
middle component in low-luminosity sources comes from their near-exponential 
profile (median $n = 0.92$; Table~2), their relatively high ellipticity 
(median $e = 0.27$), and the frequent detection of dust-like small-scale 
features in their residual images.  It is important to recognize that this 
embedded disky component is not a minor perturbation to the overall structure 
of low-luminosity ellipticals.  On average it accounts for $\sim 20\%$ of the 
$V$-band luminosity.  With a typical effective radius of $R_e \approx 2.5$ 
kpc, which for an $n = 1$ \ser\ function translates to a scale length of 
$h=R_e/1.678 = 1.5$ kpc, these disky structures should not be confused with 
nuclear or circumnuclear disks identified through dust absorption or line 
emission in \emph{HST} images (e.g., Harms et al. 1994; Tomita et al. 2000; 
Tran et al. 2001), which are much more compact.  Nor are they equivalent to 
the extended disks familiar in S0 galaxies, which have $h \approx 1-10$ kpc
(Laurikainen et al. 2010), with a median value of 3.2 kpc.  Instead, the 
disk-like component isolated in our study is none other than that responsible 
for producing the disky isophotes that have long been known to be common in 
ellipticals of lower luminosity (e.g., Carter 1978; Lauer 1985; Bender \& 
M\"ollenhoff 1987; Bender 1988; Franx et al. 1989; Peletier et al. 1990).  Our 
decomposition enables this component to be extracted and studied 
quantitatively.  The most straightforward interpretation for these structures 
is that they are remnants of dissipative mergers, along the lines discussed by 
Hopkins et al. (2009a,b,c).

We believe that the middle component in high-luminosity galaxies has a very 
different physical origin.  In terms of gross properties, it is physically 
larger (median $R_e = 3.5$ kpc), rounder ($e = 0.20$), more centrally 
concentrated ($n \approx 1.6$), and occupies $\sim 20$\% of total luminosity.
It delineates a tight sequence in the photometric scaling relations and in the 
stellar mass-size plane.  As discussed in S. Huang et al. 
(in preparation), in many ways the combination of the inner and middle 
components for high-luminosity ellipticals resembles the compact, massive 
galaxies dubbed ``red nuggets'' at high redshifts (e.g., Damjanov et al. 2009).

\subsection{Outer Component: Extended Envelope}

As mentioned earlier, in its most extreme form extended envelopes have long 
been known to exist in brightest cluster galaxies.  The CGS sample, however,
has but two brightest cluster galaxies---NGC~1399 in Fornax and NGC~4696 in the
Centaurus cluster.  Most of our 
ellipticals are indeed relatively luminous, but they are representative of the 
overall local elliptical galaxy population and are in no way biased by extreme 
objects.  Yet, we find that present-day ellipticals almost always contain a 
large-scale, ``puffed up'' outer component.  We detect this component in at 
least 75\% of our sample, essentially independent of luminosity.  The average 
properties of the extended emission in the two luminosity groups are 
qualitatively similar, but differ in detail.  Although the two subsamples 
exhibit very comparable luminosity fractions ($f \approx 0.6$), the 
high-luminosity objects have median $R_e = 17.8$ and $n = 1.6$, to be compared 
with $R_e = 6.6$ and $n = 0.7$ for the low-luminosity sources.  The envelopes 
of the more luminous objects are also systematically more flattened, as 
discussed in Section 3.5 and illustrated in Figure~21.  However, both 
subsamples form a more-or-less 
continuous sequence that sits distinctly offset from the global relations,  
from the sequences defined by the other two subcomponents in the photometric 
scaling relations (Figure~25), and in the stellar mass-size relation 
(Figure~28).  Now, the scatter for the outer component in these diagrams is 
notably larger. Part of this surely can be blamed on the extra uncertainty 
associated with these difficult measurements (representative error bars are 
given in all the plots), but most of the observed increased scatter is 
probably intrinsic. 

Inspection of Figure~18 from Kormendy \& Bender (2012)
reveals something intriguing: the distribution of points for the disks of
S0 galaxies bears an uncanny resemblance to those for the outer component
of our ellipticals.  Is this apparent similarity just a coincidence or is
it telling us something fundamental about common formation physics?

Much recent research draws attention to the dramatic size 
evolution experienced by elliptical galaxies.  Observations (Bezanson et 
al. 2009; van~Dokkum et al. 2010) show that massive early-type galaxies 
typically double in mass and triple in size from $z \approx 1$ to $z = 0$.  
Most of the size growth takes place in the outskirts, through a series of 
minor, dissipationless mergers and hot-phase accretion (Naab et al. 2007, 
2009; Oser et al. 2010, 2012).  In the next paper of this series, we 
critically examine this model in light of our observations.

\section{Future Directions}

The perspective afforded by the multi-component nature of elliptical galaxies
gives us a new window to investigate several other issues related to 
massive galaxies.

\begin{itemize}

\item Many recent studies suggest that the massive galaxy population has 
  undergone strong size evolution since its formation epoch.  The most 
  massive galaxies at $z \approx 2.5$ appear old, dead, and much more compact 
  than present-day galaxies of similar mass (e.g., Cimatti et al. 2004; 
  McCarthy et al. 2004; Bezanson et al. 2009; van~Dokkum et al. 2010).  Minor, 
  dissipationless mergers may govern how these ``red nuggets'' accumulate 
  outer stellar envelopes so that their density profiles match those of local 
  giant ellipticals (Naab et al. 2009; Oser et al. 2010).  We explore this 
  issue in a companion paper (S. Huang et al. in preparation).

\item The mass of supermassive black holes scales tightly with the properties 
  of the spheroidal (bulge) component of the host galaxy, especially its 
  stellar velocity dispersion (Gebhardt et al. 2000; Ferrarese \& Merritt 
  2000), luminosity (Kormendy \& Richstone 1995), and mass (Magorrian et al. 
  1998; H\"aring \& Rix 2004).  The observations of McConnell et al. (2011, 
  2012) suggest that the largest black holes in the most massive local 
  ellipticals may deviate systematically from lower mass objects in their 
  black hole mass-host relations.  In light of the substructure identified in 
  this study, it is interesting to ask whether any of the subcomponents in 
  giant ellipticals actually correlates better with the black hole mass than 
  the entire galaxy as a whole.  To address this issue, we are in the 
  processing of applying our 2-D decomposition technique to deep images of all 
  ellipticals with dynamical detections of black holes.

\item Integral-field spectroscopy has revealed the great diversity of the 
  kinematic properties of the central regions of early-type galaxies, which can 
  be broadly classified as fast or slow rotators (Emsellem et al. 2007; 
  Cappellari et al. 2011a).  At the same time, the velocity structure in the 
  far outer regions of ellipticals is also beginning to be explored with a 
  variety of kinematic tracers (e.g., Coccato et al. 2009; 
  Murphy et al. 2011; Pota et al. 2012; Romanowsky et al. 2012;).
  What is the connection between the kinematic and photometric substructure?

\item Local elliptical galaxies are generally red and have gentle color 
  gradients (e.g., La~Barbera et al. 2010).  If the subcomponents trace 
  distinct episodes or modes of star formation or assembly, they may imprint 
  measurable features in the global color distribution.  An upcoming work will 
  examine this problem using the rest of the CGS multi-band data.

\item Globular cluster systems also carry important clues about the formation 
  and evolutionary processes of their host galaxies.  Bimodal or multi-modal 
  color distributions are frequently observed, as are radial variations in 
  their relative number density (Brodie \& Strader 2006, and references 
  therein).  It would be fruitful to investigate the detailed 2-D spatial 
  distribution of globular clusters to ascertain whether there is any 
  connection between different cluster populations and the structural 
  components identified in this work.

\item Our 2-D decomposition has identified several possible misclassified S0 
  galaxies.  It is interesting to note that a four-component model always 
  provides a reasonably good description of their structure.  We do not 
  discuss the physical properties and nature the subcomponents here, but we 
  will return to this issue in a systematic analysis of all S0 galaxies in CGS. 
      
\end{itemize}


\section{Summary}

We present a comprehensive structural analysis of 94 representative, nearby 
elliptical galaxies spanning a range of environments and stellar masses 
($M_* \approx 10^{10.2}-10^{12.4}$ \solmass).  We use \galfit\ 3.0 to perform 2-D 
multi-component decomposition of relatively deep, moderately high-resolution 
$V$-band images acquired as part of the Carnegie-Irvine Galaxy Survey.  
Extensive experiments were performed to devise an optimal strategy for 
determining the sky level and its uncertainty, which is one of the most 
important factors that influences the decomposition, to explore the 
consequences of PSF blurring in the nuclear regions, and to verify the 
robustness of our fitting strategy.  The final models describe as much of the 
visible structures as possible using the minimum number of components with 
reasonable \ser\ parameters, while simultaneously minimizing the residuals and 
accounting for all the geometric constraints.

Our analysis challenges the conventional notion that the main body of giant 
ellipticals follows a single structure described by a high \ser\ index (e.g., 
$n$ \gax\ $3-4$).  We propose that the global light distribution of the 
majority (\gax\ 75\%) of ellipticals is best described by three \ser\ 
components: a compact, inner core with typical effective radius $R_e$ \lax\ 1 
kpc and luminosity fraction $f \approx 0.1-0.15$; an intermediate-scale, middle 
component with $R_e \approx 2.5$ kpc and $f \approx 0.2-0.25$; and an extended, outer 
envelope with $R_e \approx 10$ kpc and $f \approx 0.6$.  The subcomponents 
have relatively low \ser\ indices, in the range $n \approx 1-2$.  The middle 
components of lower luminosity sources typically exhibit the lowest values of 
$n$, consistent with other evidence (e.g., high ellipticity, disk-like scaling 
relations, dust features) that these systems often contain embedded stellar 
disks.  The outer envelopes of high-luminosity ($M_V \leq -21.3$ mag), high-mass 
($M_*$ \gax\ $10^{11}$ \solmass) objects are notably more flattened than the 
inner portions of the galaxy.

The three principal subcomponents identified in our image decomposition are 
robust features in bright, local elliptical galaxies.  Our belief in their 
physical reality is bolstered by the fact that the individual subcomponents 
share a set of common, relatively homogeneous observational characteristics, 
that their structural parameters obey well-defined (and in some cases rather 
tight) photometric scaling relations, and that they follow distinct tracks on 
the stellar mass-size relation.  Future papers in this series will take 
advantage of the perspective gained from this work to further shed light on 
the formation and evolutionary pathway of massive galaxies.  

\acknowledgements
We thank the anonymous referee for helpful comments that improved this paper.
This work was supported by the Carnegie Institution for Science (LCH), the UC 
Irvine School of Physical Sciences (AJB), the China Scholarship Council (SH, 
Z-YL), and under the National Natural Science Foundation of China under grant
11133001 and 11273015 (SH). SH thanks Prof. Q.-S. Gu and the School of Space
Science and Astronomy in Nanjing University for providing long-term support.
Funding for the SDSS and SDSS-II has been provided by the Alfred P. Sloan 
Foundation, the Participating Institutions, the National Science Foundation, the 
U.S. Department of Energy, the National Aeronautics and Space Administration,
the Japanese Monbukagakusho, the Max Planck Society, and the Higher Education
Funding Council for England. The SDSS Web site is {\tt http://www.sdss.org}.
  

\appendix

\section{A. The PSF and the Red Halo Effect} 

Michard (2002) and Wu et al. (2005) discuss the so-called red halo effect for 
images obtained using thinned CCDs, such as that employed for CGS.  This 
instrumental effect adds an extended, halo-like feature to the wings of the 
PSF. The amplitude of this feature varies significantly with wavelength and is 
particularly pronounced in redder bandpasses (e.g., $I$ band).  The red halo
effect imprints low-level structure on the scale of the extended PSF wings on 
any sufficiently peaked and strong source, including the bright cores of 
galaxies.  Not taking this effect into account in the analysis can potentially 
imprint spurious features on the brightness profiles of galaxies, on angular
scales corresponding to the size of the PSF halo.  Although the effect is most 
noticeable in the color profile, in principle it can also have some low-level 
impact on the surface brightness distribution.  While it is unlikely that 
this PSF effect can strongly influence our image decomposition analysis, it 
would be prudent, if possible, to avoid this problem altogether because our 
main scientific goal hinges on measuring subtle internal substructure in 
elliptical galaxies.

We study the red halo effect using a method similar to that of Michard (2002). 
To create a PSF with sufficiently high signal-to-noise ratio to reveal its
faint, extended wings, it is necessary to create a composite PSF image by 
combining the unsaturated wings of a very bright star with the unsaturated 
core of a fainter star.  We searched for appropriate candidates from the large 
collection of CGS images, selecting candidate stars from regions of the image 
sufficiently far from the main galaxy to be unaffected by it.  Figure~A1 shows 
composite PSFs that extend out to a radius of $\sim 100$\asec, selected from 
three observations obtained under 0\farcs8, 1\farcs2, and 1\farcs8 seeing.   
The PSFs for the four CGS bands have been normalized at a radius of 2\asec.

The red halo effect is immediately obvious in the $I$-band PSF, which has a 
prominent, extended wing that extends from $\sim 10$\asec$-$60\asec.  To first 
order, both the amplitude and the shape of the red halo are independent of the 
seeing and, implicitly, of time because the three observations were taken 
over widely separated observing runs.  However, due to the uncertainty of the 
sky background subtraction and the precision of PSF profile matching, it is 
not clear whether or not we can build a composite PSF model robust enough to 
be 

\vskip 0.3cm
\begin{figure*}[t]
\figurenum{A1}
\centerline{\psfig{file=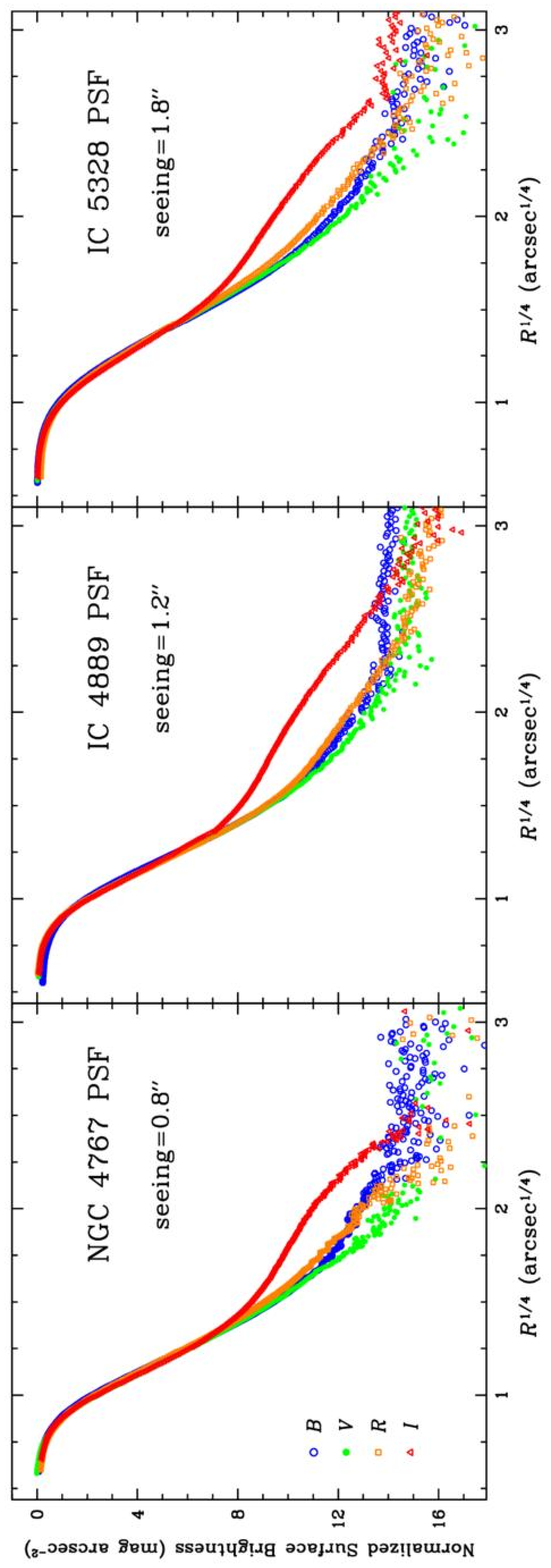,width=18.0cm,angle=270}}
\figcaption[figA1.eps]{
The radial profiles of the combined PSF for three stars (in different 
galaxies) that represent three different seeing conditions. The profiles from 
different bands are shown in different colors and symbols, after being 
normalized at 2\asec.  The red halo effect is clearly seen in the $I$-band PSF.
\label{figA1}}
\end{figure*}
\vskip 0.3cm

\noindent 
applicable to all the $I$-band observations.  For this reason, we decided 
to omit the $I$-band images from the current analysis.  No clear evidence of a
similar halo effect is seen in the PSFs of the other three filters.  Both 
the $B$-band and $R$-band PSFs show only a mildly extended wing, which seems 
altogether absent in the $V$ band.  We have thus chosen to focus only on 
the $V$-band images for the current analysis.

\section{B. Sky Background Estimation}

We estimate the level of the sky background by performing a multi-component 
2-D fit to the entire image.  The galaxy is modeled with a series of \ser\ 
components (up to five), and the sky is included as a free parameter to 
be determined by the fit.  After examining the suite of acceptable models, we 
choose as the final sky level the average of the sky values from all the 
acceptable fits. 

An important factor that affects the accuracy with which the sky background 
can be estimated is the size of the image relative to the size of the galaxy.
This is especially critical for elliptical galaxies whose outer light profile
is usually very extended.  When the galaxy extent occupies a significant 
fraction of the image, the sky determination will be strongly model-dependent.
This problem has been tested by Yoon et al. (2011; see their Figure~23) using 
simulated galaxy images and single-component \ser\ models.  These authors 
performed a comprehensive statistical analysis to quantify the dependence of 
the uncertainty of the sky background on the relative image size (in units of 
$R_{50}$). The uncertainty increases with decreasing relative image size, 
especially when $R_{\rm img}$ \lax\ $5 R_{50}$. 

Different from Yoon et al. (2011), we propose that in real images the level of 
the sky uncertainty should be directly compared to the intrinsic fluctuation 
of the background because the sky background is usually not flat, which 
translates into a measurement uncertainty. To measure the background 
fluctuation, we apply object masks and an aggressive mask for the main galaxy 
to the image and select a large number of small, unmasked, non-overlapping, 
box regions. As mentioned above, the sky is not flat and has fluctuations that 
cannot be masked out perfectly. In light of the non-flatness of the sky, each 
box serves as an independent measurement of the true sky.  Under that 
assumption, the effective uncertainty of the sky background is the RMS over 
all the measurements divided by the number of boxes: $\sigma_{\rm sky,eff} 
\approx \sigma_{\rm box} / \sqrt{n_{\rm box}}$, where $\sigma_{\rm box}$ is 
the standard deviation of the mean pixel values from all the boxes and 
$n_{\rm box}$ is the number of useful boxes.

Factors such as the size of the boxes, the size of the mask for the central 
galaxy, and the number of boxes can affect these measurements.  We test the 
impact of different choices.  In the end, we choose to use $1.8 R_{80}$ as 
the size for the galaxy mask, with an elliptical shape fixed to the axis ratio 
and position angle given in Paper~I. This value was selected based on a series 
of tests to ensure that the mask covers the most visible parts of the galaxy and 
leaves enough space for background estimation at the same time. We generate 120 
random boxes for each image and then remove the ones that overlap; the final 
number of useful boxes is always around 100.  When $R_{\rm img} \ge 6 R_{50}$, 
we choose the box size to be $30 \times 30$ pixels; for smaller relative image 
sizes we pick a box size of $20 \times 20$ pixels to ensure a comparable number
of boxes for both large and small images. 

The sky estimates based on the technique proposed here should 
converge with the model-independent values in the limit when the image is 
large enough to permit the sky to be measured directly. We test this using a 
subsample of 30 ellipticals that satisfy $R_{\rm img} > 10 R_{50}$, for which 
direct sky values are available from Paper~I.  We artificially reduce the 
relative image size by keeping the galaxy fixed at the center and progressively
cropping the outer regions of the image to produce simulated images where 
$R_{\rm img} = 10, 8, 6, 5$ and $4 R_{50}$.  For each of these modified 
images, we fit models with 1 to 5 \ser\ components plus a sky background 
term.   As our purpose here is to achieve the most realistic representation 
of the galaxy, the different \ser\ components are allowed to have different 
centers, and the $m=1$ Fourier mode is added to each of them. 

It was immediately apparent from these tests that a single-component model is 
inadequate to recover a reliable sky background. The sky value from the 
single-component models deviates systematically from the true value, even when the 
image is very large.  This just reflects the fact that a single \ser\ function 
gives a poor description of the light distribution of elliptical galaxies. For 
multi-component models, as shown in Figure~B1, the scatter of the sky 
uncertainty strongly increases with decreasing relative image size 
($R_{\rm img}/R_{50}$). However, for sufficiently large images multi-component 
models can recover the intrinsic sky value with good accuracy.  It is 
reassuring that there is little systematic offset even when the image is 
very small; the median uncertainties of the sky are all very close to zero, for 
each image size bin and for models with different numbers of components.
Careful examination reveals that the handful of extreme outliers in Figure~B1
can be traced to models that are clearly incorrect.  For example, when the 
effective radius or \ser\ index of one of the components is extremely large 
(e.g., $R_e > R_{\rm img}$ or $n > 10$), the sky value tends to be severely 
underestimated.  This betrays an important disadvantage of our model-dependent 
method of sky estimation.  Nevertheless, this weakness can be overcome by 
exploring a wide suite of models to seek out the subset that produces a 
consistent, stable solution for the sky background. 

\vskip 0.3cm
\begin{figure*}[tb]
\figurenum{B1}
\centerline{\psfig{file=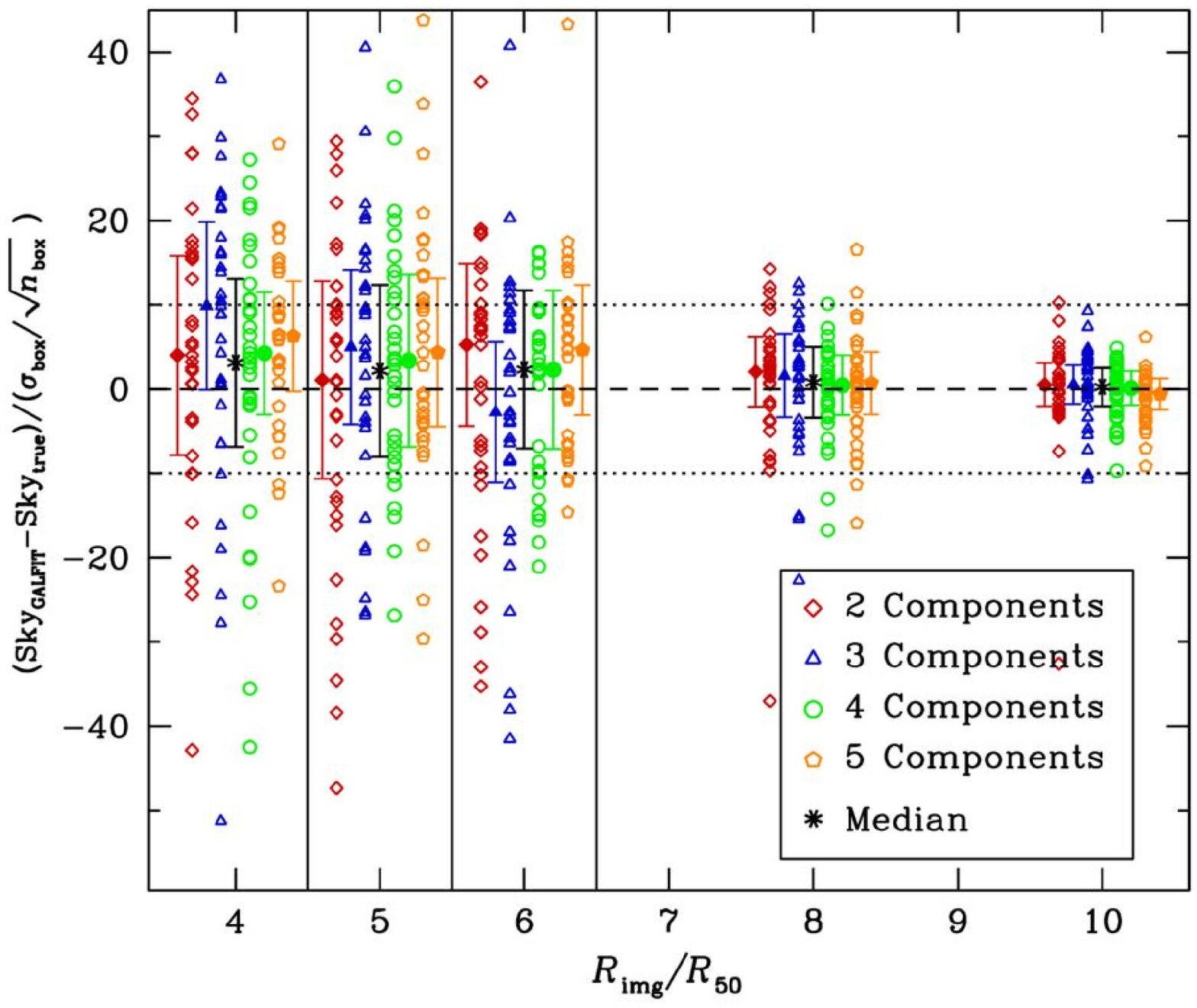,width=11.0cm}}
\figcaption[figB1.eps]{
The dependence of the relative uncertainty of the sky value, 
expressed in units of the background fluctuation (denoted by 
$\sigma_{\rm box}/\sqrt{n_{\rm box}}$; see Appendix~A), on image size 
($R_{\rm img}$), normalized to the half-light radius of the galaxy 
(${\it R}_{50}$).  The uncertainty on the sky is empirically defined to be the 
difference between the model-dependent value obtained from our multi-component 
2-D fit (${\rm Sky}_{\rm GALFIT}$) and the model-independent, presumed true 
value (${\rm Sky}_{\rm true}$) directly measured from the outer regions of 30 
angularly small galaxies.  The open colored symbols denote models with 
different numbers of components ($2-5$); their median values and corresponding 
scatter are given by the solid symbols with error bars.  For clarity, the 
different models are offset slightly along the X-axis.  We analyze simulated
images in five bins of relative image size: ${\it R}_{\rm img}/{\it R}_{50}$ = 
4, 5, 6, 8, and 10.  
The black asterisks (and corresponding error bars) represent the median value
(and its scatter) of the four different models in each size bin.  The black 
horizontal dash line and the flanking dotted lines mark 
$({\rm Sky}_{\rm GALFIT}-{\rm Sky}_{\rm true})/(\sigma_{\rm box}/
\sqrt{n_{\rm box}}) = 0.0\pm10$.
\label{figB1}}
\end{figure*}
\vskip 0.1cm

We apply this new procedure to the $V$-band images of the entire elliptical 
galaxy sample to obtain a uniform estimate of their sky background.  Guided 
by the above tests, we no longer consider the single-component \ser\ models.  
The final sky value of each galaxy is taken to be the average of the best-fit 
sky values of all acceptable models, after excluding problematic cases, if any.

Since we express the average level of uncertainty of the sky in units of the 
background fluctuation (Figure~B1), the corresponding uncertainty on the sky for 
each galaxy, in units of counts pixel$^{-1}$, can be obtained by simply 
multiplying the measured background fluctuation in the image with the average 
sky calibration uncertainty for its image size bin.  To obtain the sky 
uncertainty, we use both the background fluctuation measured for the individual 
galaxy and the average value for all ellipticals in the same image size bin.
As shown in Figure~B2, the resulting sky uncertainties agree well for both 
choices of background fluctuation values.  The expected trend with relative 
image size is also very clear: the sky uncertainty increases dramatically with 
decreasing image size, especially when $R_{\rm img}$ \lax\ $6 R_{50}$. 

Figure~B3 compares the new sky estimates and uncertainties with those 
originally calculated in Paper~I.  For the sky values, the difference between 
the new and old values shows no trends with the relative image size of the 
galaxy.  However, the old values of the sky uncertainty for small galaxies 
($R_{\rm img}$ \lax\ $6 R_{50}$) are on average systematically lower than the 
new values, by $\sim$1.9 counts pixel$^{-1}$; for large galaxies, the trend is 
reversed, such that the original sky uncertainties are systematically 
overestimated (by approximately the same magnitude). 

\section{C. External Checks with SDSS}

As in Paper~II, we performed external checks of our 2-D fits using images from 
the SDSS.  The spatial resolution, seeing conditions, and photometric depth of 
the SDSS data are different from those of CGS.  Six of the CGS ellipticals are 
in the SDSS DR8 catalog (Aihara et al. 2011). We retrieved their $g$-band 
images, the closest analogs to the CGS $V$-band data, from SDSS DR8 Data 
Access\footnote{{\tt http://www.sdss3.org/dr8/data\_access.php}}. 
The SDSS images are already sky-subtracted.  We created objects masks and 
empirical PSF images using the same method for 

\vskip 0.1cm
\begin{figure*}
\figurenum{B2}
\centerline{\psfig{file=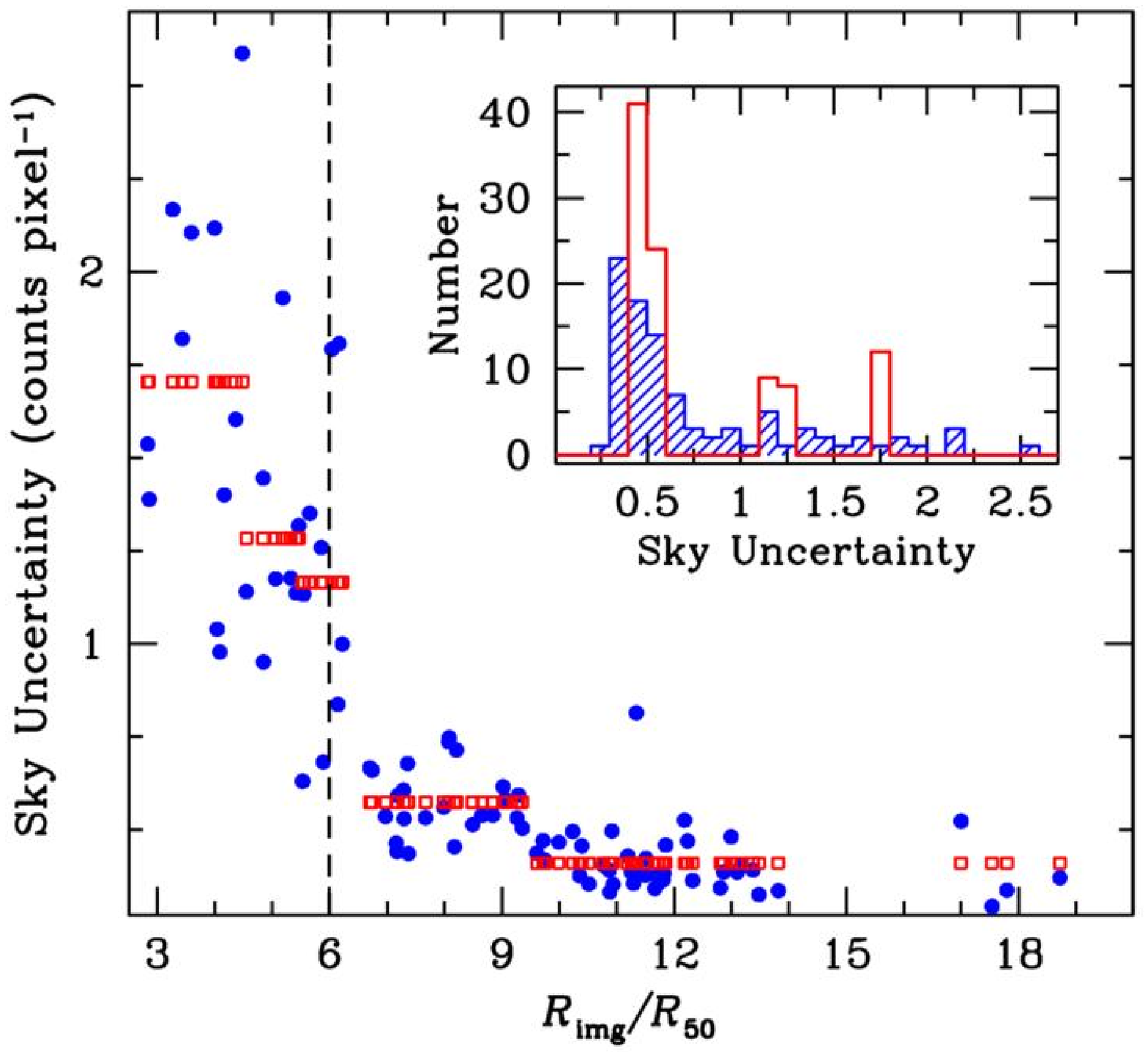,width=8.75cm}}
\figcaption[figB2.eps]{
The relation between the new sky uncertainties and the relative image 
size.  The sky uncertainties are calculated using the background fluctuation 
measured for each individual galaxy (blue dots) and the average background
fluctuation in each image size bin (red squares). The distribution of the 
uncertainties are plotted in the inset figure; hatched blue histograms 
correspond to the individually measured background fluctuations, whereas the
dashed red histograms denote the average values based on image size.  The
vertical dashed line marks ${\it R}_{\rm img}/{\it R}_{50}=6$, below which 
the sky uncertainties increase dramatically.
\label{figB2}}
\end{figure*}

\begin{figure*}
\vskip 0.3cm
\figurenum{B3}
\centerline{\psfig{file=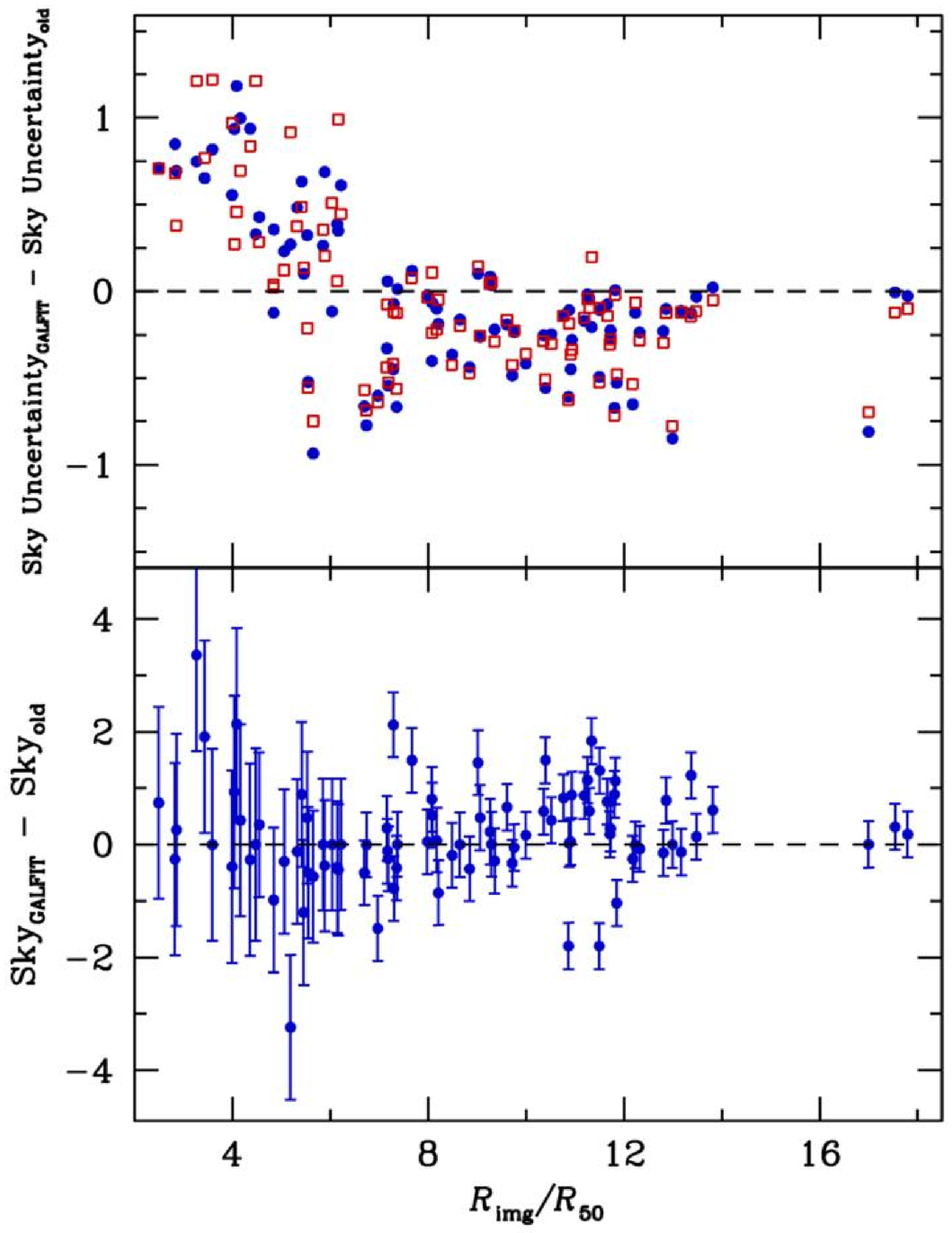,width=8.75cm}}
\figcaption[figB3.eps]{
Comparison of the new, GALFIT-based sky values and uncertainties with 
the old values given in Paper~I, plotted as a function of relative image size, 
${\it R}_{\rm img}/{\it R}_{50}$. The unit for both sky values and their 
uncertainties are counts. The top panel shows sky uncertainties calculated 
using the background fluctuation measured for each individual galaxy 
(blue dots) and the average background fluctuation in each image size bin 
(red squares).  The error bars in the bottom panel reflect the typical sky 
uncertainty, calculated following our new method.
\label{figB3}}
\end{figure*}
\vskip 0.3cm

\vskip 0.3cm
\figurenum{C1}
\centerline{\psfig{file=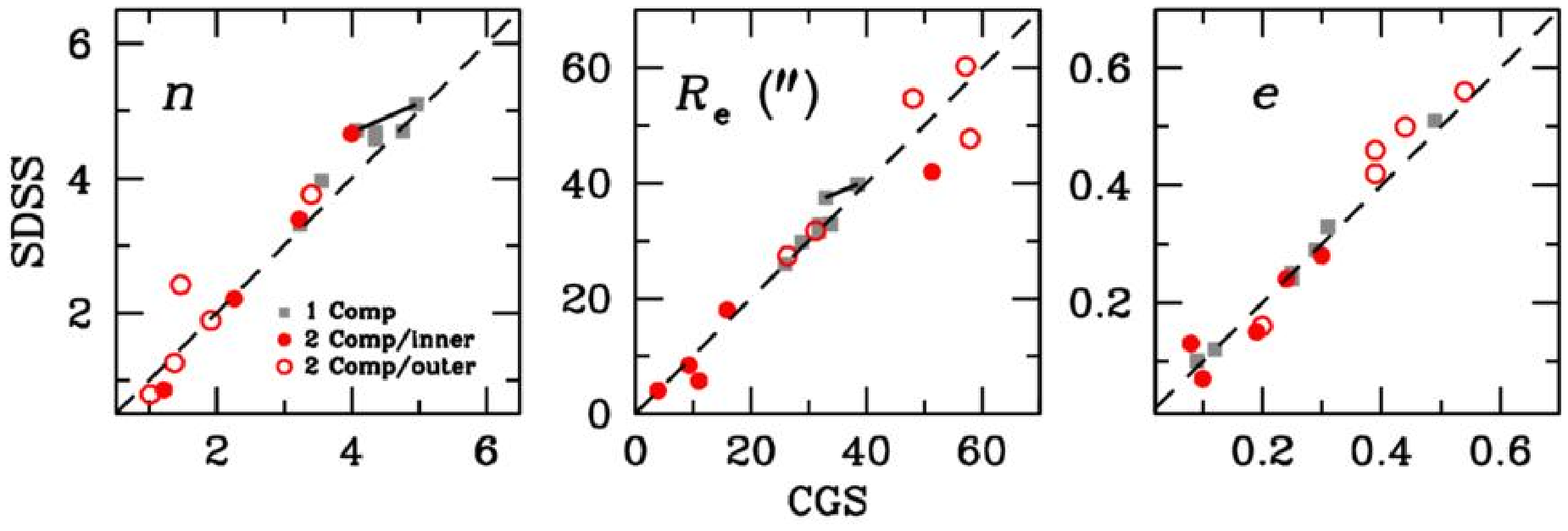,width=17.0cm}}
\figcaption[figC1.eps]{
Comparison of one- and two-component models for six CGS galaxies 
that are in common with SDSS.  The three panels show the fitted values of 
the \ser\ index, effective radius, and axis ratio; the dashed line indicates 
equality. Grey filled pentagons are from the one-component fits, and red 
filled circles and blue open circles represent the inner and outer components 
of the two-component fits, respectively.   The two grey filled pentagons 
connected by a solid line are for NGC~1199; they represent fits
with and without its central dust lane masked out.
\label{figC1}}
\vskip 0.3cm

\noindent
CGS.  We fit the SDSS images 
with one- and two-component \ser\ models, starting with different initial 
parameters but otherwise following our standard procedures for CGS.  We 
refrain from fitting three-component models to the SDSS data because they have
lower resolution.  Despite the very different characteristics of the two data 
sets, we find that the fits produce very similar and consistent results, both in 
terms of the image residuals and in recovering basic parameters, as illustrated
in Figure~C1.

\section{D. Central Mask}

To decide on the appropriate size for the central mask, a series of 
single-component \ser\ models is fit to each elliptical galaxy, adopting central 
masks with different sizes ranging from 0.5 to 8 times the FWHM of the PSF. The 
result is shown in Figure~D1, where the relative change of the \ser\ index is 
plotted against the size of the central mask. It is apparent that for most 
elliptical galaxies the change of the \ser\ index is on average much weaker for 
radii beyond $2 \times$FWHM, indicating a transition to the regime outside of 
which the seeing disk has much reduced influence.  Based on this, we adopt a 
radius of $2 \times$FWHM for the central mask.  It is also interesting to note 
that different galaxies respond differently to the central mask, and that the 
\ser\ index ``profile'' for a large fraction of the objects is not flat 
outside the seeing disk. Both of these phenomena point to the intrinsic 
complexity of the central regions for nearby elliptical galaxies.

\vskip 0.3cm
\figurenum{D1}
\centerline{\psfig{file=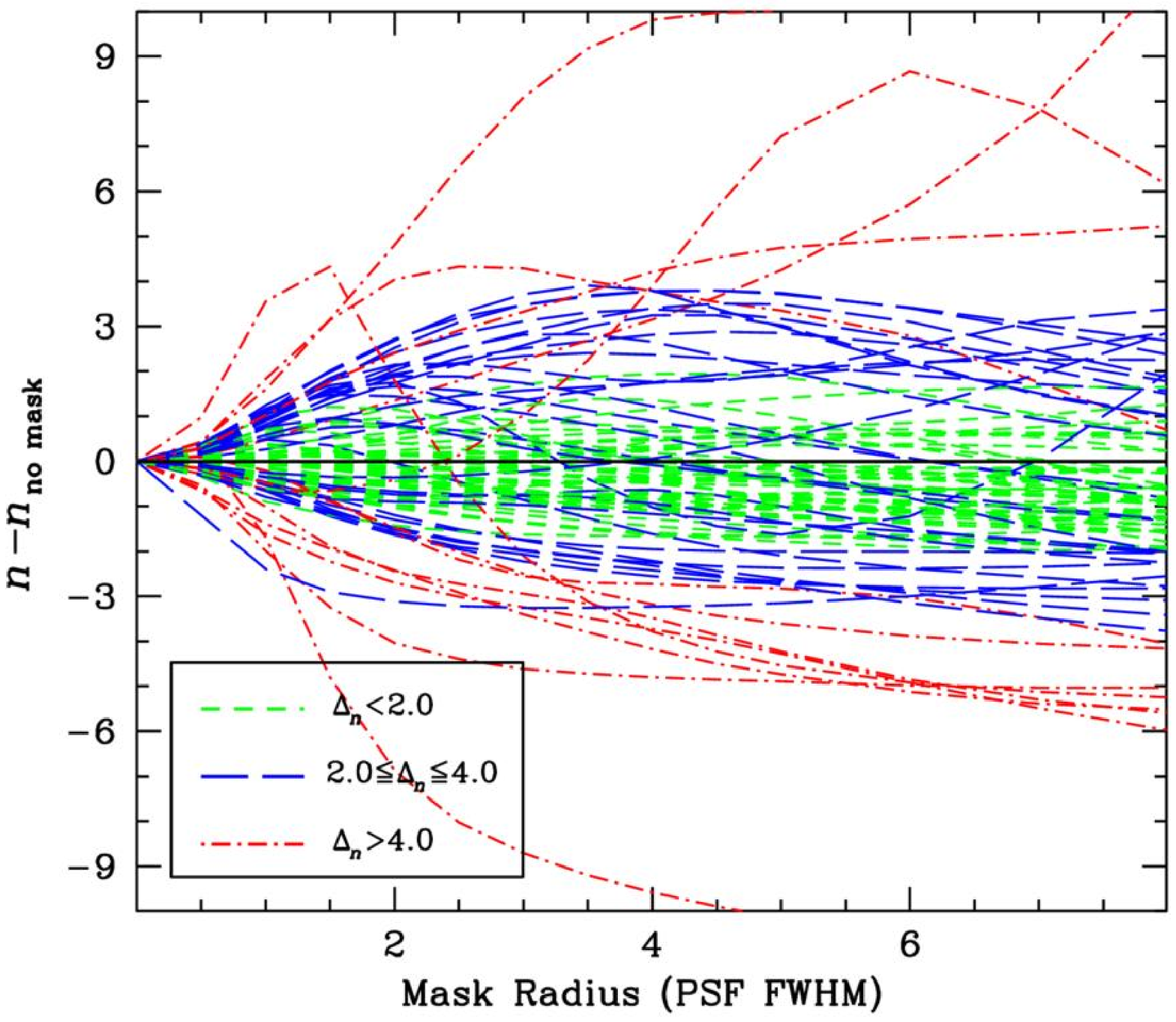,width=11.0cm}}
\figcaption[figD1.eps]{
The effect of a central mask on the best-fitting \ser\ index for 
single-component fits of the elliptical galaxy sample.  Each object is 
fit with circular masks of different radii, ranging from 0.8 to 8 times 
the PSF FWHM.  While large systematic differences exist between models 
with and without a central mask, in general the fits are stable for masks with 
radii \gax 2 times FWHM of the PSF.
\label{figD1}}
\vskip 0.3cm

\section{E. Atlas of Final Fits}

Figures~E1--E94 present the atlas of the final fits for the 94 elliptical 
galaxies in CGS.  The format and labeling conventions are the same as those
in Figures 5 and 6.  The right panels
display, from top to bottom, images of the original data, the best-fit model,
and the residuals; each panel has a dimension of $2 R_{50} \times 2 R_{50}$ and 
is centered on the nucleus.  The left panels compare the 1-D profiles of the 
data and the model; from top to bottom, we show the isophotal shape parameters
$A_4$ and $B_4$, $e$, PA, $\mu$, and the residuals between the data and the 
model.  The observed $\mu$ profile is extracted with $e$ and PA fixed to the 
average value of the galaxy (Paper~II), which is indicated by the dotted line 
in the $e$ and PA panels.  Individual subcomponents, each extracted using the 
geometric parameters generated by {\tt GALFIT}, are plotted with different line 
types and colors; the PSF is plotted with arbitrary amplitude.  Colored squares 
on the $e$, PA, and $\mu$ plots mark the ellipticity, position angle, and effective
surface brightness of each component.  The same color scheme is used in the
greyscale images, where the ellipses trace $R_{50}$, $e$, and PA of each
component.  In the residual plot, the three black dashed lines indicate the
position of zero residuals and the level of photometric error.  We only 
display three sample pages for illustration; the full set of figures is 
available in the electronic version of the paper, as well as on the project 
Web site {\tt http://cgs.obs.carnegiescience.edu}.

\begin{figure*}[t]
\figurenum{E1}
\centerline{\psfig{file=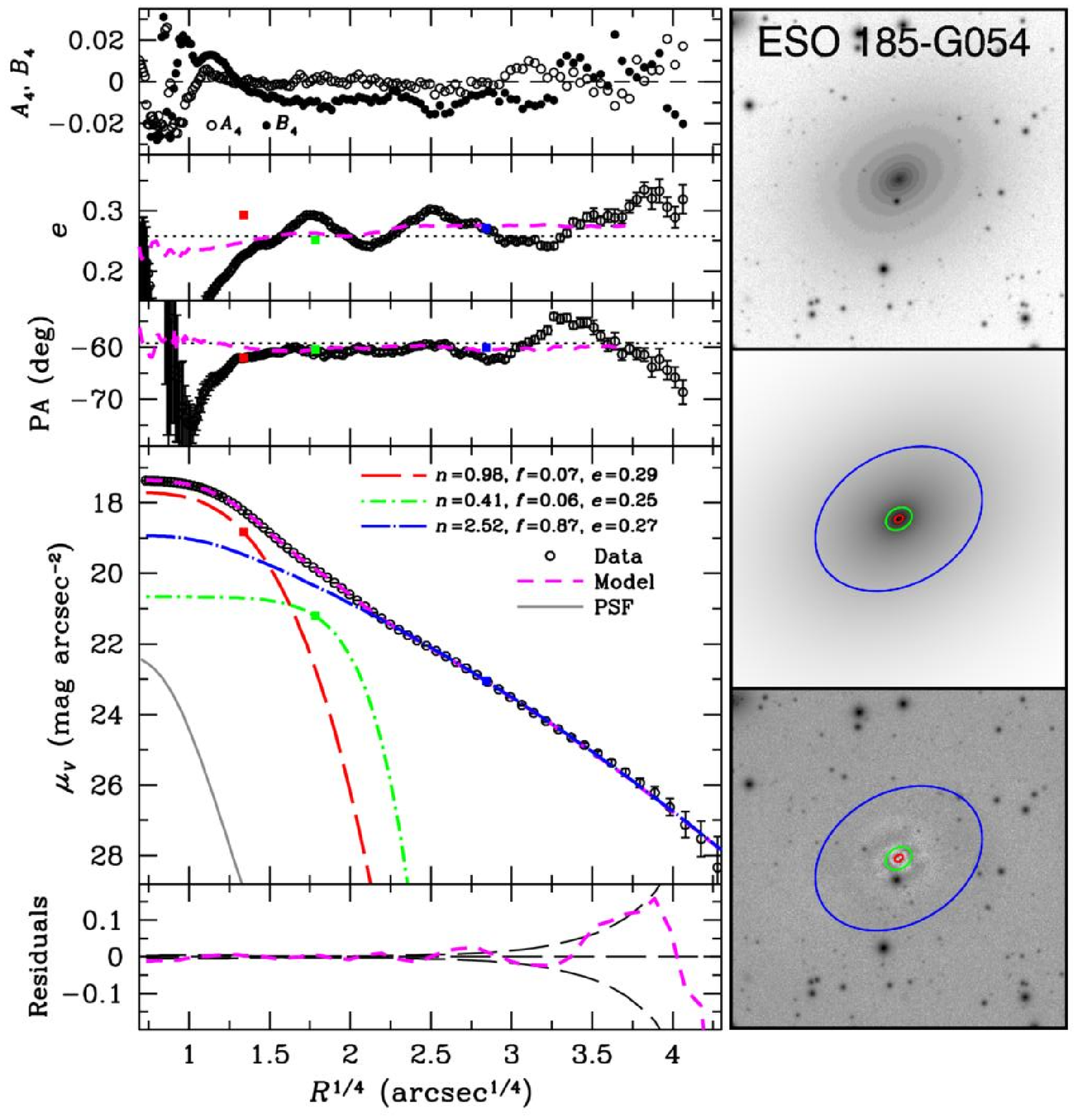,width=18.5cm,angle=270}}
\figcaption[figE1.eps]{Best-fit model of ESO~185-G054. See Figure 6 for details.
\label{figE1}}
\end{figure*}
\vskip 0.3cm

\begin{figure*}[t]
\figurenum{E2}
\centerline{\psfig{file=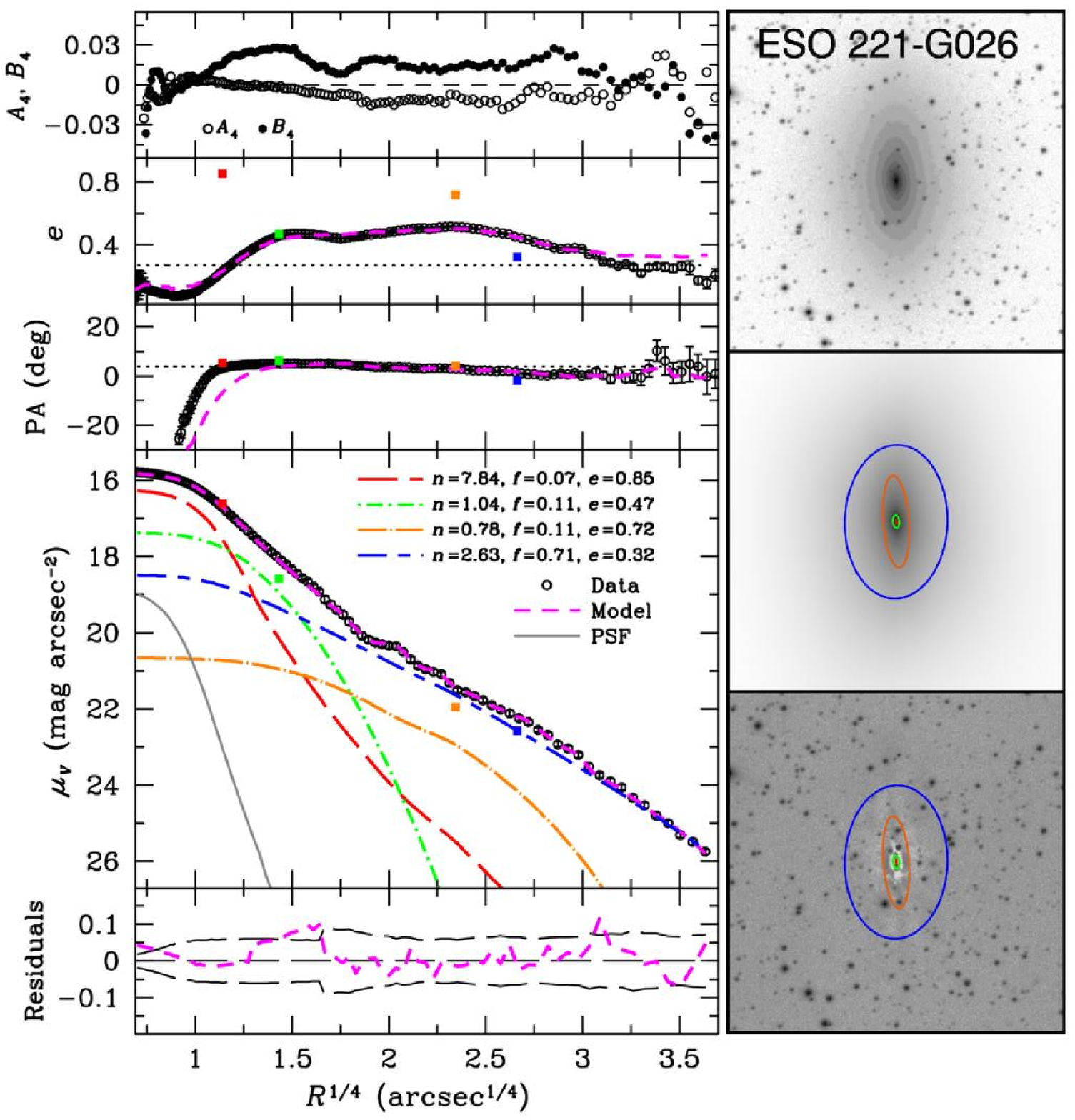,width=18.5cm,angle=270}}
\figcaption[figE2.eps]{Best-fit model of ESO~221-G026. See Figure 6 for details.
\label{figE2}}
\end{figure*}
\vskip 0.3cm

\begin{figure*}[t]
\figurenum{E3}
\centerline{\psfig{file=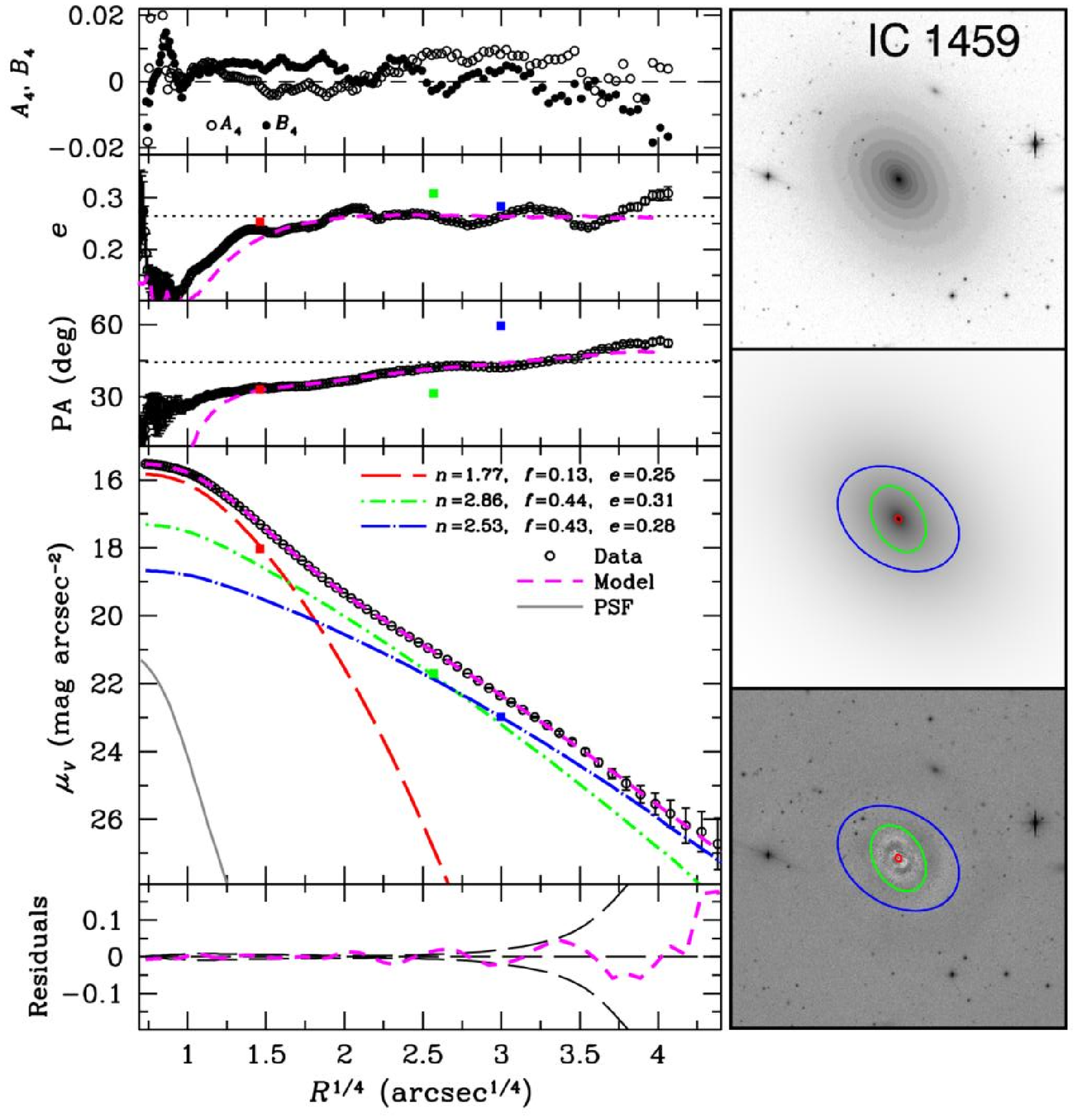,width=18.5cm,angle=270}}
\figcaption[figE3.eps]{Best-fit model of IC~1459. See Figure 6 for details.
\label{figE3}}
\end{figure*}
\vskip 0.3cm

\end{CJK*}

\end{document}